\documentclass[a4paper,11pt]{article}
\usepackage{amsmath,amsfonts,amssymb,graphicx,bezier,citesort}
\newcommand{\nc}{\newcommand}
\nc{\rnc}{\renewcommand}
\nc{\nn}{\nonumber}
\nc{\ch}{{\rm{ch}}}
\nc{\sh}{{\rm{sh}}}
\rnc{\Im}{{\rm{Im}\,}}
\rnc{\Re}{{\rm{Re}\,}}
\nc{\tnh}{{\rm{th}}}
\nc{\mfa}{{\mathfrak{a}}}
\nc{\mfab}{\overline{\mfa}}
\nc{\mfA}{{\mathfrak{A}}}
\nc{\mfAb}{\overline{\mfA}}
\nc{\thetab}{\overline{\theta}}
\nc{\mfsl}{{\mathfrak{sl}}}
\nc{\db}{\displaybreak[0]\\}
\nc{\sotimes}{\mathop{\otimes}_{s}}
\nc{\cl}{{C}}
\nc{\mcl}{C}
\numberwithin{equation}{section}
\def\contourgr{
 \begin{picture}(95,70)
 \put(5,58){(a)}
 \put(5,30){\vector(1,0){80}}
 \put(45,5){\vector(0,1){50}}
 \put(7.5,29){\line(0,1){2}}
 \put(82.5,29){\line(0,1){2}}
 \put(86,30){{\small $\Re v$}}
 \put(42,57){{\small $\Im v$}}
 \put(5.,24){{\small $-\frac{\pi}{2}$}}
 \put(81,24){{\small $\frac{\pi}{2}$}}
 \put(20,20){\circle*{1.5}}
 \put(25,25){\circle{1.5}}
 \thicklines
 \put(7.5,30){\line(1,0){15}}
 \put(22.5,30){\line(0,-1){5}}
 \put(25,25){\oval(5,5)[b]}
 \put(27.5,30){\line(0,-1){5}}
 \put(27.5,30){\vector(1,0){55}}
 \put(7.5,15){\line(1,0){10}}
 \put(17.5,15){\line(0,1){5}}
 \put(20,20){\oval(5,5)[t]}
 \put(22.5,20){\line(0,-1){5}}
 \put(22.5,15){\vector(1,0){60}}
 \put(7.5,45){\vector(1,0){75}}
 \put(77,6.9){\scriptsize{$\circ$: $\theta_1$}}
 \put(77,2.0){\scriptsize{$\bullet$: $\theta_0$}}
 \put(50,17){\tiny{$\cl_0$}}
 \put(50,32){\tiny{$\cl_1$}}
 \put(50,47){\tiny{$\overline{\cl}_0$}}
 \end{picture}}
\def\contourst{
 \begin{picture}(95,70)
 \put(5,30){\vector(1,0){80}}
 \put(45,5){\vector(0,1){50}}
 \put(7.5,29){\line(0,1){2}}
 \put(82.5,29){\line(0,1){2}}
 \put(86,30){{\small $\Re v$}}
 \put(42,57){{\small $\Im v$}}
 \put(5.,24){{\small $-\frac{\pi}{2}$}}
 \put(81,24){{\small $\frac{\pi}{2}$}}
 \put(20,40){\makebox(0,0){\rule{1.2mm}{1.2mm}}}
 \put(24.25,34.25){\framebox(1.5,1.5){}}
 \put(70,20){\circle*{1.5}}
 \put(65,25){\circle{1.5}}
 \thicklines
 \put(7.5,30){\line(1,0){15}}
 \put(22.5,30){\line(0,1){5}}
 \put(25,35){\oval(5,5)[t]}
 \put(27.5,30){\line(0,1){5}}
 \put(27.5,30){\line(1,0){35}}
 \put(65,25){\oval(5,5)[b]}
 \put(62.5,30){\line(0,-1){5}}
 \put(67.5,30){\line(0,-1){5}}
 \put(67.5,30){\vector(1,0){15}}
 \put(7.5,15){\line(1,0){60}}
 \put(67.5,15){\line(0,1){5}}
 \put(70,20){\oval(5,5)[t]}
 \put(72.5,20){\line(0,-1){5}}
 \put(72.5,15){\vector(1,0){10}}
 \put(7.5,45){\line(1,0){10}}
 \put(17.5,45){\line(0,-1){5}}
 \put(20,40){\oval(5,5)[b]}
 \put(22.5,40){\line(0,1){5}}
 \put(22.5,45){\vector(1,0){60}}
 \put(61,6.9){\scriptsize{$\circ$: $\theta_1$}}
 \put(61,2.0){\scriptsize{$\bullet$: $\theta_0$}}
 \put(72,6.9){\scriptsize{$\square$: $-\theta_1$}}
 \put(72,2.0){\scriptsize{$\blacksquare$: $-\theta_0$}}
 \put(50,17){\tiny{$\cl_0$}}
 \put(50,32){\tiny{$\cl_1$}}
 \put(50,47){\tiny{$\overline{\cl}_0$}}
 \end{picture}}
\def\contourdd{
 \begin{picture}(95,70)
 \put(5,58){(a)}
 \put(5,30){\vector(1,0){80}}
 \put(45,5){\vector(0,1){50}}
 \put(7.5,29){\line(0,1){2}}
 \put(82.5,29){\line(0,1){2}}
 \put(86,30){{\small $\Re v$}}
 \put(42,57){{\small $\Im v$}}
 \put(5.,24){{\small $-\frac{\pi}{2}$}}
 \put(81,24){{\small $\frac{\pi}{2}$}}
 \put(20,40){\makebox(0,0){\rule{1.2mm}{1.2mm}}}
 \put(24.25,34.25){\framebox(1.5,1.5){}}
 \put(70,20){\circle*{1.5}}
 \put(65,25){\circle{1.5}}
 \thicklines
 \put(7.5,30){\line(1,0){15}}
 \put(22.5,30){\line(0,1){5}}
 \put(25,35){\oval(5,5)[t]}
 \put(27.5,30){\line(0,1){5}}
 \put(27.5,30){\line(1,0){35}}
 \put(65,25){\oval(5,5)[b]}
 \put(62.5,30){\line(0,-1){5}}
 \put(67.5,30){\line(0,-1){5}}
 \put(67.5,30){\vector(1,0){15}}
 \put(7.5,15){\vector(1,0){75}}
 \put(7.5,45){\vector(1,0){75}}
 \put(51,6.9){\scriptsize{$\circ$: $\theta_1$}}
 \put(51,2.0){\scriptsize{$\bullet$: $\theta_0$}}
 \put(62,6.9){\scriptsize{$\square$: $-\theta_1$}}
 \put(62,2.0){\scriptsize{$\blacksquare$: $-\theta_0$}}
 \put(50,17){\tiny{$\cl_0$}}
 \put(50,32){\tiny{$\cl_1$}}
 \put(50,47){\tiny{$\overline{\cl}_0$}}
 \end{picture}}
\def\contourspt{
 \begin{picture}(95,70)
 \put(5,58){(b)}
 \put(5,30){\vector(1,0){80}}
 \put(45,5){\vector(0,1){50}}
 \put(7.5,29){\line(0,1){2}}
 \put(82.5,29){\line(0,1){2}}
 \put(86,30){{\small $\Re v$}}
 \put(42,57){{\small $\Im v$}}
 \put(5.,24){{\small $-\frac{\pi}{2}$}}
 \put(81,24){{\small $\frac{\pi}{2}$}}
 \put(20,20){\makebox(0,0){\rule{1.2mm}{1.2mm}}}
 \put(24.25,24.25){\framebox(1.5,1.5){}}
 \put(70,20){\circle*{1.5}}
 \put(65,25){\circle{1.5}}
 \thicklines
 \put(7.5,30){\line(1,0){15}}
 \put(22.5,30){\line(0,-1){5}}
 \put(25,25){\oval(5,5)[b]}
 \put(27.5,25){\line(0,1){5}}
 \put(27.5,30){\line(1,0){35}}
 \put(65,25){\oval(5,5)[b]}
 \put(62.5,30){\line(0,-1){5}}
 \put(67.5,30){\line(0,-1){5}}
 \put(67.5,30){\vector(1,0){15}}
 \put(7.5,15){\line(1,0){10}}
 \put(17.5,15){\line(0,1){5}}
 \put(20,20){\oval(5,5)[t]}
 \put(22.5,20){\line(0,-1){5}}
 \put(22.5,15){\line(1,0){45}}
 \put(67.5,15){\line(0,1){5}}
 \put(70,20){\oval(5,5)[t]}
 \put(72.5,20){\line(0,-1){5}}
 \put(72.5,15){\vector(1,0){10}}
 \put(7.5,45){\vector(1,0){75}}
 \put(61,6.9){\scriptsize{$\circ$: $\theta_1$}}
 \put(61,2.0){\scriptsize{$\bullet$: $\theta_0$}}
 \put(72,6.9){\scriptsize{$\square$: $-\theta^{\ast}_1$}}
 \put(72,2.0){\scriptsize{$\blacksquare$: $-\theta^{\ast}_0$}}
 \put(50,17){\tiny{$\cl_0$}}
 \put(50,32){\tiny{$\cl_1$}}
 \put(50,47){\tiny{$\overline{\cl}_0$}}
 \end{picture}}
\def\contoursl{
 \begin{picture}(95,70)
 \put(5,58){(b)}
 \put(5,30){\vector(1,0){80}}
 \put(45,5){\vector(0,1){50}}
 \put(7.5,29){\line(0,1){2}}
 \put(82.5,29){\line(0,1){2}}
 \put(86,30){{\small $\Re v$}}
 \put(42,57){{\small $\Im v$}}
 \put(5.,24){{\small $-\frac{\pi}{2}$}}
 \put(81,24){{\small $\frac{\pi}{2}$}}
 \put(20,40){\makebox(0,0){\rule{1.2mm}{1.2mm}}}
 \put(24.25,34.25){\framebox(1.5,1.5){}}
 \put(20,20){\circle*{1.5}}
 \put(25,25){\circle{1.5}}
 \thicklines
 \put(7.5,30){\line(1,0){15}}
 \put(22.5,30){\line(0,-1){5}}
 \put(25,25){\oval(5,5)[b]}
 \put(27.5,25){\line(-1,2){5}}
 \put(7.5,30){\line(1,0){15}}
 \put(25,35){\oval(5,5)[t]}
 \put(27.5,30){\line(0,1){5}}
 \put(27.5,30){\vector(1,0){55}}
 \put(7.5,15){\vector(1,0){75}}
 \put(7.5,45){\vector(1,0){75}}
 \put(61,6.9){\scriptsize{$\circ$: $\theta_1$}}
 \put(61,2.0){\scriptsize{$\bullet$: $\theta_0$}}
 \put(72,6.9){\scriptsize{$\square$: $\theta^{\ast}_1$}}
 \put(72,2.0){\scriptsize{$\blacksquare$: $\theta^{\ast}_0$}}
 \put(50,17){\tiny{$\cl_0$}}
 \put(50,32){\tiny{$\cl_1$}}
 \put(50,47){\tiny{$\overline{\cl}_0$}}
 \end{picture}}
\def\contourspsh{
 \begin{picture}(95,70)
 \put(5,58){(a)}
 \put(5,30){\vector(1,0){80}}
 \put(45,5){\vector(0,1){50}}
 \put(7.5,29){\line(0,1){2}}
 \put(82.5,29){\line(0,1){2}}
 \put(86,30){{\small $\Re v$}}
 \put(42,57){{\small $\Im v$}}
 \put(0.,24){{\small $-\frac{\pi}{2}$}}
 \put(83,24){{\small $\frac{\pi}{2}$}}
 \multiput(7.5,5)(0,1){50}{\line(0,1){0.5}}
 \multiput(82.5,5)(0,1){50}{\line(0,1){0.5}}
 \put(20,20){\makebox(0,0){\rule{1.2mm}{1.2mm}}}
 \put(24.25,24.25){\framebox(1.5,1.5){}}
 \put(70,20){\circle*{1.5}}
 \put(65,25){\circle{1.5}}
 \put(6.4,7){$\star$}
 \put(6.4,33){$\ast$}
 \thicklines
 \put(7.5,30){\line(1,0){15}}
 \put(22.5,30){\line(0,-1){5}}
 \put(25,25){\oval(5,5)[b]}
 \put(27.5,25){\line(0,1){5}}
 \put(27.5,30){\line(1,0){35}}
 \put(65,25){\oval(5,5)[b]}
 \put(62.5,30){\line(0,-1){5}}
 \put(67.5,30){\line(0,-1){5}}
 \put(67.5,30){\vector(1,0){15}}
 \put(10,15){\line(1,0){7.5}}
 \put(10,15){\line(0,-1){6}}
 \put(7.5,9){\oval(5,5)[br]}
 \put(17.5,15){\line(0,1){5}}
 \put(20,20){\oval(5,5)[t]}
 \put(22.5,20){\line(0,-1){5}}
 \put(22.5,15){\line(1,0){45}}
 \put(67.5,15){\line(0,1){5}}
 \put(70,20){\oval(5,5)[t]}
 \put(72.5,20){\line(0,-1){5}}
 \put(72.5,15){\line(1,0){7.5}}
 \put(80,15){\vector(0,-1){6}}
 \put(82.5,9){\oval(5,5)[bl]}
 \put(10,45){\line(1,0){70}}
 \put(10,45){\line(0,-1){10}}
 \put(7.5,35){\oval(5,5)[br]}
 \put(80,45){\vector(0,-1){10}}
 \put(82.5,35){\oval(5,5)[bl]}
 \put(18,6.9){\scriptsize{$\circ$: $\theta_1$}}
 \put(18,2.0){\scriptsize{$\bullet$: $\theta_0$}}
 \put(28,6.9){\scriptsize{$\square$: $-\theta^{\ast}_1$}}
 \put(28,2.0){\scriptsize{$\blacksquare$: $-\theta^{\ast}_0$}}
 \put(48,6.9){\scriptsize{$\star$: $\widetilde{\theta}_0$}}
 \put(48,2.0){\scriptsize{$\ast$: $\widetilde{\theta}_0+i\gamma$}}
 \put(50,17){\tiny{$\cl_0$}}
 \put(50,32){\tiny{$\cl_1$}}
 \put(50,47){\tiny{$\overline{\cl}_0$}}
 \end{picture}}
\def\contourspsl{
 \begin{picture}(95,70)
 \put(5,58){(b)}
 \put(5,30){\vector(1,0){80}}
 \put(45,5){\vector(0,1){50}}
 \put(7.5,29){\line(0,1){2}}
 \put(82.5,29){\line(0,1){2}}
 \put(86,30){{\small $\Re v$}}
 \put(42,57){{\small $\Im v$}}
 \put(0.,24){{\small $-\frac{\pi}{2}$}}
 \put(83,24){{\small $\frac{\pi}{2}$}}
 \multiput(7.5,5)(0,1){50}{\line(0,1){0.5}}
 \multiput(82.5,5)(0,1){50}{\line(0,1){0.5}}
 \put(7.5,25){\makebox(0,0){\rule{1.2mm}{1.2mm}}}
 \put(24.25,27.25){\framebox(1.5,1.5){}}
 \put(7.5,7){\circle*{1.5}}
 \put(65,28){\circle{1.5}}
 \put(6.4,17){$\star$}
 \put(6.4,41){$\ast$}
 \thicklines

 \put(12.5,30){\line(0,-1){5}}
 \put(7.5,25){\oval(10,10)[br]}
 \put(80,45){\vector(0,-1){20}}
 \put(82.5,25){\oval(5,5)[bl]}

 \put(12.5,30){\line(1,0){10}}
 \put(22.5,30){\line(0,-1){2}}
 \put(25,28){\oval(5,5)[b]}
 \put(27.5,28){\line(0,1){2}}
 \put(27.5,30){\line(1,0){35}}
 \put(65,28){\oval(5,5)[b]}
 \put(62.5,30){\line(0,-1){2}}
 \put(67.5,30){\line(0,-1){2}}
 \put(67.5,30){\line(1,0){10}}
 \put(77.5,30){\vector(0,-1){5}}
 \put(82.5,25){\oval(10,10)[bl]}
 \put(7.5,15){\vector(1,0){75}}
 \put(10,45){\line(1,0){70}}
 \put(10,45){\line(0,-1){20}}
 \put(7.5,25){\oval(5,5)[br]}
 \put(80,45){\vector(0,-1){20}}
 \put(82.5,25){\oval(5,5)[bl]}
 \put(18,6.9){\scriptsize{$\circ$: $\theta_1$}}
 \put(18,2.0){\scriptsize{$\bullet$: $\theta_0$}}
 \put(28,6.9){\scriptsize{$\square$: $-\theta^{\ast}_1$}}
 \put(28,2.0){\scriptsize{$\blacksquare$: $\overline{\theta}_0$}}
 \put(48,6.9){\scriptsize{$\star$: $\widetilde{\theta}_0$}}
 \put(48,2.0){\scriptsize{$\ast$: $\widetilde{\theta}_0+i\gamma$}}
 \put(50,17){\tiny{$\cl_0$}}
 \put(50,32){\tiny{$\cl_1$}}
 \put(50,47){\tiny{$\overline{\cl}_0$}}
 \end{picture}}
\begin{document}
%
\title{Finite temperature
correlations for the\\
$U_q(\mfsl(2|1))$-invariant
generalized Hubbard model\footnote{Dedicated to 
E.~M\"uller-Hartmann on the occasion of his 60th birthday}}
\author{
Andreas Kl\"umper\thanks{E-mail address:
kluemper@printfix.physik.uni-dortmund.de}\ \ and 
Kazumitsu Sakai\thanks{E-mail address:
sakai@printfix.physik.uni-dortmund.de}\\\\
\emph{Theoretische Physik I, Universit\"at Dortmund,} \\ 
\emph{Otto-Hahn-Str. 4, D-44221 Dortmund, Germany}} 
\date{May 21, 2001}
\maketitle
%
%
\begin{abstract}
We study an integrable model of one-dimensional
strongly correlated electrons at finite 
temperature by explicit calculation of the correlation lengths
of various correlation functions. 
The model is invariant with respect to the
quantum superalgebra $U_q(\mfsl(2|1))$ and 
characterized by the Hubbard interaction, correlated 
hopping and pair-hopping terms.
Using the integrability,
the graded quantum transfer matrix is constructed.
{}From the analyticity of its eigenvalues, 
a closed set of non-linear integral equations is derived
which describe the thermodynamical quantities and 
the finite temperature correlations.
The results show a crossover from a regime with dominating
density-density correlations to a regime with dominating 
superconducting pair correlations.
Analytical calculations in the low temperature limit are
also discussed.
\end{abstract}
%
%
%
\section{Introduction}
%
In recent decades, integrable models in one-dimensional 
strongly correlated electron systems have attracted 
considerable attention in relation to high-$T_c$ 
superconductivity.
In general, integrable models are solvable by means of the 
Bethe ansatz. 
The critical exponents of the 
correlation functions depend only on the 
$R$-matrix of the underlying infinite 
dimensional symmetry and the geometry of the Fermi sea 
of the ground state
(see the book~\cite{KBIbook} and
references therein). 
This important fact shows that all integrable models
with a given $R$-matrix exhibit the same 
universal critical behavior.

On the other hand (1+1)-dimensional critical 
phenomena are connected with  conformal 
field theory (CFT)~\cite{BPZ}. 
The central charge and conformal dimensions 
are calculated by the scaling behavior of 
the ground state and the low-lying excitations~\cite{Card}. 
The conformal dimensions determine the long distance asymptotics of
all correlation functions of any local operators.
For integrable models, one obtains the conformal dimensions by use of
the Bethe ansatz in terms of the so called dressed charge function
satisfying a linear integral equation.
Thus the Bethe ansatz together with CFT
provide a powerful tool to study 
the correlation functions especially at 
the critical point $T=0$~\cite{KY,FK}.

For finite temperatures however, one encounters technical difficulties.
Despite the fact that the models are still solvable even
off criticality, the above powerful methods lose 
effect due to the collapse of conformal invariance.
Qualitatively, we see the long-distance behavior of the correlation
functions show exponential decay 
and calculate the low temperature asymptotics by 
extension of the conformal mapping (see for example~\cite{KBIbook}).
However we need another approach for the {\it quantitative} 
understanding of the correlation functions at finite temperatures.

As a totally different approach, the quantum 
transfer matrix (QTM) has been proposed 
recently to overcome such 
difficulties~\cite{MSuzPB,InSuz,InSuz2,Koma,SAW,SNW,DdV,TakQT,Mizu,Klu,KZeit}. 
Originally the QTM method has been developed 
as an alternative method to the traditional
thermodynamical studies~\cite{YY,G,Tak}.
Utilizing the Trotter mapping, one deals with a
two-dimensional classical system instead of 
the original one-dimensional quantum system. 
Using the underlying integrability,
one constructs a family of {\it commuting} QTMs~\cite{Klu,KZeit}.
The commuting QTMs are diagonalized by 
means of the Bethe ansatz.
Thermodynamical quantities are expressed
in terms of the sole largest eigenvalue.
Also, this procedure has remarkable advantages
for the studies of correlation 
functions, which is the main topic of our
paper. 
Explicitly, the correlation lengths can be calculated by taking ratios
of the largest eigenvalue and the sub-leading
ones~\cite{InSuz2,TakQT,KZeit,Mizu,KZeit,KSS,S3U,Sak,KCMS}.

In this paper we study a strongly correlated electron system
using the approach sketched above.
As a result, we evaluate the six crucial 
correlation lengths for arbitrary particle densities and
finite temperatures: 
(i) the longitudinal and (ii) transversal spin-spin 
correlations, (iii) the density-density correlations,
(iv) the one-particle Green function,
(v) the singlet and (vi) triplet superconducting pair
correlations.
These are the first quantitative
calculations for 
correlation functions of strongly correlated 
electron systems off criticality.

The Hamiltonian of the model~\cite{BKZ} on a periodic lattice of
size $L$ is defined by
\begin{align}
\mathcal{H}&=-\sum_{j=1}^{L}\sum_{\sigma=\uparrow,\downarrow}
       (c^{\dagger}_{j,\sigma}c_{j+1,\sigma}+
         c^{\dagger}_{j+1,\sigma}c_{j,\sigma})
        \exp\left(-\frac{1}{2}(\eta-\sigma \gamma)n_{j,\sigma}
       -\frac{1}{2}(\eta+\sigma\gamma)n_{j+1,-\sigma}\right) \nn \\
& \quad     +U\sum_{j=1}^{L}n_{j,\uparrow}n_{j,\downarrow}+
       t_p\sum_{j=1}^{L}(
        c^{\dagger}_{j+1,\uparrow}
        c^{\dagger}_{j+1,\downarrow}c_{j,\uparrow}c_{j,\downarrow}+
        c^{\dagger}_{j,\uparrow}
        c^{\dagger}_{j,\downarrow}c_{j+1,\uparrow}c_{j+1,\downarrow}),
\label{hamiltonian}
\end{align}
where $c_{j,\sigma}^{\dagger}$ ($c_{j,\sigma}$) denotes the fermion
creation (annihilation) operator at $j$th site satisfying the
canonical aniticommutation relations
\begin{equation}
\{c^{\dagger}_{j,\sigma},c^{\dagger}_{k,\tau}\}=
\{c_{j,\sigma},c_{k,\tau}\}=0,\quad
\{c^{\dagger}_{j,\sigma},c_{k,\tau}\}=\delta_{jk}\delta_{\sigma\tau},
\end{equation}
and $n_{j,\sigma}$ is the number operator, i.e. $n_{j,\sigma}=
c^{\dagger}_{j,\sigma}c_{j,\sigma}$.
The ground state properties and the long-distance
behavior of the correlation functions have been 
studied in~\cite{BKZ} by the Bethe ansatz and CFT.
Due to the massive spin excitations ($\gamma\not=0$), the ground state
properties are described by a one-component Tomonaga-Luttinger liquid
(Luther-Emery liquid). For this system the one-particle Green function
decays exponentially fast, the density-density and certain pair
correlations are quasi-longranged with algebraic decay.
As the model is one-dimensional it does not exhibit off-diagonal
long-range order. However, the crossover behavior from dominant density-density
correlations to dominant (singlet) pair correlations driven by the
density of particles manifests superconducting properties.  More
quantitatively, on the basis of scaling arguments and RG concepts we
would expect that higher dimensional interactions such as interchain
couplings are relevant and may lead to true condensation of Cooper
pairs.

The model is related to a trigonometric
$R$-matrix which is a solution to the graded YBE~\cite{KulSk}
associated with the four-dimensional irreducible
representation of the quantum superalgebra 
$U_q(\mfsl(2|1))$~\cite{GHLZ} (see also~\cite{BGLZ} for 
the rational case $\gamma=0$).
The diagonalization of the Hamiltonian for general $\eta$ and $\gamma$
was done in~\cite{BKZ} and the diagonalization of the 
associated row-to-row transfer 
matrix was done for the rational case $\gamma=0$ in~\cite{RamMar,PM}.
The thermodynamics have been investigated
by the traditional 
thermodynamic Bethe ansatz based on the
string hypothesis in~\cite{BF} and the present 
approach in~\cite{JKS}.
%

%
The asymptotic behavior of correlation functions of this model have
been studied so far only for the ground state where a crossover was
established from a normal to a ``superconducting'' regime driven by
the particle density. Here we want to extend this study to finite
temperatures where we are especially interested in the extension of the
``superconducting regime'' in the density-temperature phase diagram.
For this purpose we define the QTM acting on a
$\mathbb{Z}_2$-graded vector space.
As a consequence, the genuine fermionic 
statistics are built into 
the QTM algebraically 
through the Grassmann parity $p[j]$ 
(see eq.~\eqref{defQTM2}).
Regarding the thermodynamical quantities,
one might content oneself with 
the corresponding null Grassmann 
parity model\footnote
{
We obtain the null-parity model 
satisfying the ordinary Yang-Baxter
equation by multiplying the original 
graded $R$-matrix by suitable minus signs. 
},
because
the fermionic statistics does not 
affect the thermodynamical quantities directly.
However, one must be careful to keep track 
of the fermionic statistics when
considering the physical quantities 
changing the particle number.
As the auxiliary and quantum
space are the same graded vector space,
we easily obtain the eigenvalues
of the QTM by using the results from 
the algebraic Bethe ansatz for the 
row-to-row transfer matrix.
The resultant eigenvalues can be
expressed  in the dressed vacuum form (DVF)
via the Bethe ansatz equation (BAE).
The difference between the DVF of the 
QTM and the one of the row-to-row transfer
matrix consists only of their vacuum parts.
Hence, in the actual calculation to 
the DVF, one utilizes the methods
developed in the finite-size correction
problem \cite{KBP}.
In consequence, all information of the 
BAE is contained in three 
non-linear integral equations (NLIEs),
which are valid for a system with any Trotter number $N$.
In this approach, the limit $N\to\infty$
can be taken {\it analytically}.
This finite set of NLIEs allows for highly accurate
numerical calculations of physical quantities
such as the above mentioned correlation lengths.
In particular, we observe temperature dependent
crossover phenomena between regimes dominated by density-density
or (singlet) superconducting pair correlations, respectively.

At low temperatures, the three NLIEs are reduced to 
only one {\it linear} integral equation, 
which reflects the one-component Tomonaga-Luttinger 
(Luther-Emery) liquid properties.
This linear equation is evaluated analytically 
and reproduces the predictions from CFT.
The formulations in this paper are applicable
to various kinds of strongly correlated electron 
systems which have superalgebra invariance,
for example the supersymmetric extended Hubbard model
($\mfsl(2|2)$-invariance)~\cite{EKS1,EKS2}. 

The layout of this paper is as follows. 
In section~2 we define the commuting 
QTM  constructed by the graded $R$-matrix.
The QTM is diagonalized by 
the algebraic Bethe ansatz 
in analogy to the ordinary row-to-row transfer 
matrix.
In section~3 we study the eigenvalues 
by introducing certain auxiliary functions.
Investigating their analyticity,
we determine the NLIE for 
the largest and sub-leading
eigenvalues.
We obtain six sets of NLIEs
characterizing the above mentioned correlations.
The numerical results for these NLIE are discussed.
Section~4 is devoted to the analytical 
calculation of the low temperature asymptotics
of the correlations characterized
by gapless excitations.
Section 5 contains our conclusions. 
Details of technical calculations are summarized in
various appendices. 
%
\section{The graded quantum transfer matrix}
%
The model~\eqref{hamiltonian} is solvable under the conditions
\begin{equation}
t_p=\frac{U}{2}=\frac{\sh\gamma}{\sh\alpha\gamma},\quad
\exp(-\eta)=\frac{\sh(\alpha+1)\gamma}{\sh\alpha\gamma}. 
\label{inteqcond}
\end{equation} 
Here the physical region of the two parameters $\alpha$ and $\gamma$
is restricted to $\gamma\ge 0$ and $\alpha>0$ (repulsive) or
$\alpha<-1$ (attractive). [This does not include the pure Hubbard
model which corresponds to $\eta=\gamma=t_p=0$ and $U\not=0$ violating
\eqref{inteqcond}.  Note however, that a unified treatment of the
$U_q(\mfsl(2|1))$-invariant generalized Hubbard model and the standard
Hubbard model might be possible on the basis of the work in
\cite{BaAlc}.]

Several prominent models are comprised
as  special cases:
the supersymmetric $t-J$ model ($\gamma=0,\alpha\to 0$),
the pure correlated-hopping model ($|\alpha|\to\infty$)
and a model of hard core particles of three species
(doubly occupied site, single electron with spin pointing either up or
down) ($\alpha\to -1$).
For any finite $\gamma$ the spinon excitations 
have a mass gap and so do single particle excitations. 
The pair excitations (singlet) 
and the particle-hole type excitations are massless. 
The former excitations become gapless for the 
repulsive case ($\alpha>0$) in the rational limit 
$\gamma=0$. 
Consequently for $\gamma>0$ the one-particle Green
function shows exponential decay and hence 
the momentum distribution function has no singularity  
at the Fermi point.

The classical counterpart of this model is given by
a trigonometric $R$-matrix which is associated with
the one parameter family of the 4-dimensional irreducible
representation of the 
quantum superalgebra $U_q(\mfsl(2|1))$~\cite{GHLZ}
(see also ref.~\cite{BGLZ} for the rational case $\gamma=0$).
Choosing the following basis 
\begin{equation}
|1\rangle\equiv|\uparrow\downarrow\rangle=
       c^{\dagger}_{\downarrow}
       c^{\dagger}_{\uparrow}|0\rangle,\quad
|2\rangle\equiv|\downarrow\rangle=c^{\dagger}_{\downarrow}|
                0\rangle,\quad
|3\rangle\equiv|\uparrow\rangle=c^{\dagger}_{\uparrow}|
                0\rangle,\quad
|4\rangle\equiv|0\rangle,
\end{equation}
with grading $|1\rangle$, $|4\rangle$ even
(bosonic) and $|2\rangle$, $|3\rangle$ odd (fermionic),
one determines the following 
36 non-zero matrix elements as
\begin{align}
R_{11}^{11}(v)&=\frac{[2\alpha-v]}{[2\alpha+v]},\quad
R_{22}^{22}(v)=R_{33}^{33}(v)=-1, \quad
R_{44}^{44}(v)=\frac{[2\alpha+2+v]}{[2\alpha+2-v]},  \nn \db
R_{12}^{12}(v)&=R_{13}^{13}(v)=R_{21}^{21}(v)=R_{31}^{31}(v)=
                 -\frac{[v]}{[2\alpha+v]},          \nn \db
R_{24}^{24}(v)&=R_{34}^{34}(v)=R_{42}^{42}(v)=R_{43}^{43}(v)=
                 \frac{[v]}{[2\alpha+2-v]},       \nn \db
R_{14}^{14}(v)&=R_{41}^{41}(v)=
          \frac{[v][v-2]}{[2\alpha+v][2\alpha+2-v]},\quad
R_{23}^{23}(v)=R_{32}^{32}(v)=\frac{[v]^2}{[2\alpha+v][2\alpha+2-v]},\nn \db
R_{12}^{21}(v)&=R_{13}^{31}(v)=\frac{e^{-\gamma/2}[2\alpha]}
                                {[2\alpha+v]},   \quad
R_{21}^{12}(v)=R_{31}^{13}(v)=\frac{e^{\gamma/2}[2\alpha]}
                                    {[2\alpha+v]},\nn \db
R_{14}^{41}(v)&=\frac{e^{-\gamma v}[2\alpha][2\alpha+2]}
                     {[2\alpha+v][2\alpha+2-v]},   \quad
R_{41}^{14}(v)=\frac{e^{\gamma v}[2\alpha][2\alpha+2]}
                     {[2\alpha+v][2\alpha+2-v]},\nn \db
R_{23}^{32}(v)&=\frac{-e^{\gamma v/2}[2][v]-[2\alpha][2\alpha+2]}
                     {[2\alpha+v][2\alpha+2-v]},   \quad
R_{32}^{23}(v)=\frac{-e^{-\gamma v/2}[2][v]-[2\alpha][2\alpha+2]}
                     {[2\alpha+v][2\alpha+2-v]},\nn \db
R_{24}^{42}(v)&=R_{34}^{43}(v)=\frac{e^{-\gamma v/2}[2\alpha+2]}
                     {[2\alpha+2-v]},            \quad   
R_{42}^{24}(v)=R_{43}^{34}(v)=\frac{e^{\gamma v/2}[2\alpha+2]}
                     {[2\alpha+2-v]},\nn \db
R_{14}^{23}(v)&=-R_{32}^{41}(v)=
-\frac{\sqrt{[2\alpha][2\alpha+2]}e^{-\gamma(1+v)/2}[v]}
                   {[2\alpha+v][2\alpha+2-v]}, \nn \db
R_{14}^{32}(v)&=-R_{23}^{41}(v)=
\frac{\sqrt{[2\alpha][2\alpha+2]}e^{\gamma(1-v)/2}[v]}
                   {[2\alpha+v][2\alpha+2-v]}, \nn \db
R_{23}^{14}(v)&=-R_{41}^{32}(v)=
-\frac{\sqrt{[2\alpha][2\alpha+2]}e^{\gamma(1+v)/2}[v]}
                   {[2\alpha+v][2\alpha+2-v]}, \nn \db
R_{32}^{14}(v)&=-R_{41}^{23}(v)=
\frac{\sqrt{[2\alpha][2\alpha+2]}e^{-\gamma(1-v)/2}[v]}
                   {[2\alpha+v][2\alpha+2-v]},
\label{bw}
\end{align}
where $R_{\alpha\beta}^{\gamma\delta}(v)$
are defined by
\begin{equation}
R(v)|\alpha\rangle\sotimes|\beta\rangle=
    |\gamma\rangle\sotimes|\delta\rangle R_{\alpha\beta}^{\gamma\delta}(v)
\end{equation}
with $[v]:=\sh(\gamma v/2)$ and the graded tensor product satisfies
$P|\alpha\rangle\sotimes|\beta\rangle:=
(-)^{p[\alpha]p[\beta]}|\beta\rangle\sotimes|\alpha\rangle$ where
$P$ denotes permutation.
This $R$-matrix is known to satisfy the graded Yang-Baxter
equation (GYBE) \cite{KulSk}
\begin{equation}
R_{12}(u-v)R_{13}(u)R_{23}(v)=R_{23}(v)R_{13}(u)R_{12}(u-v),
\label{gybe}
\end{equation}
which explicitly reads in terms of the matrix elements
\begin{multline}
R_{\gamma_1\gamma_2}^{\beta_1\beta_2}(u-v)
R_{\alpha_1\gamma_3}^{\gamma_1\beta_3}(u)
R_{\alpha_2\alpha_3}^{\gamma_2\gamma_3}(v)
               (-)^{(p[\alpha_1]+p[\gamma_1])p[\gamma_2]}  \\
=
R_{\gamma_2\gamma_3}^{\beta_2\beta_3}(v)
R_{\gamma_1\alpha_3}^{\beta_1\gamma_3}(u)
R_{\alpha_1\alpha_2}^{\gamma_1\gamma_2}(u-v)
               (-)^{(p[\beta_1]+p[\gamma_1])p[\gamma_2]},
\label{YBE2}
\end{multline}
where the Grassmann parity $p[j]$ 
is defined by $p[1]=p[4]=0$ (bosonic) $p[2]=p[3]=1$ (fermionic).
In eq.~\eqref{YBE2} and later we sum over repeated
indices.
Note that the parity of the $R$-matrix is even, i.e.
\begin{equation} 
p[R]\equiv p[\alpha]+p[\beta]+p[\gamma]+p[\delta]\equiv 0 \mod 2.
\label{parity}
\end{equation}
To relate the model~\eqref{hamiltonian} to the above
classical model, we construct the graded transfer matrix 
\begin{equation}
T(v)={\rm Str}_{a}\{R_{aL}(v)\dots R_{a2}(v)R_{a1}(v)\}.
\end{equation}
Using the $R$-matrix parity~\eqref{parity},
one writes explicitly the transfer matrix
by means of the matrix elements \eqref{bw}
\begin{equation}
T_{\alpha_1\alpha_2\dots\alpha_L}^{\beta_1\beta_2\dots\beta_L}(v)
=(-)^{p[a]+\sum_{j=2}^{L}(p[\alpha_j]+p[\beta_j])\sum_{k=1}^{j-1}p[\beta_k]}
 R_{\gamma_{L-1}\alpha_L}^{a\,\,\,\,\,\,\,\,\,\,\,
 \beta_L}(v)\dots
 R_{\gamma_{1}\alpha_2}^{\gamma_2\beta_2}(v)
 R_{a\,\,\alpha_1}^{\gamma_1\beta_1}(v).
\end{equation}
Due to the GYBE \eqref{gybe}, the transfer matrices $T(v)$
commute for different spectral 
parameters $v$ and $v^{\prime}$ 
\begin{equation}
[T(v),T(v^{\prime})]=0.
\end{equation}
Using the fact that the local Hamiltonian $\mathcal{H}_{j j+1}$
can be expressed as the  derivative of $R_{j j+1}(v)$ 
\begin{align}
H_{j j+1}&=-\frac{2\sh(\alpha+1)\gamma}{\gamma}
          P_{j j+1}\frac{d}{dv}R_{j j+1}(v) \biggl|_{v=0} \nn \\
         &\quad+
          2\ch(\alpha+1)\gamma-2(n_{j,\uparrow}+n_{j,\downarrow}+
           n_{j+1,\uparrow}+n_{j+1,\downarrow})\ch(\alpha+1)\gamma,
\end{align}
we can expand the transfer matrix $T(v)$ with respect to $v$ as
\begin{equation}
T(v)=T(0)\left(1-\frac{\gamma v}{2\sh(\alpha+1)\gamma}
              \mathcal{H}^{\prime}+O(v^2)\right).
\label{rT}
\end{equation}
Here we introduced the superpermutation operator $P_{j j+1}$
whose matrix elements are
\begin{equation}
[P_{j j+1}]_{\alpha\beta}^{\gamma\delta}=
[R_{j j+1}(0)]_{\alpha\beta}^{\gamma\delta}=
            (-)^{p[\alpha]p[\beta]}
                \delta_{\alpha\delta}\delta_{\beta\gamma}.
\label{permutation}
\end{equation}
Note that $\mathcal{H}^{\prime}$ denotes 
the Hamiltonian~\eqref{hamiltonian} 
with shifted chemical potential and 
ground state energy 
\begin{equation}
\mathcal{H}^{\prime}=\mathcal{H}-
                     2L\ch(\alpha+1)\gamma+2\mathcal{N}_{\rm e}
                       \ch(\alpha+1)\gamma,
\end{equation}
where $\mathcal{N}_{\rm e}$ is the electron number operator.

In order to treat the finite temperature case,
we consider another $R$-matrix defined by
\begin{equation}
\overline{R}_{\alpha\beta}^{\gamma\delta}(v)=
(-)^{p[\beta](p[\alpha]+p[\gamma])}R_{\beta\gamma}^{\delta\alpha}(v).
\label{bw2}
\end{equation}
This matrix $\overline{R}$ satisfies the following type 
of GYBE
\begin{equation}
\overline{R}_{12}(v-u)\overline{R}_{13}(-u)R_{23}(v)
=R_{23}(v)\overline{R}_{13}(-u)\overline{R}_{12}(v-u),
\label{gybe2}
\end{equation}
which reads explicitly
\begin{multline}
\overline{R}_{\gamma_1\gamma_2}^{\beta_1\beta_2}(v-u)
\overline{R}_{\alpha_1\gamma_3}^{\gamma_1\beta_3}(-u)
R_{\alpha_2\alpha_3}^{\gamma_2\gamma_3}(v)
               (-)^{(p[\alpha_1]+p[\gamma_1])p[\gamma_2]}  \\
=
R_{\gamma_2\gamma_3}^{\beta_2\beta_3}(v)
\overline{R}_{\gamma_1\alpha_3}^{\beta_1\gamma_3}(-u)
\overline{R}_{\alpha_1\alpha_2}^{\gamma_1\gamma_2}(v-u)
               (-)^{(p[\beta_1]+p[\gamma_1])p[\gamma_2]}.
\label{MYBE2}
\end{multline}
Using this, one constructs the transfer matrix
\begin{equation}
\overline{T}(v)={\rm Str}_{a}\{\overline{R}_{aL}(v)\dots 
               \overline{R}_{a2}(v)\overline{R}_{a1}(v)\},
\end{equation}
explicitly reading
\begin{equation}
\overline{T}_{\alpha_1\alpha_2\dots\alpha_L}^{\beta_1\beta_2\dots\beta_L}(v)
=(-)^{p[a]+\sum_{j=2}^{L}(p[\alpha_j]+p[\beta_j])\sum_{k=1}^{j-1}p[\beta_k]}
 \overline{R}_{\gamma_{L-1}\alpha_L}^{a\,\,\,\,\,\,\,\,\,\,\,
 \beta_L}(v)\dots
 \overline{R}_{\gamma_{1}\alpha_2}^{\gamma_2\beta_2}(v)
 \overline{R}_{a\,\,\alpha_1}^{\gamma_1\beta_1}(v).
\end{equation}
We expand this transfer matrix in the same manner
as \eqref{rT},
\begin{equation}
\overline{T}(v)=
\overline{T}(0)\left(1-\frac{\gamma v}{2\sh(\alpha+1)\gamma}
              \mathcal{H}^{\prime}+O(v^2)\right).
\label{lT}
\end{equation}
Combining the above relations and
using $\overline{T}(0)T(0)=1$ 
(note that $T(0)$ and
$\overline{T}(0)$ denote the right and left
shift operators, respectively), one obtains
\begin{equation}
\overline{T}(u)T(u)=1-\frac{\gamma u}{\sh(\alpha+1)\gamma}
              \mathcal{H}^{\prime}+O(v^2).
\end{equation}
{}From the above equation, it follows that
\begin{equation}
\exp(-\beta\mathcal{H}^{\prime})=\lim_{N\to\infty}\bigl[
\overline{T}(u_N)T(u_N)\bigr]^{N/2},\quad
u_{N}=\frac{2\beta\sh(\alpha+1)\gamma}{N \gamma}.
\end{equation}
Here $\beta$ is the reciprocal temperature; $\beta=1/T$
and the height $N$ of the
fictitious underlying square lattice is even and
referred to as the Trotter number.
Hence the free energy $f$ of the 
original quantum system~\eqref{hamiltonian} is given by
\begin{equation}
f=-\lim_{L\to \infty}\lim_{N\to \infty}
 \frac{1}{L \beta}\ln\left({\rm Tr}
             \bigl[\overline{T}(u_N)T(u_N)\bigr]^{N/2}
             \right)+
 2(1-n_{\rm e})\ch(\alpha+1)\gamma,
\label{freepre}
\end{equation}
where $n_{\rm e}$ denotes the particle density of the
system.
Calculating  
${\rm Tr}\bigl[\overline{T}(u_N)T(u_N)\bigr]^{N/2}$ from the
eigenvalues of $\overline{T}(u_N)T(u_N)$ is a 
serious problem, because the spectrum is infinitely degenerate
in the limit of infinite Trotter number
$u_{N}{\rightarrow} 0$ (for ${N\to\infty}$).
To avoid this difficulty, we transform 
${\rm Tr}[\overline{T}(u_N)T(u_N)]^{N/2}$ as
\begin{align}
{\rm Tr}&\prod_{k=1}^{N/2}
{\rm Str}_{a_{2k} a_{2k-1}}\{\overline{R}_{a_{2k} L}(u_N)
\dots\overline{R}_{a_{2k} 1}(u_N) 
R_{a_{2k-1} L}(u_N)\dots R_{a_{2k-1} 1}(u_N)\}   \nn \\
&\quad={\rm Str}\prod_{j=1}^{L}{\rm Tr}_j
  \prod_{k=1}^{N/2}\overline{R}_{a_{2k} j}(u_N)R_{a_{2k-1} j}(u_N).
\end{align}
Now we introduce the QTM,
\begin{equation}
T_{Q}(v)={\rm Tr}_j\prod_{k=1}^{N/2}
         \overline{R}_{a_{2k} j}(u_N+v)R_{a_{2k-1} j}(u_N-v),
\end{equation}
explicitly
\begin{align}
{T_{Q}^{}}_{\alpha_1\alpha_2\dots\alpha_N}^{\beta_1\beta_2\dots\beta_N}(v)
&=(-)^{\sum_{j=2}^{N}(p[\alpha_j]+p[\beta_j])\sum_{k=1}^{j-1}p[\beta_k]} \nn \\
&\quad \times
 \overline{R}_{\alpha_N\gamma_{N-1}}^{\beta_N a}(u_N+v)
 R_{\alpha_{N-1}\gamma_{N-2}}^{\beta_{N-1}\gamma_{N-1}}(u_N-v)\dots
 \overline{R}_{\alpha_2\gamma_{1}}^{\beta_2\gamma_2}(u_N+v)
 R_{\alpha_1 a}^{\beta_1\gamma_1}(u_N-v).
\label{defQTM2}
\end{align}
Due to the GYBE~\eqref{gybe} and~\eqref{gybe2},
the QTM is commutative, i.e.
\begin{equation}
[T_Q(u_N,v),T_Q(u_N,v^{\prime})]=0.
\end{equation}
We will see that the largest eigenvalue is separated by a gap from the rest
of the spectrum for any $N$ persisting in the limit
$N\to\infty$ as long as $T>0$.
Since the two limits in eq.~\eqref{freepre} can be
interchanged, for a proof cf.~\cite{MSuzPB,InSuz}, we take the limit
$L\to\infty$ first.
As there exists a finite gap
between the largest eigenvalue $\Lambda^{\rm max}(0)$ and
the sub-leading eigenvalues $\Lambda^{\rm sub}(0)$,
we can write 
\begin{equation}
f=-\lim_{N\to\infty}\frac{1}{\beta}\ln\Lambda^{\rm max}(0)
  +2(n_{\rm e}-1)\ch(\alpha+1)\gamma.
\label{free}
\end{equation}
In this approach, the thermodynamical completeness 
$-\lim_{\beta\to 0}\beta f=\ln 4$ follows easily
from $\lim_{N\to\infty}\Lambda^{\rm max}(0)=4$ which is obvious from
the definition of $R(v)$~\eqref{bw} 
and $\overline{R}(v)$~\eqref{bw2}.

Most significantly the present method makes it
possible to calculate various correlation lengths at
finite temperature through the relation
\begin{equation}
\frac{1}{\xi}=-\lim_{N\to\infty}
\ln\biggr|\frac{\Lambda^{\rm sub}(0)}
          {\Lambda^{\rm max}(0)} \biggr|.
\label{correlation}
\end{equation}

The QTM can be diagonalized by means of the algebraic 
Bethe ansatz method. 
Due to the fact that auxiliary and quantum space
are the same $\mathbb{Z}_2$-graded vector space, one
can utilize the results for the algebraic Bethe ansatz for
the row-to-row transfer matrix case~\cite{RamMar,PM}\footnote{
We remark that the formulation in~\cite{RamMar} is based on the
non-graded rational ($\gamma=0$) 
$R$-matrix which is 
connected to our graded case by 
$R_{\alpha\beta}^{\gamma\delta}\to
(-)^{p[\gamma]p[\delta]}R_{\alpha\beta}^{\gamma\delta}$. 
The basic difference between the standard and
graded formulation lies in the existence of 
extra phase factors in the dressed vacuum form and 
the Bethe ansatz equation.}. 
Replacing the  parameters 
$u_N$ and $v$ by
\begin{equation}
\frac{\gamma}{2}v\to iv,\quad \frac{\gamma}{2}u_N\to
u_N=\frac{\beta\sh(\alpha+1)\gamma}{N},
\end{equation}
and choosing the state $(|1\rangle\sotimes |4\rangle)^{\sotimes N/2}$
as the pseudo vacuum state (reference state),
one  writes the eigenvalue $\Lambda(v)$ of the QTM 
in the dressed vacuum form (DVF)  
\begin{align}
\Lambda(v)&=\phi_1(v)\frac{q_1(v+\frac{i}{2}\gamma(2\alpha+1))}
                           {q_1(v+\frac{i}{2}\gamma)}e^{2\mu\beta}+
            \phi_2(v)\frac{q_1(v+\frac{i}{2}\gamma(2\alpha+1))}
                           {q_1(v+\frac{i}{2}\gamma)}
                      \frac{q_2(v+i \gamma)}{q_2(v)}e^{\mu\beta} \nn \\
        &\quad+\phi_2(v)\frac{q_1(v+\frac{i}{2}\gamma(2\alpha+1))}
                           {q_1(v-\frac{i}{2}\gamma)}
                      \frac{q_2(v-i \gamma)}{q_2(v)}e^{\mu\beta}+
          \phi_3(v)\frac{q_1(v+\frac{i}{2}\gamma(2\alpha+1))}
                           {q_1(v-\frac{i}{2}\gamma)},
\label{DVF}
\end{align}
where the functions $\phi_1(v)$, $\phi_2(v)$ and $\phi_3(v)$ are
defined by
\begin{align}
\phi_1(v)&=\left(\frac{\sh(iv+u_N)\sh(iv+u_N-\gamma)
                       \sh(iv-u_N+\alpha\gamma)}
                     {\sh(iv+u_N-(\alpha+1)\gamma)
                      \sh(iv+u_N+\alpha\gamma)
                      \sh(iv-u_N-\alpha\gamma)}\right)^{\frac{N}{2}}, \nn \\
\phi_2(v)&=\left(\frac{\sh(iv-u_N)\sh(iv+u_N)}
                       {\sh(iv-u_N-\alpha\gamma)
                        \sh(iv+u_N-(\alpha+1)\gamma)}
                        \right)^{\frac{N}{2}},  \\
\phi_3(v)&=\phi_1(-v)|_{\alpha\to-\alpha-1}, \nn
\end{align}
and the chemical potential $\mu$ (shifted by
$\mu+2\ch(\alpha+1)\gamma\to \mu$, see Appendix~A)
have been introduced.
The functions $q_1(v)$ and $q_2(v)$ are written in the form
\begin{equation}
q_1(v)=\prod_{j=1}^{n}\sin(v-v_j^{(1)}), \,\,\,\,
q_2(v)=\prod_{j=1}^{m}\sin(v-v_j^{(2)}),    
\end{equation}
where the unknown parameters $\{v_j^{(1)}\}_{j\in\{1,2,\cdots, n\}}$ 
and  $\{v_j^{(2)}\}_{j\in\{1,2,\cdots, m\}}$  are the
Bethe ansatz rapidities determined from the Bethe ansatz equation (BAE)
\begin{equation}
\frac{q_2(v_j^{(1)}+\frac{i}{2}\gamma)}
     {q_2(v_j^{(1)}-\frac{i}{2}\gamma)}=
-\frac{\phi_1(v_j^{(1)}-\frac{i}{2} \gamma)}
            {\phi_2(v_j^{(1)}-\frac{i}{2} \gamma)} e^{\beta \mu},\quad
\frac{q_1(v_j^{(2)}+\frac{i}{2}\gamma)}
     {q_1(v_j^{(2)}-\frac{i}{2}\gamma)}=
-\frac{q_2(v_j^{(2)}+i \gamma)}
      {q_2(v_j^{(2)}-i\gamma)}.
\label{BAE}
\end{equation}

At the end of this section, we emphasize that
our QTM is based on the {\it graded} formulations,
which means the genuine fermionic statistics of the 
model are properly built into the algebraic structure 
of the QTM through the Grassmann parity $p[j]$ 
(see~\eqref{defQTM2}). 
Nevertheless the DVF and the BAE have quite 
simple forms.
This indicates that the differences between
the null and non-null Grassmann parity models
in one dimension are embedded into 
differences of boundary conditions.
%
\section{Non-linear integral equations}
%
In this section we analyze the eigenvalue $\Lambda(v)$ in DVF 
form~\eqref{DVF} by
selecting certain auxiliary functions including
the information of the BAE. 
{}From now on we consider the repulsive case $\alpha>0$.
For the largest eigenvalue,
the auxiliary functions and 
NLIE have been introduced in~\cite{JKS}.
For some special eigenvalues with certain symmetries including the
largest eigenvalue, one needs only two auxiliary functions~\cite{JKS}.

In order to treat the general case,
we introduce the following 
three auxiliary functions.
\begin{alignat}{3}
&\mfa_0(v)=\frac{\lambda_1(x)(\lambda_3(x)+\lambda_4(x))}
               {\lambda_2(x)(\lambda_1(x)+\lambda_2(x)+\lambda_3(x)+
                \lambda_4(x))}\Biggr|_{x=v+\frac{i}{2}\alpha\gamma}, &\quad&
&\mfA_0(v)=1+\mfa_0(v), \nn \db
&\mfab_0(v)=\frac{\lambda_2(x)}{\lambda_3(x)+\lambda_4(x)}
                              \Biggr|_{x=v+\frac{i}{2}\alpha\gamma}, &\quad&
&\mfAb_0(v)=1+\mfab_0(v), \nn \db
&\mfa_1(v)=\frac{\lambda_1(x)}{\lambda_2(x)+\lambda_3(x)+\lambda_4(x)}
                              \Biggr|_{x=v+\frac{i}{2}\alpha\gamma}, &\quad&
&\mfA_1(v)=1+\mfa_1(v).
\label{aux}
\end{alignat}
where the functions $\lambda_j(x)$ denote the $j$th term on the
right hand side in the DVF \eqref{DVF}.

These auxiliary functions satisfy certain 
functional relations.
Exploring their analyticity,
one transforms these functional relations into
a closed set of NLIE.
Here we consider the NLIE for the largest eigenvalue
and some sub-leading ones which describe the
correlations.
The details of the calculations are deferred to
Appendices~B and C.
%
\subsection{Largest eigenvalues}
%
First we consider the largest eigenvalue.
As described above, one finds the NLIE for this
eigenvalue in~\cite{JKS}.
For completeness we study this case by 
using the above three auxiliary functions.

The largest eigenvalue belongs to the sector 
$n=N$ and $m=N/2$ in the BAE~\eqref{BAE}.
{}From numerical calculations with finite Trotter
number $N$, we observe the above
auxiliary functions have the following analytical 
properties:
The function $\mfA_0(v)$ ($\mfAb_0(v)$) is 
analytic and non-zero in a finite strip
in the lower (upper) half plane
$-\gamma/2<\Im v<0$ ($0<\Im v<\gamma/2$)\footnote{
In fact these auxiliary functions have
trivial zeros and poles of order $N/2$
which come from the vacuum parts in the DVF.
These zeros and poles partly appear in the above ``analyticity
strips'' and result into the
leading terms of the NLIEs.
}, 
and the function $\mfA_1(v)$ is 
analytic and non-zero along the real axis.
Hereafter we call the region $-\gamma/2<\Im v<\gamma/2$
``physical strip".
Due to this analyticity, we can transform the
functional relations satisfied by the above
auxiliary functions into NLIEs by using the
Fourier transform together with Cauchy's
theorem. In this procedure one takes the
Trotter limit $N\to\infty$ analytically
(see Appendix~B). 
Consequently, we obtain the following closed set
of NLIE.
\begin{align}
\ln\mfa_0(v)&=\beta\psi(v)+k\ast\ln\mfAb_0(v)+
               k\ast\ln\mfA_1(v)+\beta \mu,\nn\\
\ln\mfab_0(v)&=\beta\psi(-v)+\overline{k}\ast
               \ln\mfA_0(v)+
               \overline{k}\ast\ln\mfA_1(v)+\beta \mu,
\label{nlielgst}     \\
\ln\mfa_1(v)&=\beta(\psi(v)+\psi(-v))+\bar{k}\ast
               \ln\mfA_0(v)+k\ast\ln\mfAb_0(v)+k_1
               \ast\ln\mfA_1(v)+2\beta\mu, \nn
\end{align}
with the kernels and leading terms
\begin{align}
&k(v)=\frac{\sh\gamma}{2\sh(i v)\sh(i v-\gamma)},\quad
\overline{k}(v)=k(-v),\quad k_1(v)=k(v)+\overline{k}(v), \nn \\
&\psi(v)=\frac{\sh^2(\alpha+1)\gamma}
        {\sh(i v+\frac{1}{2}\alpha\gamma)
         \sh(i v-\frac{1}{2}\gamma(\alpha+2))},
\end{align}
where the symbol $*$ denotes the convolution defined by
\begin{equation}
f\ast g(v)
=\frac{1}{\pi}\int_{\cl}f(v-x)g(x) dx
=\frac{1}{\pi}\int_{v-{\cl}}f(x)g(v-x) dx.
\end{equation}
The integration contours $\cl$ for convolutions with $\ln\mfA_0$,
$\ln\mfAb_0$ and $\ln\mfA_1$ in~\eqref{nlielgst} should be taken by a
straight line ranging over a full period $\pi$ with imaginary part
$-\delta$, $+\delta$ and $0$ ($\delta$ is arbitrary but fixed in the
range $0<\delta<\gamma/2$), respectively.
Through the solution to the above NLIE, the largest eigenvalue is
expressed by
\begin{equation}
\ln\Lambda^{\rm max}(v)=
\Psi(v)+\zeta*\ln\mfA_0(v)+\overline{\zeta}*\ln\mfAb_0(v)
+(\zeta+\overline{\zeta})*\ln\mfA_1(v),
\label{largest}
\end{equation}
where the kernels $\zeta(v)$, $\overline{\zeta}(v)$ and
the leading term $\Psi(v)$ are defined by
\begin{equation}
\Psi(v)=\frac{-2\beta\sh^2((\alpha+1)\gamma)\ch((\alpha+1)\gamma)}
               {\sh(iv+(\alpha+1)\gamma)
                \sh(iv-(\alpha+1)\gamma)},   \quad
\zeta(v)=\frac{-\psi(-v)}{2\sh(\alpha+1)\gamma},
\quad \overline{\zeta}(v)=\zeta(-v).
\end{equation}
{}From eq.~\eqref{nlielgst}, we find
$\mfab_0(v)=\mfa_0(-v)$ and
the real function $\mfa_1(v)$ is symmetric
with respect to both real and imaginary axis.
Hence the threefold set of NLIEs~\eqref{nlielgst} 
can be reduced to only two NLIEs which
are identical to the ones in~\cite{JKS}.
%
\subsection{General NLIE for arbitrary eigenvalues}
%
The sub-leading eigenvalues of 
the QTM~\eqref{DVF} are treated in a way
similar to the largest eigenvalue.
The main difference lies in the
existence of additional zeros and
poles of the auxiliary functions.
Therefore one observes additional terms
in the NLIE (``sin''-terms).
Consequently all correlation functions
are characterized by the distribution patterns
of these additional zeros or poles.

Often in the case of excited
states the second factors in the convolutions are logarithms of 
periodic functions,
however with non-zero winding numbers, $\ln\mfA(\pi/2)-
\ln\mfA(-\pi/2)=2n\pi i$ ($n\in{\mathbb{Z}}$). This gives rise to a
second set of additional terms in the NLIE (``cos''-terms).

Thus we obtain the following general NLIE 
with additional terms $\ln\varphi_0(v)$, 
$\ln\varphi_1(v)$ and $\ln\overline{\varphi}_0(v)$.
\begin{align}
\ln\mfa_0(v)&=\beta\psi(v)+k\ast\ln\mfAb_0(v)+
               k\ast\ln\mfA_1(v)+\ln\varphi_0(v)+\beta \mu,\nn\\
\ln\mfab_0(v)&=\beta\psi(-v)+\overline{k}\ast
               \ln\mfA_0(v)+
               \overline{k}\ast\ln\mfA_1(v)
               +\ln\overline{\varphi}_0(v)+\beta \mu,
\label{nliegeneral}     \\
\ln\mfa_1(v)&=\beta(\psi(v)+\psi(-v))+\bar{k}\ast
               \ln\mfA_0(v)+k\ast\ln\mfAb_0(v)+k_1
               \ast\ln\mfA_1(v)+\ln\varphi_1(v)+2\beta\mu, \nn
\end{align}

The additional terms in the above general NLIE must be
determined for each eigenvalue separately.
Note that the relation 
$\varphi_1(v)=\varphi_0(v)\overline{\varphi}_0(v)$
is always valid due to certain algebraic structures of the
auxiliary functions (see Appendices B and C).
In~\eqref{nliegeneral} and later,
we take the same integration contours as in the case of the largest
eigenvalue.
Thus the corresponding NLIE for arbitrary eigenvalues
(including the largest eigenvalue) are written in the
following general form:
\begin{equation}
\ln\Lambda(v)=
\Psi(v)+\zeta*\ln\mfA_0(v)+\overline{\zeta}*\ln\mfAb_0(v)+(\zeta+\overline{\zeta})*\ln\mfA_1(v)+
        \ln\chi(v),
\label{eigengeneral}
\end{equation}
where the term $\ln\chi(v)$ is also determined
from the distribution pattern of the additional
zeros and poles.

As concrete examples,
we determine the additional terms for the eigenvalues corresponding to
the asymptotics of the:
(i) one-particle Green function,
(ii) transversal spin-spin correlations,
(iii) density-density correlations 
or longitudinal spin-spin correlations,
(iv) sub-dominant correlations of case (iii),
(v) singlet and (vi) triplet superconducting
correlations (see Appendix~C.).\\\\
%
%
\subsection{Numerical analysis of the NLIE}
\begin{figure}[hbt]
\begin{center}
\includegraphics[width=0.495\textwidth]{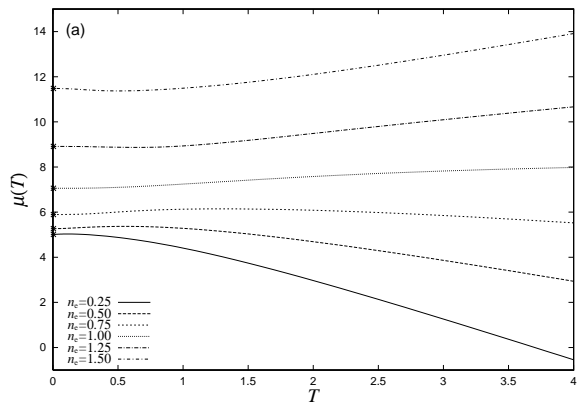}
\end{center}
\caption{Temperature dependence of the chemical potentials for
interaction parameters $\alpha=1$, $\gamma=1$ and various particle
densities. The symbol $\ast$ denotes the zero temperature chemical
potentials derived by~\eqref{defdens} and~\eqref{defchem} in
Appendix~A. In our units the Boltzmann constant is $k_B=1$.}
\label{graphchem}
\end{figure}

%
Here we evaluate the above NLIE numerically.
As seen in the above section, the NLIEs are exact
expressions for the thermodynamical quantities 
and various correlations at any finite temperature.
In contrast to the standard TBA they close at a finite level.
Hence one obtains physical quantities 
explicitly by a highly accurate numerical 
analysis.
To keep the electron density constant,
we adopt temperature dependent chemical potentials
$\mu(T)$ determined from
\begin{equation}
\frac{d\langle n_{\rm e}(T,\mu(T))\rangle}{dT}
=\frac{d}{dT}\left(\frac{\partial f}{\partial \mu}\right)_{T}=0.
\end{equation}
In Fig.~\ref{graphchem} we depict $\mu(T)$ for $\alpha=1$, $\gamma=1$ 
and various particle densities. 
We solve the NLIE by the iteration method.
In each step, convolution parts are calculated
by making use of the Fast Fourier transform.
The subsidiary conditions which determine the zeros 
$\theta_0$, $\theta_1$, etc. (see, for example, 
eq.~\eqref{subgr}) are solved by the Newton
method.
%
%
%
%
%

\begin{figure}
\begin{center}
\includegraphics[width=0.495\textwidth]{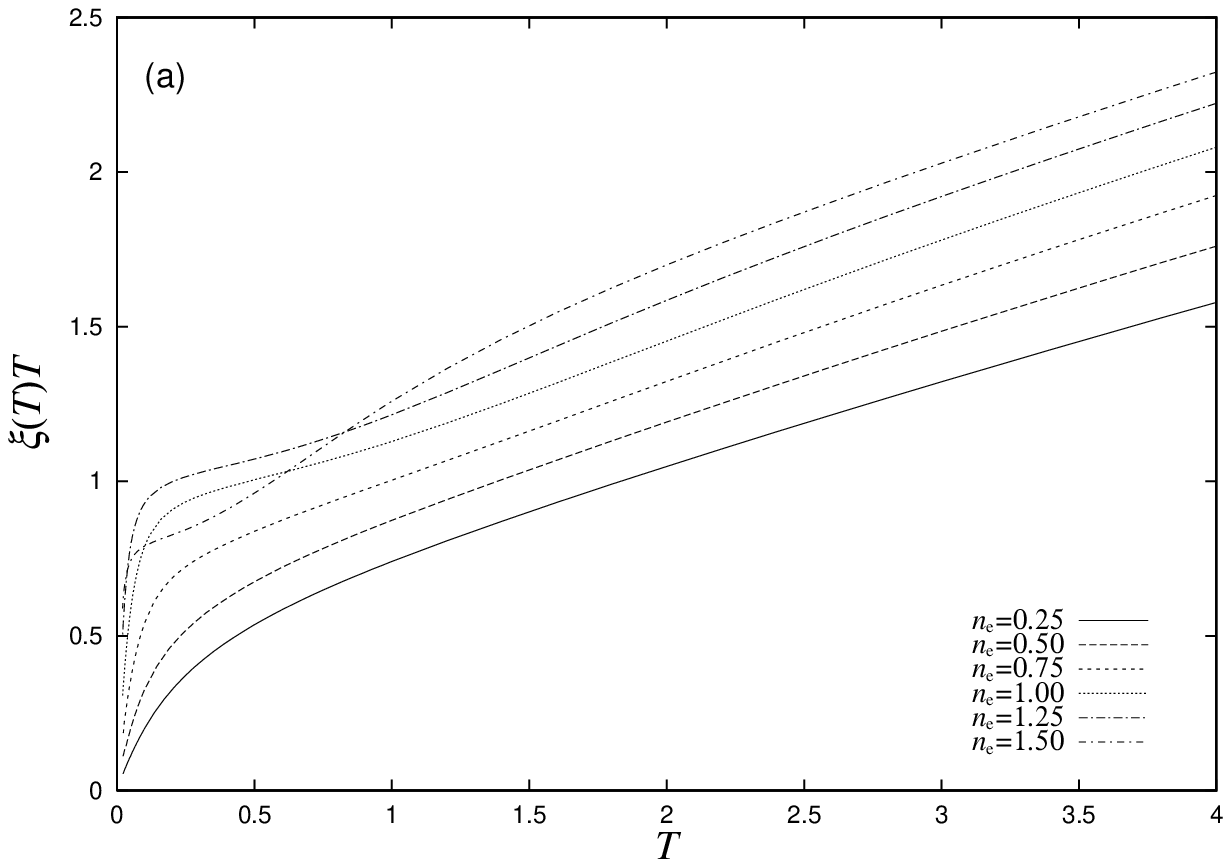}
\includegraphics[width=0.495\textwidth]{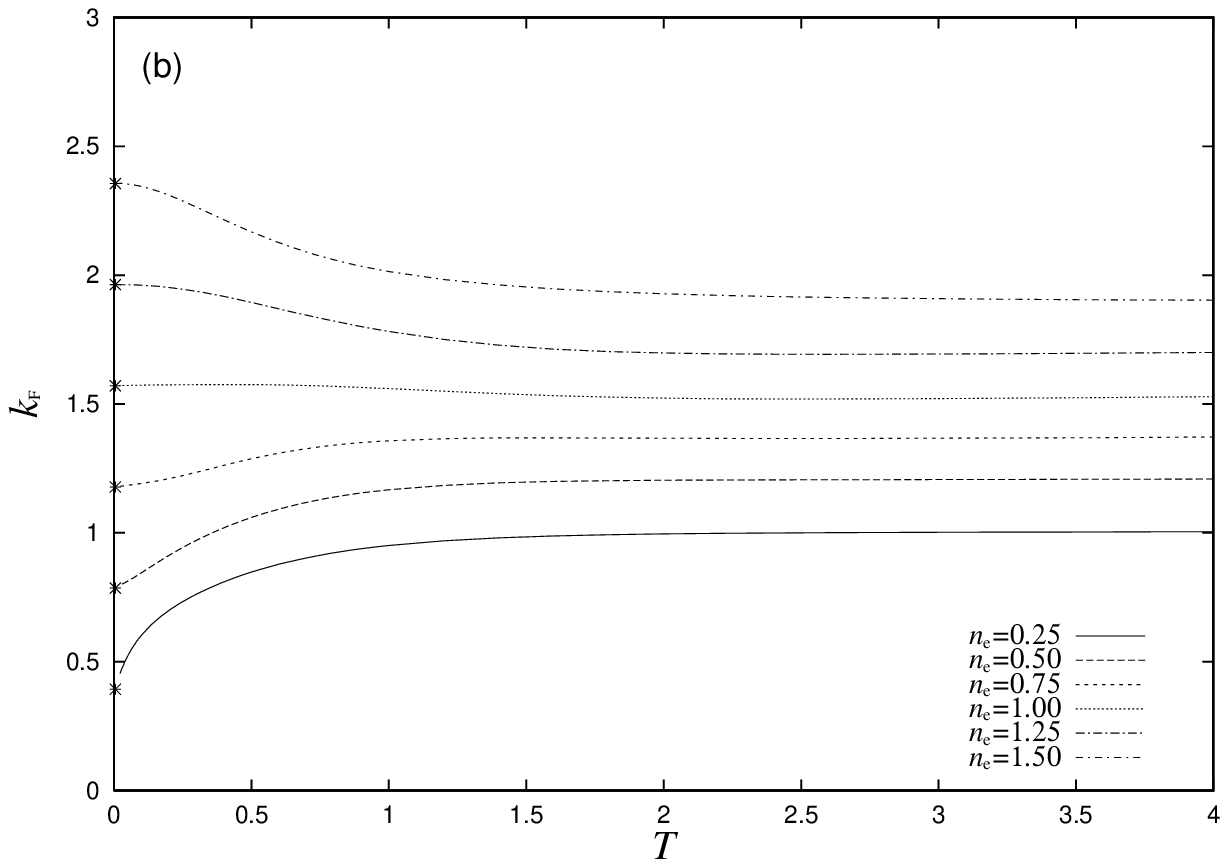}
\end{center}
\caption{(a): Temperature dependence of the scaled correlation
lengths $T\cdot\xi(T)$ for the one-particle Green function.
(b): Temperature dependence of the wave vector of the oscillation terms. 
At zero temperature
one observes the $k_{\rm F}$--oscillation 
($k_{\rm F}=\pi n_{\rm e}/2$ depicted by the symbol $\ast$)
characteristic for the one-particle Green function.}
\label{graphgr}
\end{figure}

\begin{figure}
\begin{center}
\includegraphics[width=0.495\textwidth]{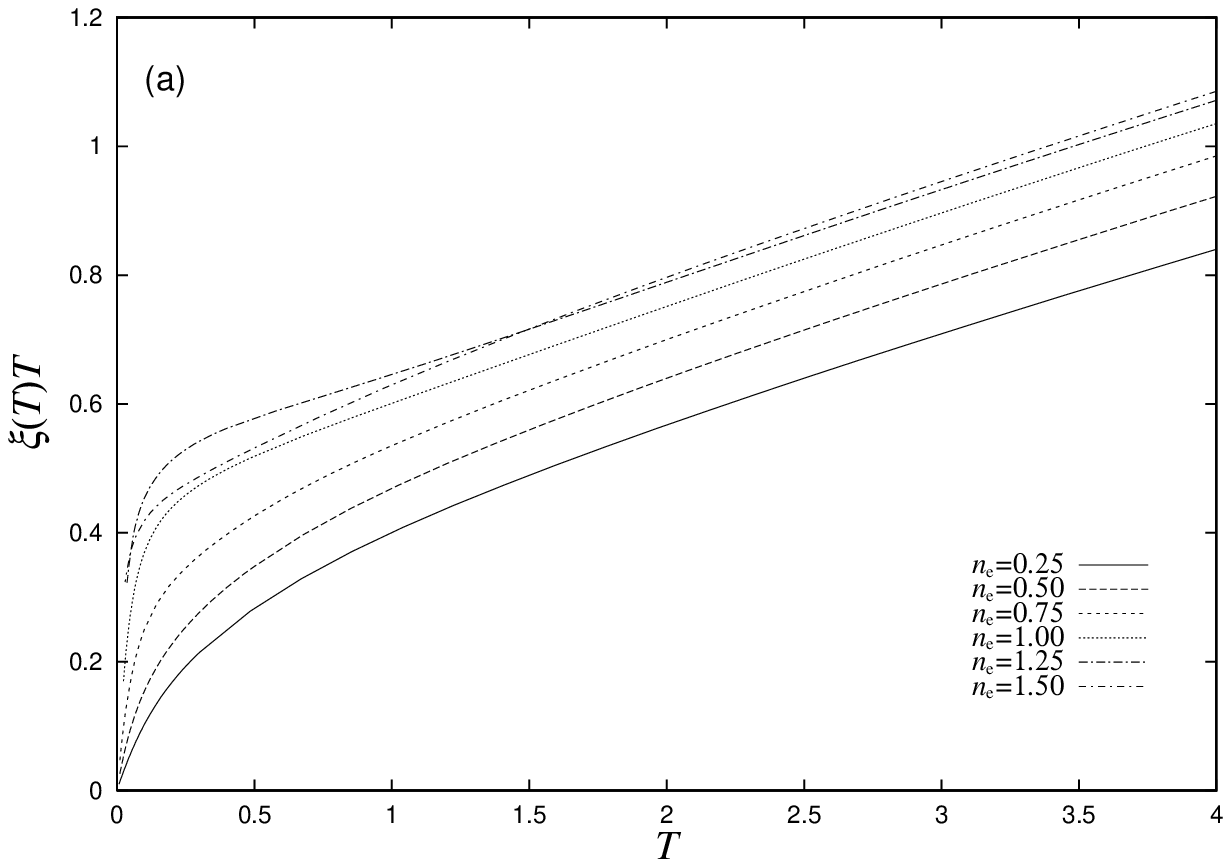}
\includegraphics[width=0.495\textwidth]{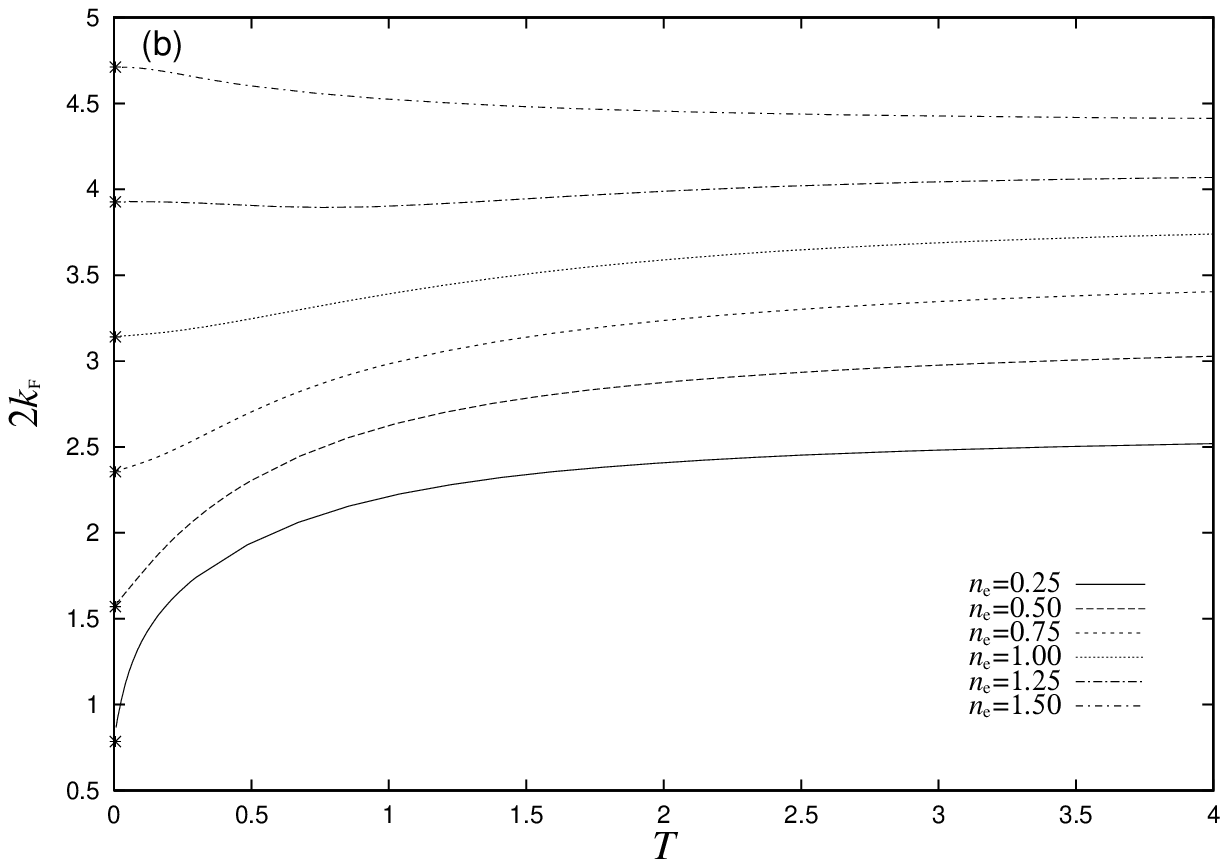}
\end{center}
\caption{(a): Temperature dependence of the correlation lengths
of the transversal spin-spin correlations.
(b): The wave vector $2k_{\rm F}$ of the oscillations
which is characteristic for the spin-spin correlations. 
At zero temperature one
observes the strict relation of wave vector with particle density
$2k_{\rm F}=\pi n_{\rm e}$ 
(analytical values are depicted by the symbol $\ast$).}
\label{graphst}
\end{figure}

\begin{figure}
\begin{center}
\includegraphics[width=0.495\textwidth]{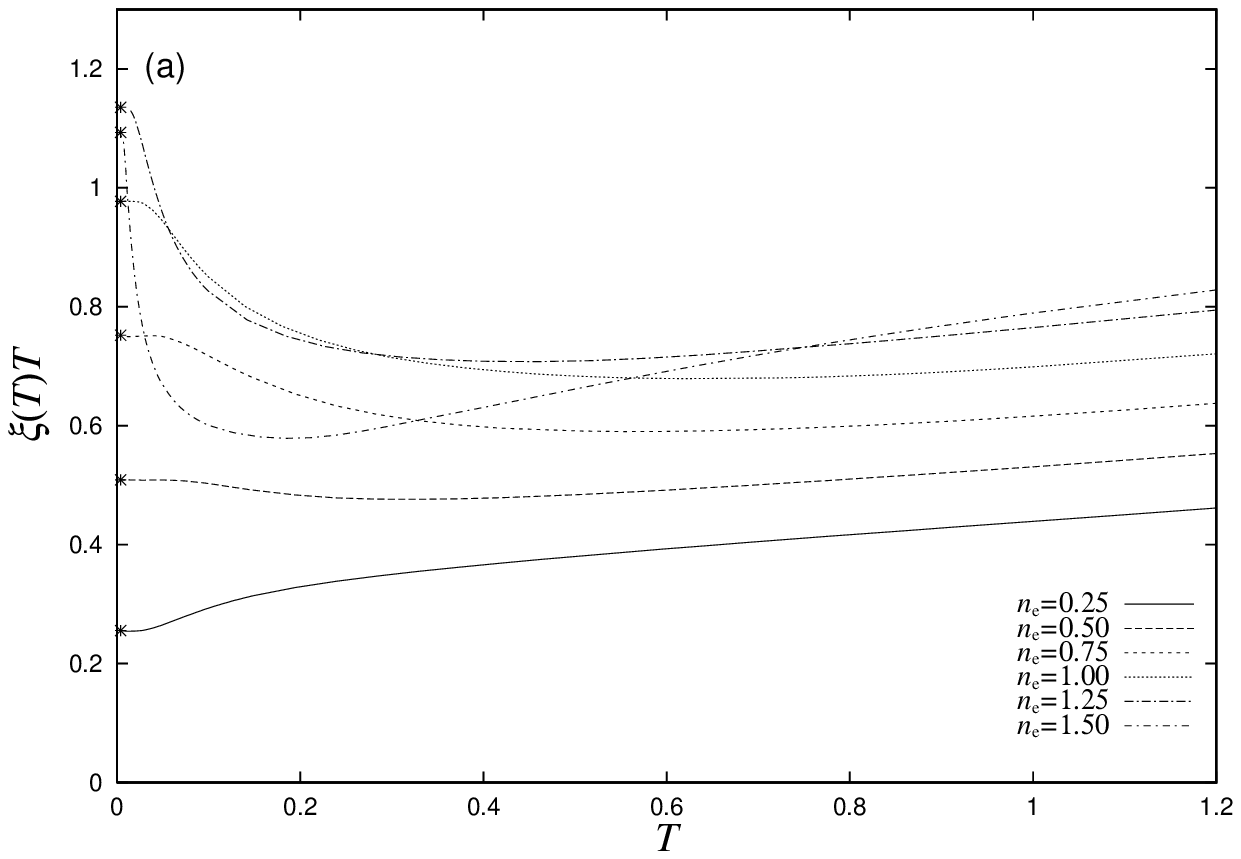}
\includegraphics[width=0.495\textwidth]{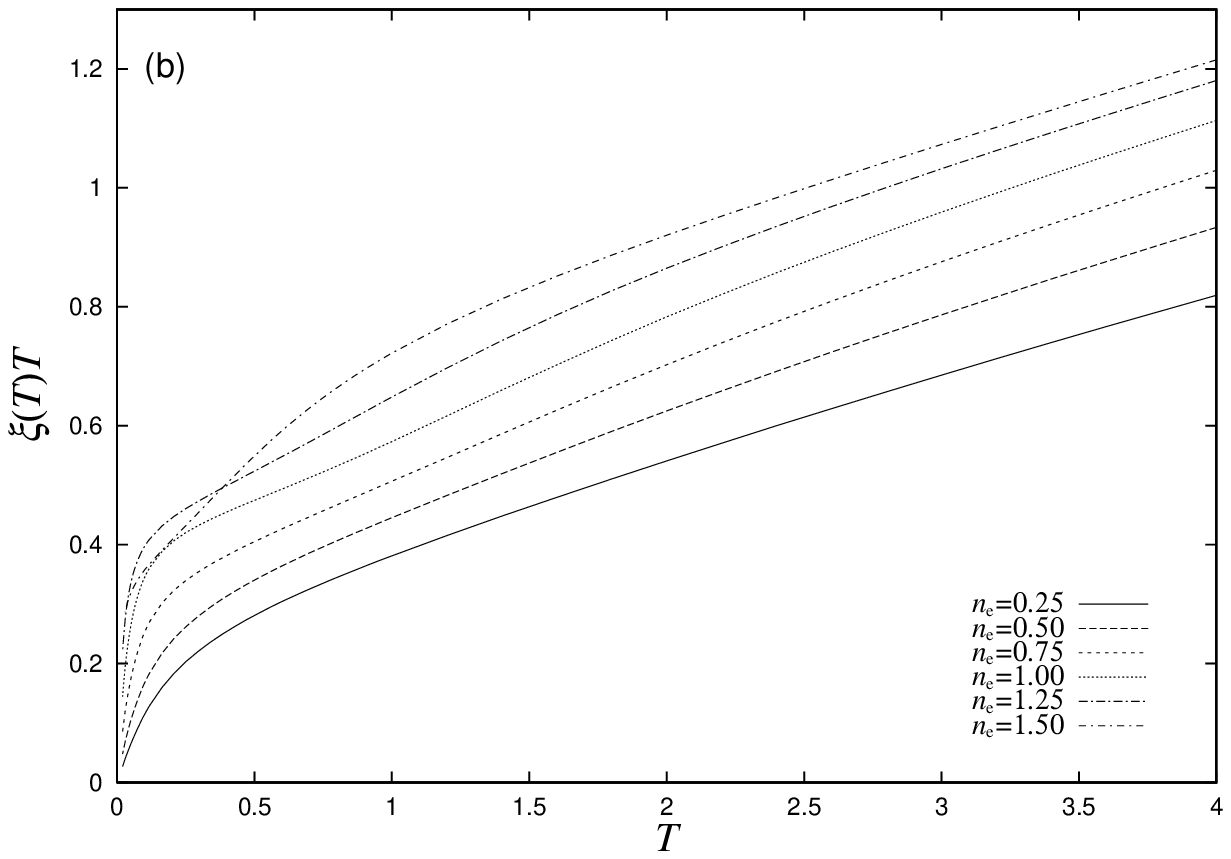}
\end{center}
\caption{Temperature dependence of the correlation lengths for
singlet superconducting pair correlations (a) and triplet
pair correlations (b). The symbol $\ast$ in (a) denotes results
of the analytical calculation in the low temperature limit
presented in section~4 confirming the predictions from CFT in Appendix~A.}
\label{graphph}
\end{figure}

\begin{figure}
\begin{center}
\includegraphics[width=0.495\textwidth]{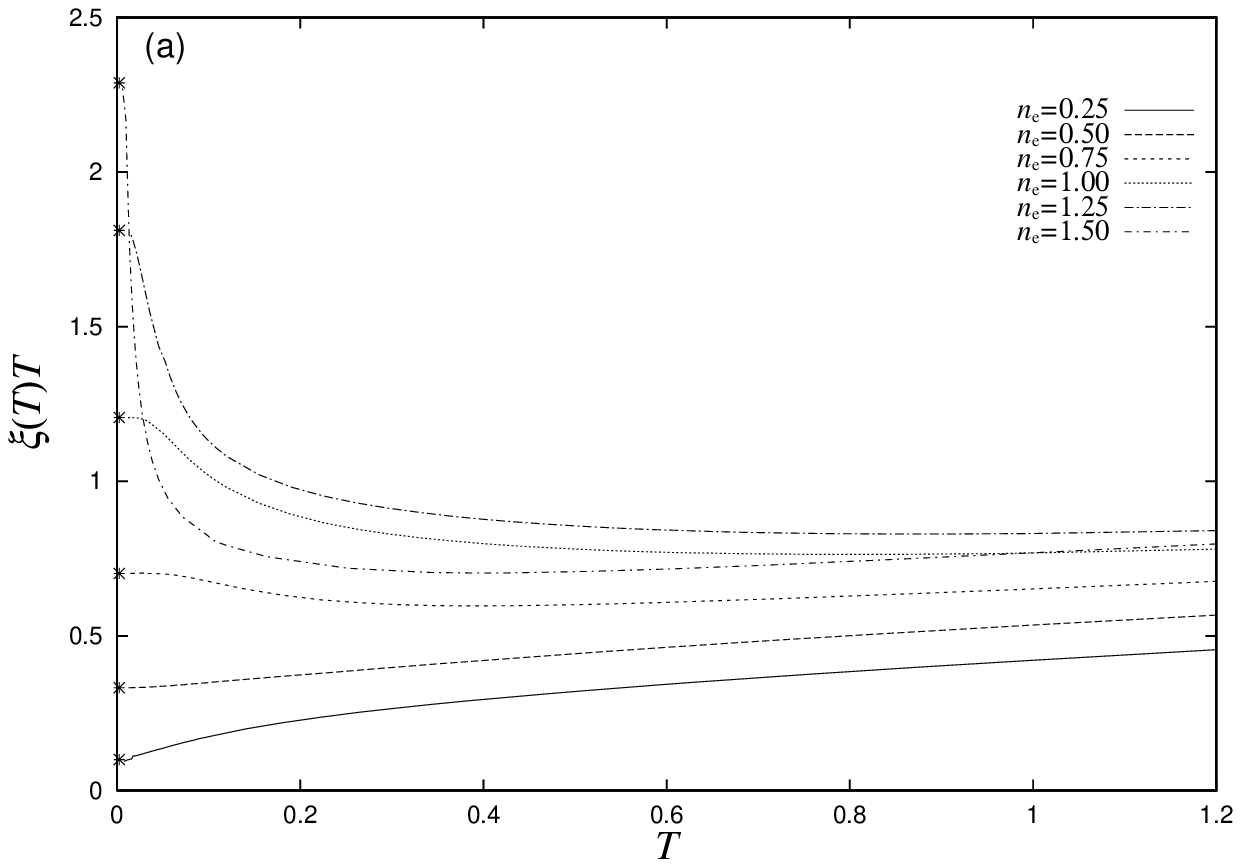}
\includegraphics[width=0.495\textwidth]{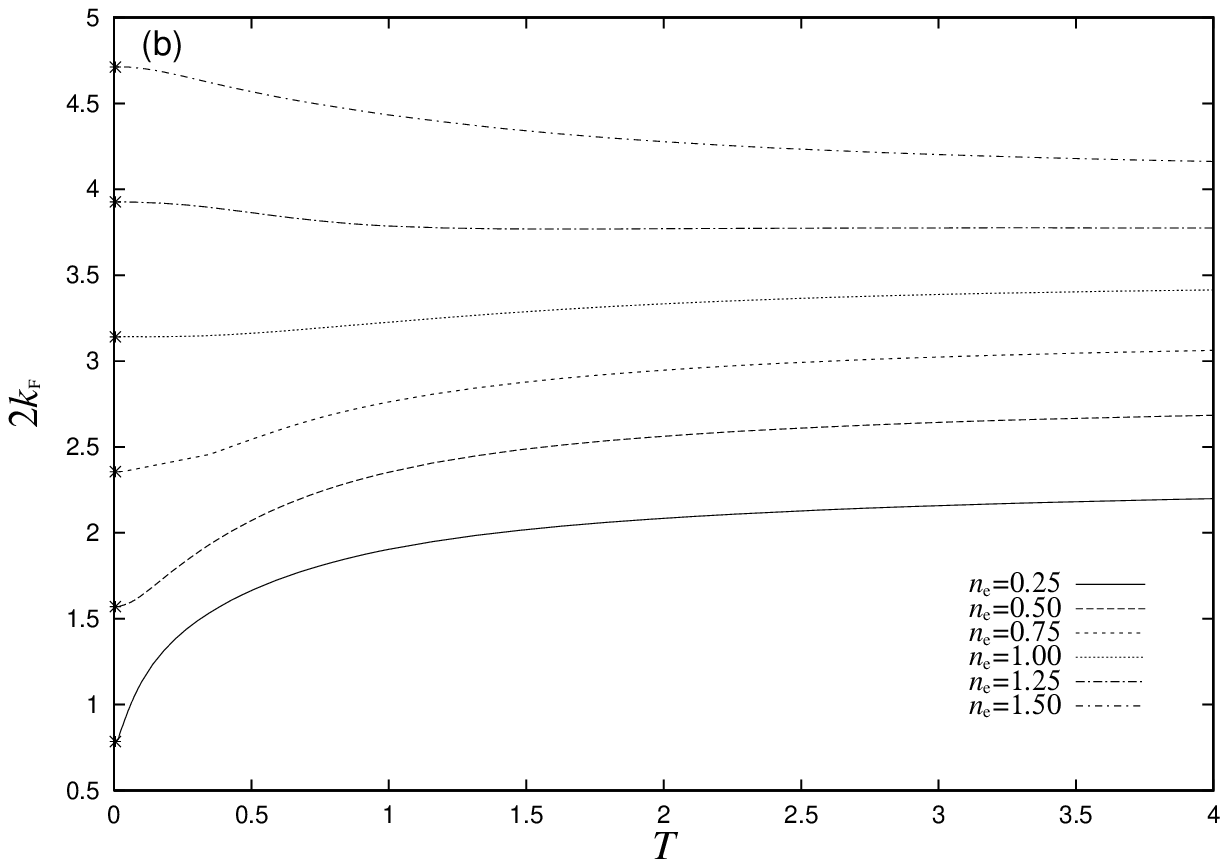}
\end{center}
\caption{(a): Temperature dependence of the correlation lengths
for the density-density correlations. The wave vector of the oscillation
term is depicted in (b). The symbol $\ast$ denotes
results of the analytical calculation in section~4 in 
the low temperature limit and the predictions from CFT in Appendix~A.}
\label{graphdd}
\end{figure}

Let us consider the NLIE characterizing the 
correlation lengths via  formula~\eqref{correlation}.
Fig.~\ref{graphgr}a shows the temperature dependent
correlation lengths (multiplied by $T$) 
of the one-particle Green functions for
various particle densities.
The case $\alpha=\gamma=1$ 
is used as a concrete example throughout this paper and is
equivalent to $U/2=t_p=1$ in~\eqref{hamiltonian}.
Due to the finite gap of single particle excitations, we observe that the
correlation length multiplied by temperature 
shows monotonous decay with decreasing temperature 
with zero limit for zero temperature.
We observe a crossover behavior of the
correlation lengths at low
temperature close to particle density $n_{\rm e}=1.5$: for 
$n_{\rm e}<1.5$ ($n_{\rm e}>1.5$) the values of $\xi$ increase (decrease)
with $n$.
A similar behavior is observed in other correlations 
described by massive excitations.
The transversal spin-spin 
correlations, the triplet
superconducting pair correlations
are depicted in Fig.~\ref{graphst}a and
in Fig.~\ref{graphph}b, respectively.
%
%
%
%
%
%
%
%
%
%
%
We also depict the temperature dependent  
oscillatory terms of the one-particle
Green functions, transversal spin-spin correlations,
and density-density correlations in Fig.~\ref{graphgr}b,
Fig.~\ref{graphst}b and Fig.~\ref{graphdd}b, respectively.
Regarding the one-particle Green functions,
one observes clearly a temperature dependence of 
the ``Fermi momentum" $k_{\rm F}$.
In the low temperature limit $T\to0$, the wave vector converges to
the expected value $k_{\rm F}=\pi n_{\rm e}/2$ which
indicates the significance of the Fermi surface for
one-particle excitations in the Tomonaga-Luttinger liquid at $T=0$.
As seen in Fig.\ref{graphchem} (and also expected) the chemical
potential grows to $+\infty$ ($-\infty$) for $T\to\infty$ if the
particle density is kept fixed at a high (low) value.  This indicates
a broadening of the momentum distribution at finite and in particular
at high temperatures. Hence, processes of particle-hole type closer to
the middle of the band become statistically dominant. Consequently,
the wave vector $k_{\rm F}$ decreases (increases) with increasing $T$
if the $T=0$ value was large (small).
As shown in Fig.~\ref{graphdd}b and Fig.~\ref{graphst}b,
oscillatory terms of the
density-density and transversal spin-spin correlations
converge to $2k_{\rm F}\to \pi n_{\rm e}$.
This behavior corresponds to the excitations 
carrying a $2k_{\rm F}$ momentum.
For the same reason as mentioned in the case of the one-particle 
Green function, this $2k_{\rm F}$  is
shifted with increasing temperature.
The quantitative difference of the 
oscillation terms for the transversal 
spin-spin correlations and the density-density 
correlations lies in the different nature of these excitations.
The density-density correlations 
are described by particle-hole excitations from 
the left to the right Fermi point without
changing the electron and spin numbers.
On the other hand, the transversal spin-spin 
correlations are characterized by massive excitations
changing the spin number.
%
%

\begin{figure}[h]
\begin{center}
\includegraphics[width=0.495\textwidth]{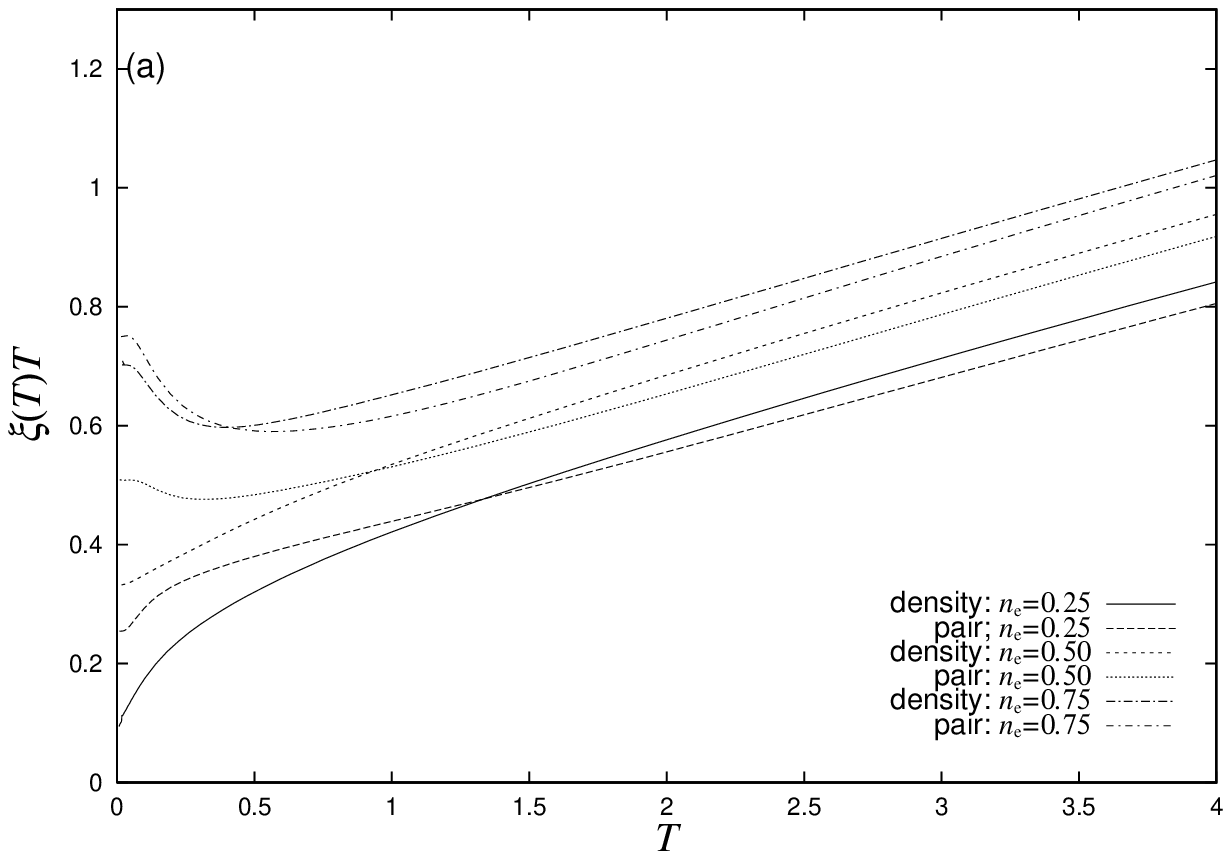}
\includegraphics[width=0.495\textwidth]{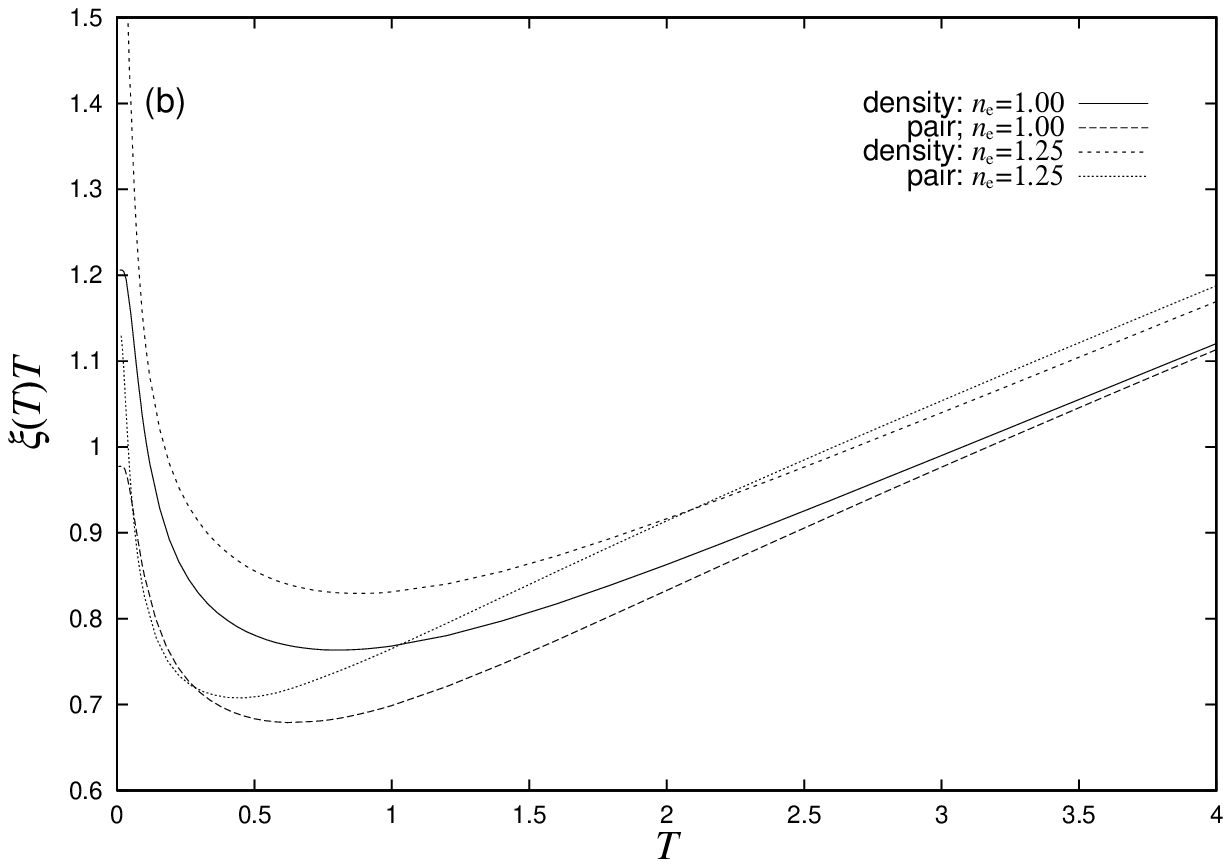}
\end{center}
\caption{(a) Low particle density: crossover from dominant 
(superconducting) singlet pair correlations at low temperature
to dominant density-density correlations at high temperature. (b)
High particle density: opposite to scenario in (a) for density $n=1.25$
(upper two curves). For $n=1.0$ the crossing lies outside the shown
temperature window.}
\label{graphcross}
\end{figure}

Fig.~\ref{graphdd}a and Fig.~\ref{graphph}a show
the density-density and the singlet
superconducting pair correlation lengths, respectively.
Due to massless excitations, the scaled correlation lengths $T\xi(T)$ 
are finite at zero temperature. 
The low temperature asymptotics agree with
the analytic calculations in  section~4
and the prediction from CFT in Appendix~A
(see Fig.~\ref{lowtemp}).
At low temperatures one observes a crossover
from dominant density-density correlations
to dominant singlet superconducting
pair correlations driven by the
particle density 
(see also Fig.~\ref{lowtemp} in Appendix~A):
below (above) a certain critical density $\rho_c$, 
the dynamics is dominated by singlet superconducting
pair correlations (density-density correlations).
This behavior is changed at higher temperatures. Here at
low (high) density $\rho$ the density-density correlations
(singlet superconducting pair correlations) dominate,
see Fig.~\ref{graphcross}.
Of course, these findings for high particle density and temperature 
do not imply any tendency towards superconductivity. It should be born 
in mind that the overall length scale of the density-density and
singlet superconducting pair correlations is rather short in comparison
to the lattice constant. Furthermore, neither of these correlations
dominate the dynamics with increasing temperature as the one-particle 
Green function takes over. This implies a picture of rather uncorrelated
one-particle dynamics at higher temperature.
%
%
%
\begin{figure}
\begin{center}
\includegraphics[width=0.495\textwidth]{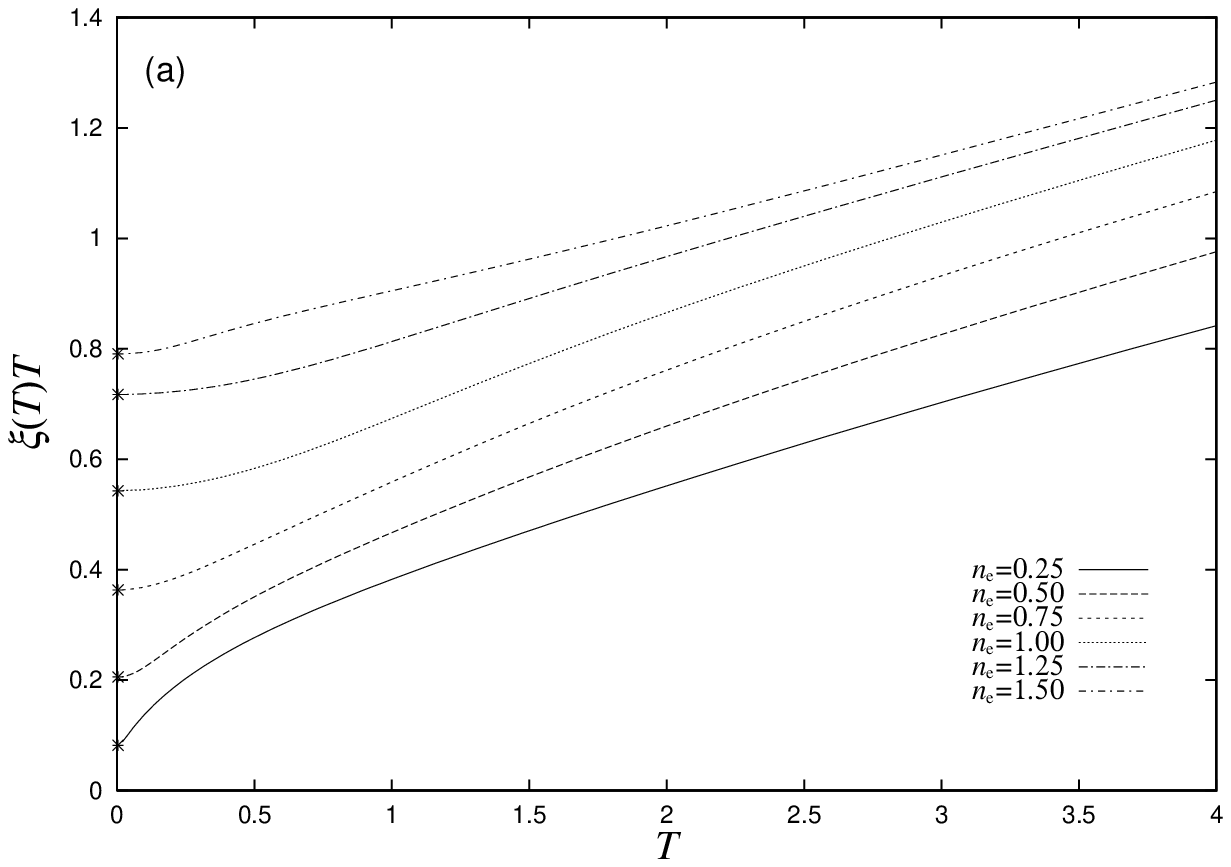}
\end{center}
\caption{Temperature dependence of the sub-dominant
density-density (longitudinal spin-spin) correlation lengths.
At low temperature there is convergence to the Fermi
velocity $v_{\rm F}/(2\pi)$ (denoted by the symbol $\ast$).}
\label{graphsl}
\end{figure}%
%
%

Finally we comment on the sub-dominant density-density
(longitudinal spin-spin) correlations in Fig.~\ref{graphsl}.
The correlations are characterized by simple
particle-hole excitations at each Fermi point.
Correspondingly,
the correlation lengths (multiplied by 
temperature $T$) approach the Fermi velocities
$v_{\rm F}/(2\pi)$ in the low temperature limit.
%
\section{Low-temperature asymptotics}
%
In this section we consider the model at low temperatures.
The largest eigenvalue and 
those sub-leading ones described by massless excitations
are calculated approximately up to order $O(1/\beta)$.
As a consequence, the NLIEs are reduced to only one 
linear integral equation and
this linear integral equation is connected directly with
the dressed functions (see Appendix~A). 

Due to the massive spin excitations,
the two auxiliary functions $\ln\mfa_0(v)$ and $\ln\mfab_0(v)$
can be neglected as $\mfa_0, \mfab_0 \sim e^{-\epsilon\beta}$ 
($\epsilon>0$).
Hence the three NLIEs characterizing the grand potential 
and correlation functions 
%
reduce to the single integral equation
%
%
\begin{equation}
\ln\mfa_1(v)=-\beta \varepsilon^{(0)}(v)+k_1
\stackrel{C_1}{*}\ln\mfA_1(v)
+\ln\tilde\varphi_1(v),
\label{asympt:nlie}
\end{equation}
where 
\begin{equation}
\varepsilon^{(0)}(v)=-\psi(v)-\psi(-v)-2\mu.
\end{equation}
%
%
and $\widetilde\varphi_1$ is a collection of only the ``cos''-terms 
of the function $\varphi_1$ listed 
in Appendix~C. The ``sin''-terms do not show up explicitly as they
are taken care of by the modified contour $\mcl_1$ (see Appendix~C). 
Throughout this section the convolution
$f*g(v)$ is understood in the manner
\begin{equation}
f*g(v):=\frac{1}{\pi}
\int_{\cl}f(v-x)g(x) dx.
\end{equation}
Numerically, one observes that the function $\mfa_1(v)$ has
a crossover behavior from $\mfa_1(v)\ll 1$ to $\mfa_1(v)\gg1$
(see Fig.~\ref{cross}).
%
%
The Fermi point $\Lambda_{\rm F}$ can be defined by
the following conditions satisfied by 
$\ln\mfa_1(v)$ for the largest eigenvalue 
($\mcl_1$ is taken by a straight line
on the real axis),
\begin{equation}
\ln\mfa_1(\pm \Lambda_{\rm F})=0 \qquad (1/\beta\ll 1),
\label{fermip}
\end{equation}
and for the excited states we demand $\Re \ln\mfa_1(\pm \Lambda_{\rm F})=0$
for $\pm \Lambda_{\rm F}$ on the real axis.
\subsection{$O(\beta)$ and $O(1)$ approximation}
\begin{figure}[t]
\begin{center}
\includegraphics[width=0.495\textwidth]{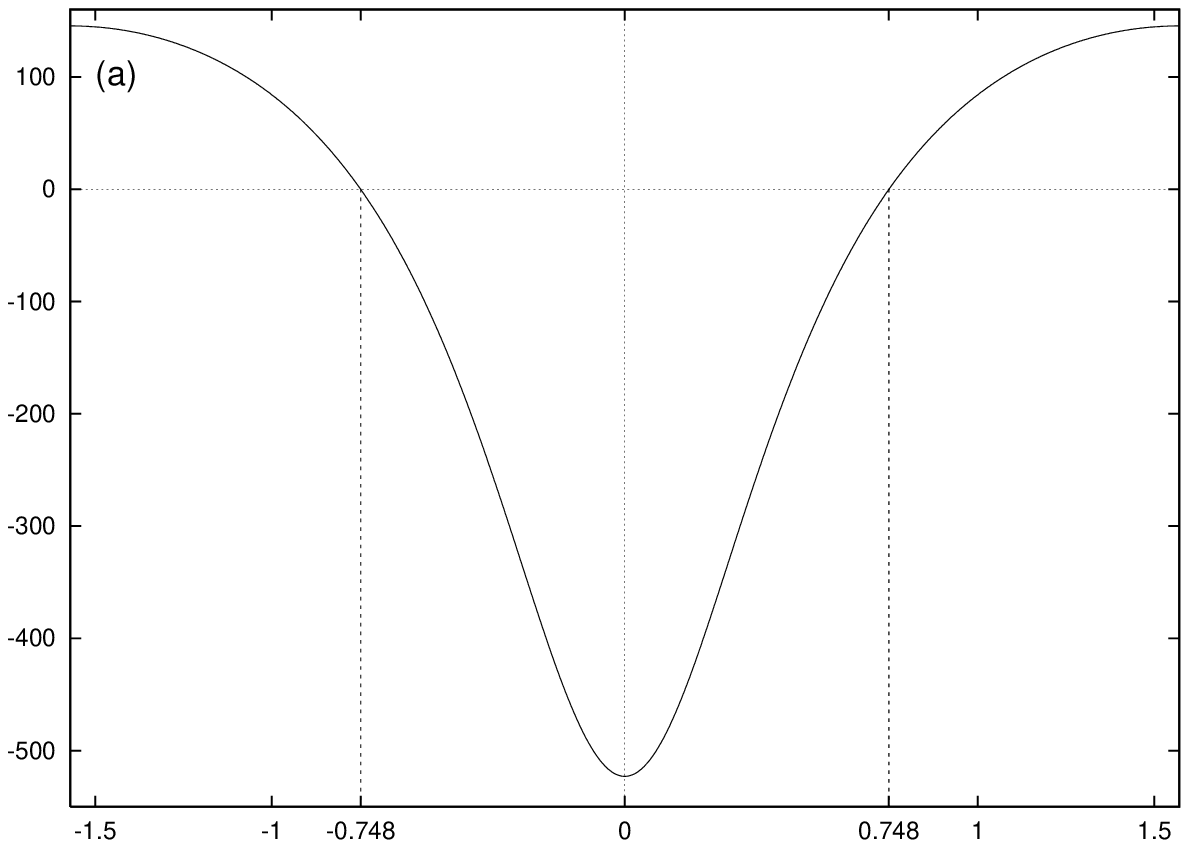}
\includegraphics[width=0.495\textwidth]{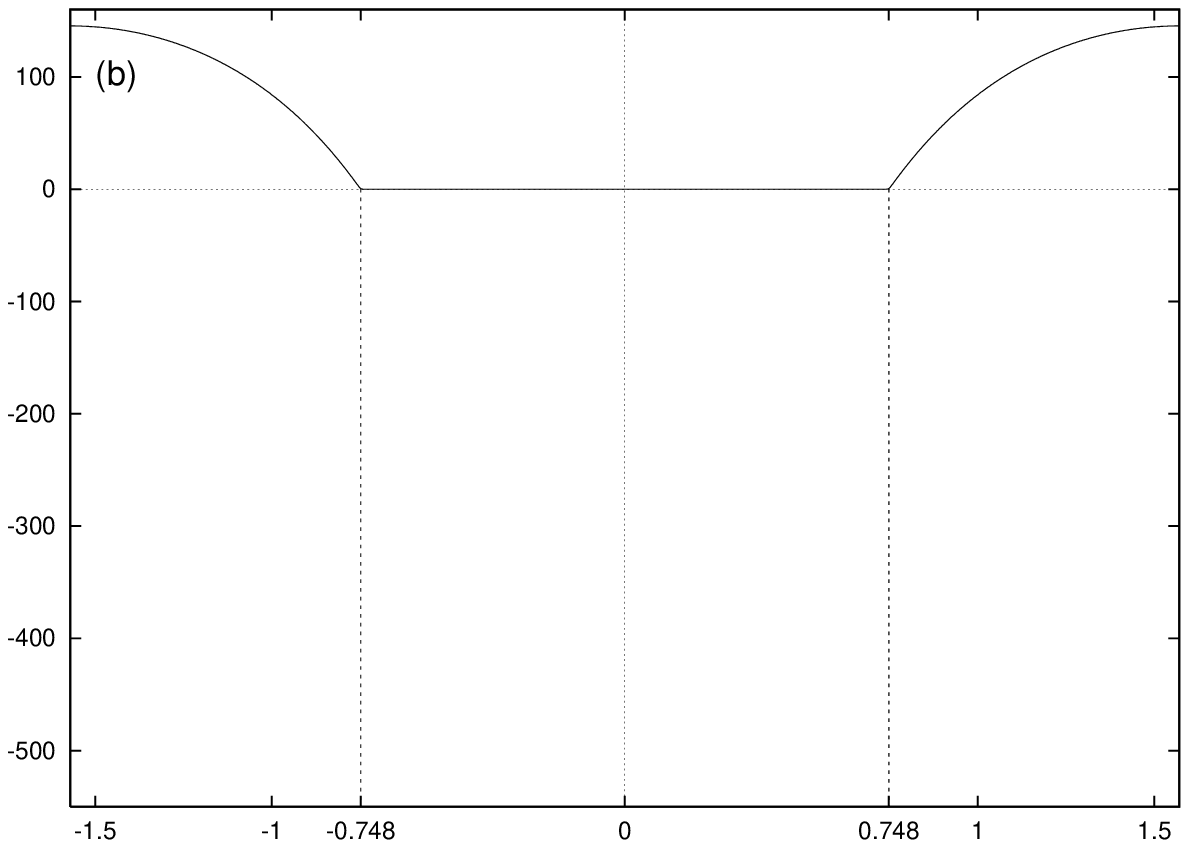}
\end{center}
\caption{Crossover behavior of the functions 
$\ln\mfa_1(v)$ and $\ln\mfA_1(v)$ (real parts)
for $\alpha=\gamma=1$, $n_{\rm e}=1.00$ and 
$\beta=50$. 
The Fermi point 
$\Lambda_{\rm F}$ defined by~\eqref{fermip}
is found to be  $\Lambda_{\rm F}=\pm0.748$.}
\label{cross}
\end{figure}

First we calculate order $O(\beta)$ and $O(1)$ approximations
for the largest and those sub-leading ones 
characterized by gapless excitations:
(i) particle-hole excitations at each of the left and 
right Fermi points
(define $n^{+}$ and $n^{-}$ as corresponding quantum numbers),
(ii) particle-hole excitations of $d_{\rm c}$
charges from the left to right Fermi point, but no change
of the total charge number and total spin number, and
(iii) singlet pair excitations near the Fermi points;
annihilate or create 
$\varDelta N_c=2 n_{c}$ ($n_{\rm c}\in{\mathbb{Z}}$)
charges in a symmetric way near the Fermi points.
The longitudinal spin-spin correlations
$\langle \sigma_j^{z}\sigma_i^{z}\rangle$ and the density-density 
correlations $\langle n_j n_i \rangle$ are characterized 
by excitations (i) and (ii). 
On the other hand, the singlet superconducting pair 
correlations $\langle c_{j+1,\uparrow}c_{j,\downarrow}
 c_{i+1,\uparrow}c_{i,\downarrow}\rangle$
is described by (iii).\\\\
%
{\bf (1) The largest eigenvalue}\\\\
%
In the low temperature limit, the function $\ln\mfA_1(v)$ 
can be written in terms of  $\mfa_1(v)$ as
$\ln\mfA_1(v)\simeq \ln\mfa_1(v)$ for $|v|>\Lambda_{\rm F}$
and $\ln\mfA_1(v)\simeq 0$ for $|v|<\Lambda_{\rm F}$
(Fig.~\ref{cross}). 
Hence we have
\begin{equation}
\ln\mfa_1(v)=-\beta\varepsilon^{(0)}(v)+k_1\stackrel{C_1}{*}\ln\mfa_1(v)
              +O(1/\beta),
\end{equation}
where the symbol $\stackrel{C_1}{*}$ for the
case of the largest eigenvalue denotes
\begin{align}
&k_1\stackrel{C_1}{*}\ln\mfa_1(v)=\frac{1}{\pi}
\int_{C_1} k_1(x)\ln\mfa_1(v-x)d x\nn \\
&\quad=\frac{1}{\pi}\left(\int_{-\frac{\pi}{2}}^{-\Lambda_{\rm F}}+
 \int_{\Lambda_{\rm F}}^{\frac{\pi}{2}}\right)k_1(x)\ln\mfa_1(v-x)d x.
\end{align}
Using the definition of the dressed energy $\varepsilon(v)$
\eqref{d-energy}, we find a solution to the above equation
\begin{equation}
\ln\mfa_1(v)=-\beta \varepsilon(v)+O(1/\beta).
\end{equation}
Then we have an approximate value for $\ln \mfa_1(v)$ 
at the Fermi points
\begin{equation}
\ln\mfa_1(\pm \Lambda_{\rm F})=0+O(1/\beta).
\label{ap1}
\end{equation}
%
%
%
{\bf (2) Particle-hole excitations at each of left
 and right Fermi points}\\\\
This excitation is characterized by excited charges at the left
(right) Fermi points $-\Lambda_{\rm F}$ ($\Lambda_{\rm F}$).
In the simplest, still characteristic case, one zero is separated by
$n^-$ ($n^+$) holes from the Fermi sea of zeros at the Fermi point
$\mp\Lambda_{\rm F}$.
The holes (zeros) are to be circumvented by the deformed contour in
anti-clockwise (clockwise) manner (see Appendix~C).
The function $\ln\mfA_1(v)$ can be approximated in terms of $\mfa_1(v)$
in the same manner as case (1). Hence we obtain the
same approximate value as in~\eqref{ap1}.\\\\
%
{\bf (3) Particle-hole excitations from left to
                     right Fermi points}\\\\
%
%
These excitations are described by $d_{\rm c}$ charges
moved from the  left (right) Fermi point
to right (left) one.
This excitation entails a change of momentum $2k_{\rm F} d_{\rm c}$.
Correspondingly, the auxiliary function $\mfA_1(v)$ has $2 d_{\rm c}$
zeros near the Fermi points with behavior different from those in 
the largest eigenvalue case: $d_{\rm c}$ zeros (of
``particle type'') are distributed in the upper half plane near the
left Fermi point and the remaining $d_{\rm c}$ zeros (of ``hole type'')
are located in
the lower half plane near the right Fermi point, and vice versa.
\begin{figure}[t]
\begin{center}
\includegraphics[width=0.495\textwidth]{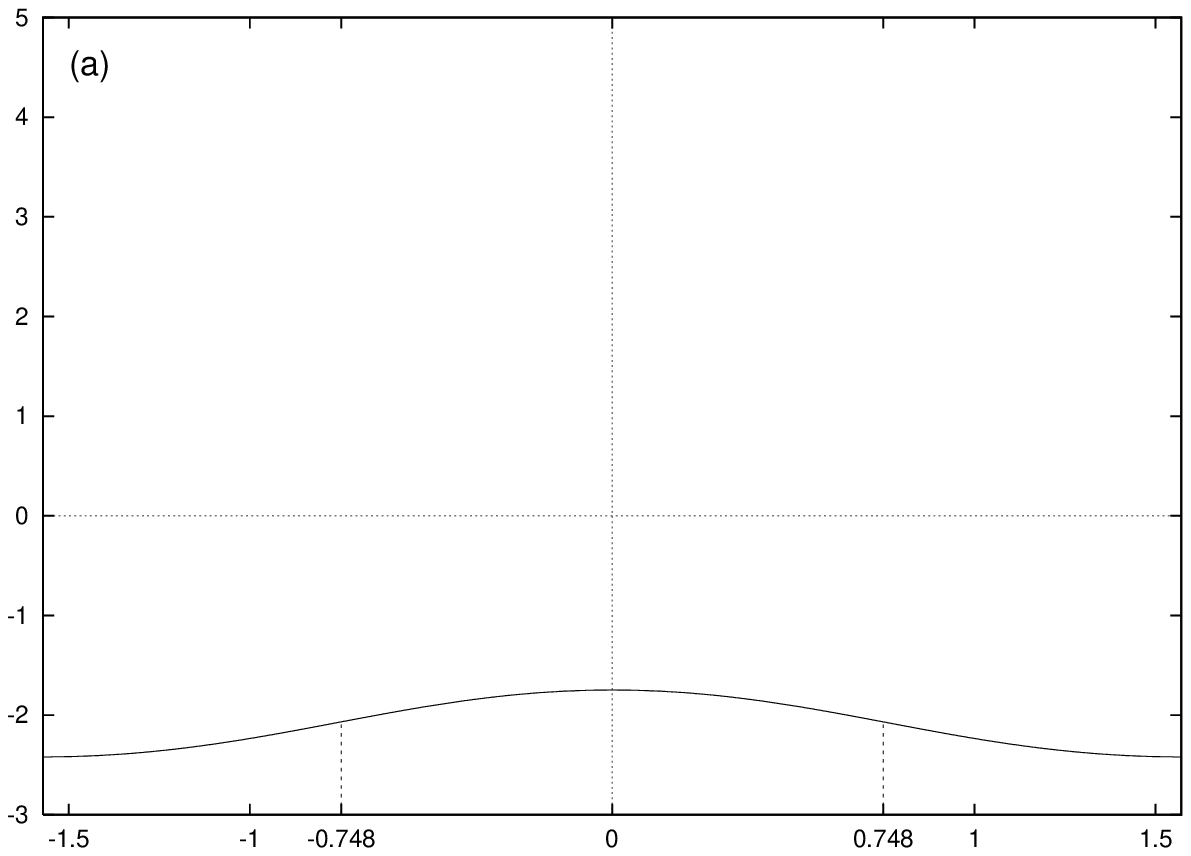}
\includegraphics[width=0.495\textwidth]{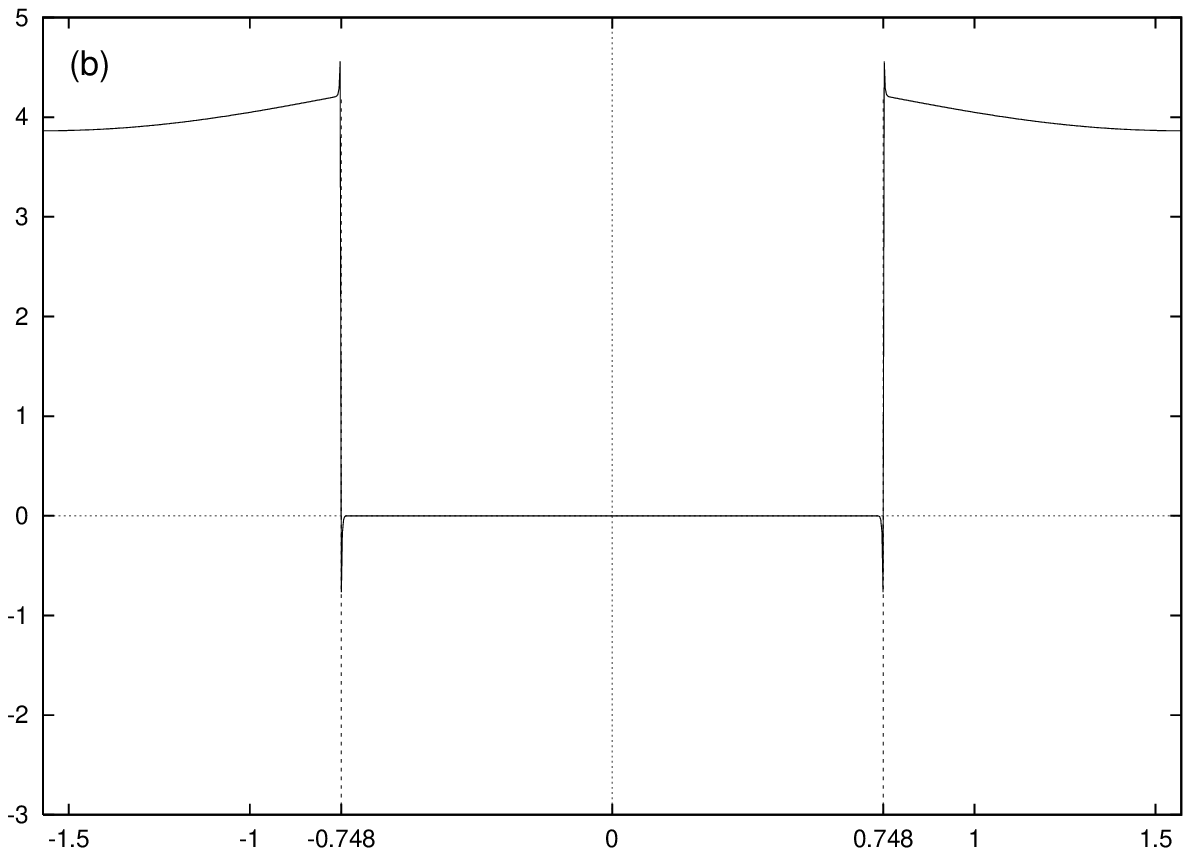}
\end{center}
\caption{Numerical results for the functions 
$\Im\ln\mfa_1(v)$ (a) and $\Im\ln\mfA_1(v)$ (b) for $\alpha=\gamma=1$, 
$n_{\rm e}=1.00$, $\beta=50$ and $d_{\rm c}=1$.
Due to the additional zeros (circumvented by the
modified integration contour $\mcl_1$), 
$\ln\mfA_1(v)$ is discontinuous
near the Fermi points $\Lambda_{\rm F}=\pm 0.748$.}
\label{crossd}
\end{figure}%

\begin{figure}[t]
\begin{center}
\includegraphics[width=0.495\textwidth]{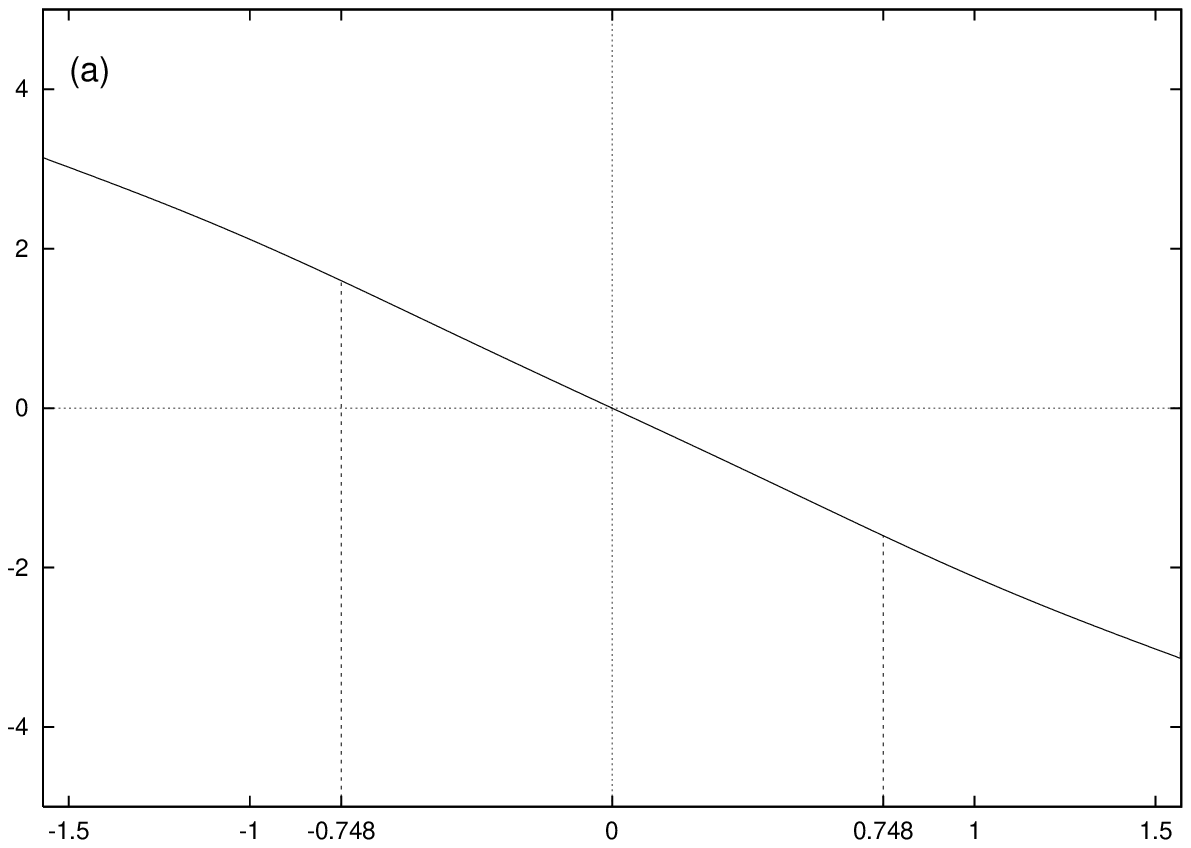}
\includegraphics[width=0.495\textwidth]{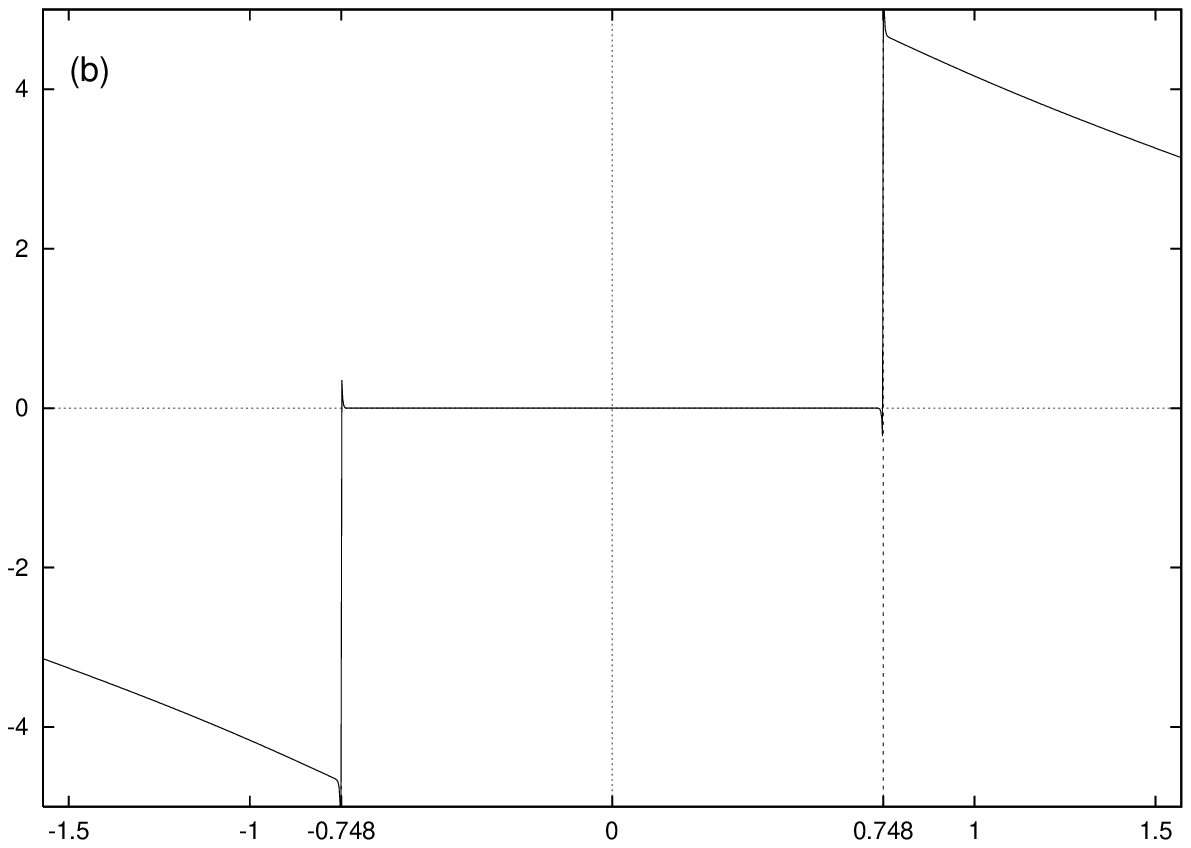}
\end{center}
\caption{
Numerical results for the functions 
$\Im\ln\mfa_1(v)$ (a) and $\Im\ln\mfA_1(v)$ (b) 
for $\alpha=\gamma=1$, 
$n_{\rm e}=1.00$, $\beta=50$ and $n_{\rm c}=1$.
%
%
One observes that the function $\ln\mfa_1(v)$ is a superposition
of a periodic function and a linearly increasing term.
The function $\ln\mfA_1(v)$ is 
discontinuous at the Fermi points 
$\Lambda_{\rm F}=\pm 0.748$.
The discontinuity derives from the additional
zeros (circumvented by the contour)
at the Fermi points.
}
\label{crossp}
\end{figure}

These zeros should be circumvented by the deformed
contours in  clockwise manner at the left Fermi 
point and anti-clockwise manner at the right Fermi point
(see Appendix~C).
At low temperatures, $\ln \mfA_1(v)$ is replaced by 
$\ln \mfA_1(v)\simeq 0$ for $|v|<\Lambda_{\rm F}$
and $\ln \mfA_1(v)\simeq \ln\mfa_1(v)+2\pi i d_{\rm c}$ for 
$|v|>\Lambda_{\rm F}$ (see Fig.~\ref{crossd} for $d_{\rm c}=1$).
Thus we have the linear integral equation characterizing
the excitation.
\begin{equation}
\ln\mfa_1(v)=-\beta\varepsilon^{(0)}(v)+
              k_1\stackrel{C_1}{*}\ln(\mfa_1(v)+2\pi i d_{\rm c})+
                                    O(1/\beta),
\end{equation}
where $C_1$ denotes the integration in the region
$|v|>\Lambda_{\rm F}$ along with the modified 
contours $\mcl_1$
defined in Appendix~C.
{}From the definition of the dressed energy 
$\varepsilon(v)$~\eqref{d-energy}
and the dressed charge $Z(v)$~\eqref{d-charge}, 
we have a solution to 
the above equation,
\begin{equation}
\ln\mfa_1(v)=-\beta \varepsilon(v)+
              \left(\frac{Z(v)}{2}-1\right)\cdot
               2\pi i d_{\rm c}+O(1/\beta).
\end{equation}
Thus we obtain the value
\begin{equation}
\ln\mfa_1(\Lambda_{\rm F})=\left(\frac{Z(\Lambda_{\rm F})}{2}-1
                   \right)\cdot 2\pi i d_{\rm c} +O(1/\beta).
\end{equation}
%
%
%
{\bf (4) Singlet pair excitations} \\\\
%
As mentioned above, singlet pair excitations are
characterized by annihilated $2 n_{\rm c}$ charges.
Correspondingly, one observes the function $\mfA_1(v)$ has
$2n_{\rm c}$ additional zeros near the Fermi points, i.e.
$n_{\rm c}$ zeros are located near the left Fermi point 
and remaining $n_{\rm c}$ zeros are located near the right
Fermi point.
All these additional zeros are distributed in the 
lower half plane and are circumvented by the deformed
contour in a manner different from case~(3),
in counter-clockwise manner 
at both Fermi points (see Appendix~C).
Simultaneously, $n_{\rm c}$ additional poles appear in the 
upper half plane and are located on the $\pi/2$-axis.
Though these poles do not appear explicitly in the NLIE,
they lead to a non-vanishing winding number of $\mfa_1(v)$, i.e.
$\ln\mfa_1(v)$ is not periodic (see Fig.~\ref{crossp})
\begin{equation}
\Im[\ln\mfa_1(\pm \pi/2)]=\mp  n_{\rm c} \pi.
\end{equation}
In the  low temperature limit, 
$\ln\mfA_1(v)$ can be replaced by 
$\ln\mfA_1(v)\simeq 0$ for $|v|<\Lambda_{\rm F}$, 
$\ln\mfA_1(v)\simeq \ln\mfa_1(v)-2\pi i n_{\rm c}$ 
for $-\pi/2<v<-\Lambda_{\rm F}$
and
$\ln\mfA_1(v)\simeq \ln\mfa_1(v)+2\pi i n_{\rm c}$ 
for $\Lambda_{\rm F}<v<\pi/2$ (see Fig.~\ref{crossp}).
%
%
%
%
%
Thus we obtain the integral equation 
\begin{equation}
\ln\mfa_1(v)=-\beta\varepsilon_0(v)+
k_1\stackrel{C_1}{*}S[\ln\mfa_1](v)+O(1/\beta),
\label{ap4}
\end{equation}
%
%
where the symbol $\stackrel{C_1}{*}$ denotes the convolution
and the functional $S[g]$ applied to an analytic function
$g(v)$ is defined stepwise by
\begin{equation}
S[g](v)=\begin{cases}
          g(v)-2\pi i n_{\rm c} & \text{for $-\pi/2<v<-\Lambda_{\rm F}$}\\
           0   & \text{for $-\Lambda_{\rm F}<v<\Lambda_{\rm F}$}\\
           g(v) + 2\pi i n_{\rm c}   & \text{for $\Lambda_{\rm F}<v<\pi/2$}
          \end{cases},
\end{equation}
with ``analytic continuation'' out of the interval $[-\pi/2,\pi/2]$
by the description $S[g](v+\pi)=S[g](v)+4\pi i n_c$.
We directly see that the real part of the solution to \eqref{ap4} is given 
by the dressed energy~\eqref{d-energy}
\begin{equation}
\ln\mfa_1(v)=-\beta \varepsilon(v)+i f(v),
\label{asympt:a}
\end{equation}
and the imaginary part $f$ has to satisfy the linear integral equation
\begin{align}
i f(v)=k_1\stackrel{C_1}{*}S[i f](v).
\label{f}
\end{align}
Obviously, the solution is unique and odd. Taking the derivative of 
\eqref{f} we find
\begin{align}
f^{\prime}(v)&=f_0^{\prime}(v)+k_1\stackrel{C_1}{*}
f^{\prime}(v), \nn \db
f_0^{\prime}(v)&=\frac{1}{\pi}\Big\{
         k_1(v-\Lambda_{\rm F})(f(\Lambda_{\rm F})+2n_{\rm c}\pi) 
+k_1(v+\Lambda_{\rm F})(-f(-\Lambda_{\rm F})+2n_{\rm c}\pi)
                \Bigr\}.      
\label{f0}  
\end{align}
The symbol $C_1$ denotes the integration 
contour 
$[-\pi/2,-\Lambda_{\rm F}]\bigcup [\Lambda_{\rm F},\pi/2]$. 
{}From the relation $f(-\pi/2)=-f(\pi/2)=n_{\rm c}\pi$, we find
for the odd function $f(v)$ 
\begin{equation}
2f(\Lambda_{\rm F})+2n_{\rm c}\pi=
-\int_{C_1}f^{\prime}(v) dv=
-\int_{C_1}\frac{Z(v)f_0^{\prime}(v)}{2} dv.
\end{equation}
Substituting \eqref{f0} and using the definition of 
the dressed charge $Z(v)$~\eqref{d-charge}, 
we arrive at
\begin{equation}
f(\Lambda_{\rm F})=\left(\frac{1}{Z(\Lambda_{\rm F})}-1\right)
\cdot2\pi  n_{\rm c}.
\end{equation}
{}From eq.~\eqref{asympt:a} and the above result, 
$\ln\mfa_1(\Lambda_{\rm F})$ is expressed by 
\begin{equation}
\ln\mfa_1(\Lambda_{\rm F})=
\left(\frac{1}{Z(\Lambda_{\rm F})}-1\right)\cdot 2\pi i n_{\rm c} .
\end{equation}
\\
{\bf (5) General cases} \\\\
%
Due to linearity and the results in (1)--(4),
one finds 
\begin{equation}
\ln\mfa_1(\Lambda_{\rm F})=\left(\frac{Z(\Lambda_{\rm F})}{2}-1
          \right)\cdot 2\pi i d_{\rm c}+
\left(\frac{1}{Z(\Lambda_{\rm F})}-1\right)\cdot 2\pi i n_{\rm c}.
\end{equation}
%
\subsection{$O(1/\beta)$ corrections for the NLIEs}
To evaluate the $O(1/\beta)$ corrections for the non-linear
integral equations \eqref{asympt:nlie}, we must calculate 
the integral term, taking into account its behavior 
near the Fermi points $\pm \Lambda_{\rm F}$.
To achieve this, we divide the integration term into three 
parts as
%
%
%
\begin{align}
\frac{1}{\pi}&\int_{-\frac{\pi}{2}}^{\frac{\pi}{2}}
k_1(x)\ln\mfA_1(v-x)dx=\int_{-\frac{\pi}{2}}^{\frac{\pi}{2}}
k_1(v-x)\ln\mfA_1(x)dx+n_c\ln\frac{\cos(v+i\gamma)}{\cos(v-i\gamma)}\nn \db
&\rightarrow\frac{1}{\pi}\int_{C_1}k_1(v-x)
(\ln\mfa_1(x)+2\pi i d_{\rm c}+2\pi i\sigma(x))d x+
n_c\ln\frac{\cos(v+i\gamma)}{\cos(v-i\gamma)}\nn \\
&\quad+\frac{1}{\pi}\int_{C_1}k_1(v-x)
\ln\left(1+\frac{1}{\mfa_1(x)}\right)d x 
+\frac{1}{\pi}\int_{|v|<\Lambda_{\rm F}}k_1(v-x)
\ln(1+\mfa_1(x))d x,
\label{int1}
\end{align}
%
%
with $\sigma(x)=-n_c, 0, n_c$ for $x\in
[-\pi/2,-\Lambda_{\rm F}],
[-\Lambda_{\rm F},\Lambda_{\rm F}],
[\Lambda_{\rm F},\pi/2]$.
Note that 
the additional ``cos"-term which depends on
the definition of the convolution (see Appendix~C)
appears in the above equation.
As we have already treated the first linear 
term in eq.~\eqref{int1}, we concentrate on the last 
two non-linear terms.
Let us split the integration intervals into two parts;
$[-\pi/2,0]\bigcup[0,\pi/2]$.
The integrations over the negative interval read
\begin{equation}
\frac{1}{\pi}\int_{-\pi/2}^{-\Lambda_{\rm F}}k_1(v-x)
                 \ln\left(1+\frac{1}{\mfa_1(x)}\right) dx+
\frac{1}{\pi}\int_{-\Lambda_{\rm F}}^{0}k_1(v-x)
                                 \ln(1+\mfa_1(x)) dx.
\label{negInt}
\end{equation}
Changing the variable to $z=-\ln\mfa_1(v)$
and using the fact that the derivative of $\ln\mfa_1(v)$
is dominated by the one of $-\beta\varepsilon(v)$, i.e.
$dz/dv\simeq-\ln^{\prime}\mfa_1(-\Lambda_{\rm F})=
\beta\varepsilon^{\prime}(-\Lambda_{\rm F})=
-\beta\varepsilon^{\prime}(\Lambda_{\rm F})$,
one obtains
\begin{equation}
\eqref{negInt}=-\frac{k_1(v+\Lambda_{\rm F})}
   {\pi\beta\varepsilon^{\prime}(\Lambda_{\rm F})}
\left(\int_{-\ln\mfa_1(-\Lambda_{\rm F})}^{\infty}\ln(1+e^{-z}) dz+
\int_{-\infty}^{-\ln\mfa_1(-\Lambda_{\rm F})}\ln(1+e^{z}) dz\right).
\end{equation}
The integration contours should be taken to circumvent
the singularities (see Appendix~C).
The resultant integration is expressed as 
\begin{equation}
\rightarrow -\frac{k_1(v+\Lambda_{\rm F})}
   {\pi\beta\varepsilon^{\prime}(\Lambda_{\rm F})} 
\left(\frac{\pi^2}{6}+\frac{1}{2}[\ln\mfa_1(-\Lambda_{\rm F})+
                      2\pi i(d_{\rm c}-n_{\rm c})]^2-
                      4\pi^2  n^{-}\right).
\end{equation}
Adding the results of the positive real axis, we obtain
\begin{align}
&\frac{1}{\pi}\int_{C_1}k_1(v-x)
\ln\left(1+\frac{1}{\mfa_1(x)}\right)d x 
+\frac{1}{\pi}\int_{|v|<\Lambda_{\rm F}}k_1(v-x)
\ln(1+\mfa_1(x))d x \nn \\
&\quad=\frac{-\pi}
   {\beta\varepsilon^{\prime}(\Lambda_{\rm F})}
\left\{
\left(\frac{1}{6}-4\varDelta^{+}\right)k_1(v-\Lambda_{\rm F})+
\left(\frac{1}{6}-4\varDelta^{-}\right)
              k_1(v+\Lambda_{\rm F})\right\},\nn \db
&\varDelta^{\pm}=\frac{1}{2}\left(\frac{n_{\rm c}}{Z(\Lambda_{\rm F})}\pm
                 \frac{d_{\rm c}}{2}Z(\Lambda_{\rm F})\right)^2+n^{\pm}.
\end{align}
Here $\varDelta^{\pm}$ is nothing but the conformal dimensions
defined in~\eqref{conformald}.
Hence we obtain the NLIE up to $O(1/\beta)$,
\begin{align}
\ln\mfa_1(v)&=-\beta\varepsilon^{(0)}(v)+k_1\stackrel{C_1}{*}
(\ln\mfa_1+2\pi i d_{\rm c}+2\pi i \sigma)(v)+
n_c\ln\frac{\cos(v+i\gamma)}{\cos(v-i\gamma)} \nn \\
&\qquad-\frac{\pi}{\beta\varepsilon^{\prime}(\Lambda_{\rm F})}
\left\{\left(\frac{1}{6}-4\varDelta^{+}\right)k_1(v-\Lambda_{\rm F})+
              \left(\frac{1}{6}-4\varDelta^{-}\right)
              k_1(v+\Lambda_{\rm F})\right\}.
\label{asympt:a2}
\end{align}
%
%
%
\subsection{$O(1/\beta)$ corrections to the eigenvalues}
%
Repeating the same argument for the eigenvalues, we obtain
\begin{align}
\ln\Lambda(0)&=2\beta\ch(\alpha+1)\gamma-\int_{C_1}
             \frac{\psi(v)+\psi(-v)}{2\pi\sh(\alpha+1)\gamma}
                (\ln\mfa_1(v)+2\pi i d_{\rm c}) d v \nn \\
&\quad+\frac{2\pi\psi(\Lambda_{\rm F})
                      (1/6-2\varDelta^{+}-2\varDelta^{-})}
        {\beta\varepsilon^{\prime}(\Lambda_{\rm F})
                \sh(\alpha+1)\gamma}.
\label{loweigen}
\end{align}
%
%
Next we apply the integral identity for bare ($a_0$, $b_0$) and
dressed functions ($a$, $b$), i.e. if $a=a_0+k\ast a$, and
$b=b_0+k\ast b$ then we have $\int a b_0=\int a_0 b$.  Applying this
to eq.~\eqref{asympt:a2} (``$a=\ln\mfa_1+2\pi i d_{\rm c}+2\pi i
\sigma$'') and the density function (``$b=\varrho$'') defined in
\eqref{density}, we obtain
\begin{align}
&-\int_{C_1}
             \frac{\psi(v)+\psi(-v)}{2\pi\sh(\alpha+1)\gamma}
                (\ln\mfa_1(v)+2\pi i d_{\rm c}) d v 
=(\beta\mu+\pi i d_{\rm c}) n_{\rm e}+
     \frac{\beta}{2} \int_{C_1}
     (\psi(v)+\psi(-v))\rho(v)dv    \nn \db
&\quad-\frac{\pi}{2\beta\varepsilon^{\prime}(\Lambda_{\rm F})} 
\left\{\left(\frac{1}{6}-4\varDelta^{+}\right)
      \int_{C_1}
      k_1(v-\Lambda_{\rm F})\varrho(v) dv 
+\left(\frac{1}{6}-4\varDelta^{-}\right)
        \int_{C_1}
      k_1(v+\Lambda_{\rm F})\varrho(v) dv\right\}. 
\label{int}
\end{align}
Substituting~\eqref{int} for \eqref{loweigen} and
using~\eqref{density}, we arrive at
\begin{align}
\ln\Lambda(0)&=-\beta\bigl\{\varepsilon_0+
             2(n_{\rm e}-1)\ch(\alpha+1)\gamma-
                   \mu n_{\rm e}\bigr\}
  -\frac{\pi^2 \varrho(\Lambda_{\rm F}) 
        (1/6-2\varDelta^{+}-2\varDelta^{-})}
        {\beta\varepsilon^{\prime}(\Lambda_{\rm F})}+
        \pi i d_{\rm c} n_{\rm e} \nn \\
&=-\beta\bigl\{\varepsilon_0+2(n_{\rm e}-1)\ch(\alpha+1)\gamma-
                   \mu n_{\rm e}\bigr\}
   +\frac{\pi(1/6-2\varDelta^{+}-2\varDelta^{-})}
        {\beta v_{\rm F}}+\pi i d_{\rm c} n_{\rm e},
\label{asympteigen}
\end{align}
where $\varepsilon_0$ and $v_{\rm F}$ denote
the ground state energy per site~\eqref{gs} and 
the Fermi velocity~\eqref{fermiv}, respectively.
Using~\eqref{asympteigen} and~\eqref{correlation}, 
one obtains the low temperature asymptotics of the correlation
lengths:
\begin{equation}
\xi=\frac{\beta v_{\rm F}}{2\pi(\varDelta^{+}+\varDelta^{-})}.
\label{asymptcor1}
\end{equation}

{}From the above calculation and the numerical 
observation, we can describe the low temperature behavior 
of the grand potential~\eqref{free}
and the correlation lengths~\eqref{correlation} 
by selecting the ``quantum numbers" 
$(n_{\rm c},d_{\rm c},n^{+},n^{-})$.\\\\
%
{\bf (a) Free energy} \\\\
%
Selecting the quantum numbers  
$(n_{\rm c},d_{\rm c},n^{+},n^{-})
=(0,0,0,0)$, we obtain the $O(1/\beta)$ correction 
for the grand potential per site~\eqref{free},
\begin{equation}
f=\bigl\{\varepsilon_0+2(n_{\rm e}-1)\ch(\alpha+1)\gamma+
                   \mu n_{\rm e}\bigr\}
   -\frac{\pi}{6 \beta^2 v_{\rm F}}.
\end{equation}
This asymptotics agrees with CFT with 
central charge $c=1$.
\\
\\\\
{\bf (b) Density-density and longitudinal spin 
         correlations} \\\\
%
To obtain the low temperature asymptotics
of the correlation lengths $\xi_{\rm d}$ 
for the density-density correlations 
$\langle n_j n_i \rangle$, the selection rules require to 
set the quantum numbers to
$(n_{\rm c},d_{\rm c},n^{+},n^{-})
=(0,\pm 1,0,0)$.
Thus we obtain
\begin{equation}
\xi_{\rm d}=\frac{2\beta v_{\rm F}}{\pi Z(\Lambda_{\rm F})^2}.
\end{equation}
In addition to the above correlation lengths, one observes a 
$2k_{\rm F}$ oscillation term in the correlation functions
\begin{equation}
2k_{\rm F}=\pi n_{\rm e}.
\end{equation}
In the case of the correlation lengths of the
longitudinal spin-spin correlations $\xi_{\rm sl}$,
we have to choose the same quantum numbers as for 
the density-density correlations.
Hence we have
\begin{equation}
\xi_{\rm sl}=\xi_{\rm d}
\end{equation}
%
{\bf (b) Sub-dominant density-density
(longitudinal spin) correlations} \\\\
%
One also obtains the sub-dominant correlation 
lengths for the density-density or longitudinal spin-spin 
correlations which are described by the quantum numbers
$(n_{\rm c},d_{\rm c},n^{+},n^{-})
=(0,0,1,0)$ or $(0,0,0,1)$. 
The resultant asymptotics of the correlation lengths
are given by
\begin{equation}
\xi^{\rm sub}_{\rm d}=\xi^{\rm sub}_{\rm sl}=
\frac{\beta v_{\rm F}}{2\pi}.
\end{equation}
\\
{\bf (d) Singlet superconducting pair correlations} \\\\
%
Finally, we consider the low temperature behavior
of the correlation lengths $\xi_{\rm sp}$ for 
the singlet superconducting pair correlations
$\langle c_{j+1,\uparrow}^{\dagger}c_{j,\downarrow}^{\dagger}
c_{i+1,\uparrow}c_{i,\downarrow}\rangle$ determined from
the quantum numbers
$(n_{\rm c},d_{\rm c},n^{+},n^{-})
=(\pm 1,0,0,0)$. 
The result reads
\begin{equation}
\xi_{\rm s p}=\frac{\beta v_{\rm F}Z(\Lambda_{\rm F})^2}{2 \pi}.
\end{equation}

All these analytical calculations of 
the low temperature properties are 
consistent with the numerical results in the previous
section and the predictions from CFT in Appendix~A.
%
\section{Summary and discussion}
%
In this paper we have investigated  the finite temperature 
correlations of the strongly correlated electron system in
%
%
one dimension with $U_q(\mfsl(2|1))$-invariance
and
various types of interaction such as Hubbard
and correlating hopping terms. 
As a consequence we observe 
a competition of normal versus ``superconducting'' correlations.
By concrete calculations of the corresponding correlation 
lengths we have found dominant pair correlations for low particle
densities $n$ and temperatures $T$. In detail, we have found that singlet pair
correlation lengths dominate over density-density correlation lengths 
for low $n_{\rm e}$ and $T$, but also for high $n$ and $T$. In the latter case,
however, the one-particle Green function dominates over all other correlations.

The computational task was achieved by a first principles approach based on
a lattice path integral formulation of the Hamiltonian at finite temperature
and the subsequent diagonalization of a suitably defined
quantum transfer matrix
describing transfer in chain direction. The Bethe ansatz equations for
the leading and next-leading eigenvalues were reformulated in terms
of a finite set of non-linear integral equations where the limit
of infinite Trotter number could be taken analytically. The NLIEs
were solved numerically for arbitrary finite temperature and various 
particle densities. In the low-temperature limit the analytic investigation
confirmed the CFT picture and reproduced the dressed energy and dressed
charge formalism known from the finite size analysis of the $T=0$
problem.

The QTM was based on ${\mathbb{Z}}_2$-grading
formulations, which reflect the proper fermionic
statistics.
The formulations are applicable directly to
other strongly correlated electron systems.

It is an interesting problem to compare our results
for the fermionic
system with the ones of the corresponding spin system.
The eigenvalues of the QTM constructed by the (null 
Grassmann parity) $R$-matrix which satisfies the ordinary 
YBE are written as
\begin{align}
\Lambda^{\rm spin}(v)&=\phi_1(v)\frac{q_1(v+\frac{i}{2}\gamma(2\alpha+1))}
                           {q_1(v+\frac{i}{2}\gamma)}e^{2\mu\beta} 
+(-)^{N-n}\phi_2(v)\frac{q_1(v+\frac{i}{2}\gamma(2\alpha+1))}
                           {q_1(v+\frac{i}{2}\gamma)} 
                      \frac{q_2(v+i \gamma)}{q_2(v)}e^{\mu\beta} \nn \\
&\quad+(-)^{N-n}\phi_2(v)\frac{q_1(v+\frac{i}{2}\gamma(2\alpha+1))}
                           {q_1(v-\frac{i}{2}\gamma)}
                      \frac{q_2(v-i \gamma)}{q_2(v)}e^{\mu\beta}
+\phi_3(v)\frac{q_1(v+\frac{i}{2}\gamma(2\alpha+1))}
                           {q_1(v-\frac{i}{2}\gamma)}.
\end{align}
Due to the extra factor $(-)^{N-n}$ ($N$ denotes the
Trotter number and $n$ is the number of eloctrons), 
the one-particle Green
function for the original system no longer corresponds 
to the one for the spin system.
Consequently the characteristic $k_{\rm F}$-oscillations
for the one-particle Green function disappear and 
instead $2k_{\rm F}$ 
oscillations will enter the corresponding correlation 
function in the spin system.

In our work we have provided the framework for a systematic
investigation of the correlation lengths at finite temperature
with an application to a characteristic set of interaction parameters. 
We have not yet
performed a comprehensive study of the phase diagrams for a wider
range of parameters. This remains to be done in a future publication. 
Also, the effect of a non-vanishing magnetic field remains to be explored.
We expect that for the isotropic limit, i.e. $\gamma\to 0$, the 
``superconducting regime" in the density-temperature phase diagram
will shrink to a point. Also the effect of a magnetic field will lead
to similar NLIEs where the chemical potential terms $\beta\mu$ have to be
replaced by $\beta(\mu\pm h/2)$ and $2\beta\mu$ remains fixed. The solutions 
to these NLIEs are expected to break the symmetries of the Bethe 
ansatz patterns with respect to reflections at the imaginary and/or real axis
leading to independent Fermi momenta of spin up and spin down electrons.
\section*{Acknowledgment}
The authors acknowledge financial support by the Deutsche 
Forschungsgemeinschaft under grant No.~Kl 645/3-3.
The authors would like to thank F. G\"ohmann, N. Kawakami,
C. Scheeren,
M. Shiroishi and J. Suzuki for useful discussions. 
%
\rnc{\theequation}{A.\arabic{equation}}\setcounter{equation}{0}
%
\section*{Appendix A\,\, Low temperature behavior from CFT}
%
The ground state properties and the long-distance
behavior of some correlation functions have 
already been investigated in~\cite{BKZ} with the aid of
the root density method and conformal field theory.
Here we slightly modify the formulation of~\cite{BKZ} 
for direct comparison of our results with 
the prediction from CFT.
The ground state properties and low-lying excitations
considered here are characterized by the BAE 
\begin{align}
&\Biggl[\frac{\sin(\lambda_j+(\frac{\alpha}{2}+1)\gamma i)
             \sin(\lambda_j+\frac{\alpha}{2}\gamma i)}
            {\sin(\lambda_j-(\frac{\alpha}{2}+1)\gamma i)
             \sin(\lambda_j-\frac{\alpha}{2}\gamma i)}
             \Biggr]^L =\prod_{k=1}^{N_{\downarrow}}
            \frac{\sin(\lambda_j-\Lambda_k+i \gamma)}
                 {\sin(\lambda_j-\Lambda_k-i \gamma)}, \nn \\
&\prod_{j=1}^{N_{\rm e}/2}
            \frac{\sin(\Lambda_k-\lambda_j+i \gamma)}
                 {\sin(\Lambda_k-\lambda_j-i \gamma)}=
-\prod_{l=1}^{N_{\downarrow}}
            \frac{\sin(\Lambda_k-\Lambda_l+i \gamma)}
                 {\sin(\Lambda_k-\Lambda_l-i \gamma)},
\label{RBAE}
\end{align}
where $N_{\rm e}$ and $N_{\downarrow}$ denote the
total number of electrons and the number of
down-spin electrons, respectively.
We derive directly the above equation
from the fact that the ground state energy is described
by 2-string electron rapidities, and hence
shifting the rapidities $\lambda_j\to\lambda_j+ i \gamma/2$,
$\lambda_{j+1}\to\lambda_j-i\gamma/2$ in eq.(6) in~\cite{BKZ}.
The  ground state energy can be written as
\begin{align}
E_0&=-2\sum_{j=1}^{N_{\rm e}/2}[\cos(k(\lambda_{j}+i \gamma/2))+
                             \cos(k(\lambda_{j}-i \gamma/2))] \nn \\
 &=-2\sum_{j=1}^{N_{\rm e}/2}
 \Biggl[2\ch(\alpha+1)\gamma+\frac{\psi(\lambda_j)+
        \psi(-\lambda_j)}{2}\Biggr],  
\end{align}
where
\begin{equation}
k(v)=\pi-2 \arctan\left(\frac{\tan v}
                  {\tnh((\alpha+1)\gamma/2)}\right).
\end{equation}
By solving the spin rapidities $\Lambda_{k}$ in terms
of the charge rapidities $\lambda_j$, we obtain the
density function $\varrho(v)$ in the thermodynamic limit,
\begin{equation}
\varrho(v)=-\frac{\psi(v)+\psi(-v)}{\pi\sh(\alpha+1)\gamma}+
         \frac{1}{\pi}\int_{C_1}k_1(v-x)\varrho(x)dx,
\label{density}
\end{equation}
where the Fermi points $\pm \Lambda_{\rm F}$ are
determined from the subsidiary condition for 
the total electron density $n_{\rm e}$
\begin{equation}
\int_{C_1}\varrho(x) dx=n_{\rm e}.
\label{defdens}
\end{equation}
Hence the ground state energy per site $\varepsilon_0$ 
is expressed in terms of the density function
\begin{equation}
\varepsilon_0=-\int_{C_1}
\left(2\ch(\alpha+1)\gamma+\frac{\psi(x)+\psi(-x)}{2}
\right)\varrho(x) dx.
\label{gs}
\end{equation}
The dressed energy $\varepsilon(v)$
\begin{equation}
\varepsilon(v)=\varepsilon^{(0)}(v)+
\frac{1}{\pi}\int_{C_1}k_1(v-x)\varepsilon(x)dx,
\label{d-energy}
\end{equation}
characterizes the low-lying excitations.
Here the bare energy function $\varepsilon^{(0)}(v)$ is
defined by (as in \eqref{bareenerg})
\begin{equation}
\varepsilon^{(0)}(v)=-\psi(v)-\psi(-v)-2\mu.
\label{bareenerg}
\end{equation}
Note that the chemical potential is shifted by
$\mu+2\ch(\alpha+1)\gamma\to\mu$.
The condition
\begin{equation}
\varepsilon(\pm \Lambda_{\rm F})=0,
\label{defchem}
\end{equation}
provides another way to define 
the Fermi points $\pm \Lambda_{\rm F}$ for given
chemical potential $\mu$.
\begin{figure}[htb]
\begin{center}
\includegraphics[width=0.495\textwidth]{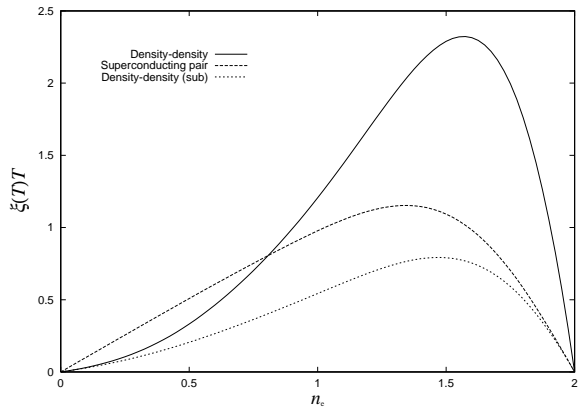}
\end{center}
\caption{Numerical results for the correlations 
based on CFT ($T\ll1$).
The density-density correlations, singlet superconducting
pair correlations and the sub-dominant density-density
correlations correspond to the quantum numbers
$(n_{\rm c}, d_{\rm c},n^{+},n^{-})=(0,\pm1,0,0)$, 
$(\pm1,0,0,0)$ and $(0,0,1,0)$ ($(0,0,0.1)$), respectively.}
\label{lowtemp}
\end{figure}%

As mentioned in section~4, 
we observe massless excitations described
by quantum numbers $(n_{\rm c},d_{\rm c},n^{+},n^{-})$.
Using the concept of CFT,
one obtains the correlation functions for 
primary fields as
\begin{equation}
\langle\phi_{\varDelta^{\pm}}(x)\phi_{\varDelta^{\pm}}(0)
\rangle=\frac{\exp(-2i d_{\rm c}k_{\rm F})}
{x^{2({\varDelta^{+}+\varDelta^{-}})}},
\label{algdecay}
\end{equation}
where $\varDelta^{\pm}$ are the conformal dimensions
of the primary operators
\begin{equation}
\varDelta^{\pm}=\frac{1}{2}\left(\frac{n_{\rm c}}{Z(\Lambda_{\rm F})}\pm
                 \frac{d_{\rm c}}{2}Z(\Lambda_{\rm F})\right)^2+n^{\pm}.
\label{conformald}
\end{equation}
Here the function $Z(\Lambda_{\rm F})$ denotes the dressed charge
determined from the integral equation
\begin{equation}
Z(v)=2+\frac{1}{\pi}\int_{C_1}k_1(v-x)Z(x)dx.
\label{d-charge}
\end{equation}

For finite temperatures in the scaling limit ($0<T\ll1$), where conformal 
invariance is valid, the correlation functions  are obtained from
the $T=0$ case by 
the following replacement~\cite{FK} in~\eqref{algdecay},
\begin{equation}
x\to\frac{v_{\rm F}\beta}{\pi}\sh\frac{\pi  x}{\beta v_{\rm F}},
\end{equation}
where $v_{\rm F}$ is the Fermi velocity 
defined by  
\begin{equation}
v_{\rm F}=\frac{\varepsilon^{\prime}(-\Lambda_{\rm F})}
             {\pi \varrho(v)}
           =-\frac{\varepsilon^{\prime}(\Lambda_{\rm F})}
             {\pi \varrho(v)}.
\label{fermiv}
\end{equation}
Thus the long-distance behavior of the correlation 
functions $G(x)$ is given by
\begin{equation}
G(x)\sim \cos(2k_{\rm F}d_{\rm c}x)\exp
\left(-\frac{2\pi (\varDelta^{+}+\varDelta^{-}) x}
     {\beta v_{\rm F}}\right),
\end{equation}
Fitting this with $G(x)\sim\exp(-x/\xi)$, one obtains
the correlation lengths $\xi$ at $T\ll 1$
\begin{equation}
\xi=\frac{\beta v_{\rm F}}{2\pi(\varDelta^{+}+\varDelta^{-})}.
\label{asymptcor2}
\end{equation}
The numerical results of~\eqref{asymptcor2} 
are shown in Fig.~\ref{lowtemp} for 
$(n_{\rm c}, d_{\rm c},n^{+},n^{-})=(0,\pm1,0,0)$, 
$(\pm1,0,0,0)$ and $(0,0,1,0)$ ($(0,0,0,1)$). 
These low temperature asymptotics of the correlation
lengths are consistent with~\eqref{asymptcor1}.
%
%
%
\rnc{\theequation}{B.\arabic{equation}}\setcounter{equation}{0}
%
\section*{Appendix B\,\, Derivation of  NLIE for 
largest eigenvalue}
%
Here we derive the NLIE for the largest eigenvalue
by using the auxiliary functions defined by~\eqref{aux}.
The derivation is applicable to the general
eigenvalues after a simple modification of the integration
contours.

The auxiliary functions are defined by certain 
combinations of $\lambda_j(x)$ ($1\le j\le4$):
\begin{alignat}{3}
&\mfa_0(v)=\frac{\lambda_1(x)(\lambda_3(x)+\lambda_4(x))}
               {\lambda_2(x)\Lambda(x)}, &\quad&
&\mfA_0(v)=\frac{(\lambda_1(x)+\lambda_2(x))
                        (\lambda_2(x)+\lambda_3(x)+\lambda_4(x))}
               {\lambda_2(x)\Lambda(x)}, \nn \db
&\mfab_0(v)=\frac{\lambda_2(x)}{\lambda_3(x)+\lambda_4(x)}, &\quad&
&\mfAb_0(v)=\frac{\lambda_2(x)+\lambda_3(x)+\lambda_4(x)}
                              {\lambda_3(x)+\lambda_4(x)}, \nn \db
&\mfa_1(v)=\frac{\lambda_1(x)}{\lambda_2(x)+\lambda_3(x)+\lambda_4(x)}, 
&\quad&
&\mfA_1(v)=\frac{\Lambda(x)}
                           {\lambda_2(x)+\lambda_3(x)+\lambda_4(x)},
\label{aux2}
\end{alignat}
where $x=v+\frac{i}{2}\alpha\gamma$.
Note that the eigenvalues are written in the following form
\begin{equation}
\Lambda(v)=\lambda_1(v)+\lambda_2(v)+\lambda_3(v)+\lambda_4(v).
\label{eigenlam}
\end{equation}
Using the definition of $\lambda_j(x)$, we can write
\begin{equation}
\lambda_1(x)+\lambda_2(x)=
\frac{q_1(x+\frac{i}{2}\gamma(2\alpha+1))}
     {q_1(x+\frac{i}{2}\gamma)}\left\{
     \phi_1(x)e^{2\mu\beta}+\phi_2(v)
     \frac{q_2(x+i\gamma)}{q_2(x)}e^{\mu\beta}\right\}.
\label{lam1}
\end{equation}
One finds that function~\eqref{lam1} has poles stemming from
$q_2(x)$.
Due to the BAE~\eqref{BAE}, the poles from $q_1(x+\frac{i}{2}\gamma)$
must be canceled out.
Hence the second factor in~\eqref{lam1}
must take the following form
\begin{align}
&\phi_1(x)e^{2\mu\beta}+\phi_2(v)
     \frac{q_2(x+i\gamma)}{q_2(x)}e^{\mu\beta} \propto 
 \frac{\sh^{\frac{N}{2}}(ix+u_N)q_1(x+\frac{i}{2}\gamma)
       q_2^{h}(x)}{\phi_{+}(x)
       \sh^{\frac{N}{2}}(ix-u_N-\alpha\gamma)q_2(x)}, \nn \db
&\phi_{+}(x)=[\sh(ix+u_N+\alpha\gamma)
          \sh(ix+u_N-(\alpha+1)\gamma)]^{\frac{N}{2}}.
\label{lam2}
\end{align}
where the function $q^{h}_2(x)$ is a certain 
analytic function which has $N+m-n$ zeros for the
largest eigenvalue case.
Substituting~\eqref{lam2} for~\eqref{lam1}, 
we obtain
\begin{equation}
\lambda_1(x)+\lambda_2(x) 
\propto \frac{\sh^{\frac{N}{2}}(ix+u_N)q_1(x+\frac{i}{2}\gamma(2\alpha+1))
       q_2^{h}(x)}
 {\phi_{+}(x)\sh^{\frac{N}{2}}(ix-u_N-\alpha\gamma)
  q_2(x)}.
\label{lam3}
\end{equation}
Similarly we find $\lambda_3(x)+\lambda_4(x)$ can be
written as
\begin{align}
&\lambda_3(x)+\lambda_4(x) 
\propto \frac{\sh^{\frac{N}{2}}(ix-u_N)q_1(x+\frac{i}{2}\gamma(2\alpha+1))
       q_2^{h}(x-i\gamma)}
 {\phi_{-}(x)\sh^{\frac{N}{2}}(ix+u_N-(\alpha+1)\gamma)q_2(x)}, \nn \db
&\phi_{-}(x)=[\sh(ix-u_N-\alpha\gamma)
              \sh(ix-u_N+(\alpha+1)\gamma)]^{\frac{N}{2}}.
\end{align}
Thanks to the BAE~\eqref{BAE}, the eigenvalues~\eqref{eigenlam}
are analytic functions (except for the trivial poles
which derive from the vacuum functions in~\eqref{DVF}).
Thus the eigenvalues~\eqref{eigenlam} must take the form
\begin{equation}
\Lambda(x)\propto
\frac{q_1(v+\frac{i}{2}\gamma(2\alpha+1))q_1^{h}(x)}
{\phi_{+}(x)\phi_{-}(x)},
\label{auxeigen}
\end{equation}
where $q_1^{h}(x)$ is an analytic function having
$2N-n$ zeros for the largest eigenvalue case.
After shifting the parameter $x\to v+\frac{i}{2}\alpha\gamma$,
we obtain the functional relations to be satisfied
by certain combinations of the auxiliary functions
\begin{align}
&\mfa_0(v)\propto\frac{[\sh(iv-u_N+\frac{\gamma}{2}\alpha)
                  \sh(iv+u_N-\frac{\gamma}{2}(\alpha+2))]^{\frac{N}{2}}
                 q_2^{h}(v+\frac{i}{2}\gamma(\alpha-2))}
                 {q^{h}_1(v+\frac{i}{2}\alpha\gamma)
                  q_2(v+\frac{i}{2}\gamma(\alpha+2))}, \nn \db
&\mfab_0(v)\propto\frac{[\sh(iv+u_N-\frac{\gamma}{2}\alpha)
                  \sh(iv-u_N+\frac{\gamma}{2}(\alpha+2))]^{\frac{N}{2}}
                 q_2(v+\frac{i}{2}\gamma(\alpha+2))}
                 {q_1(v+\frac{i}{2}\gamma(\alpha+1))
                  q_2^{h}(v+\frac{i}{2}\gamma(\alpha-2))}, \nn \db
&\frac{\mfA_1(v)}{\mfa_1(v)}\propto
                \frac{q_1(v+\frac{i}{2}\gamma(\alpha+1))
                      q_1^{h}(v+\frac{i}{2}\alpha \gamma)}
                     {\phi_{+}(v+\frac{i}{2}\alpha \gamma)
                      \phi_{-}(v+\frac{i}{2}\alpha \gamma)},  
\label{fr} \db
&\mfA_0(v)\mfA_1(v)\propto
      \frac{q_1(v+\frac{i}{2}\gamma(\alpha+1))
            q^{h}_2(v+\frac{i}{2}\gamma\alpha)}
           {[\sh(iv+u_N+\frac{\gamma}{2}\alpha)
                  \sh(iv-u_N-\frac{\gamma}{2}\alpha)]^{\frac{N}{2}}
            q_2(v+\frac{i}{2}\gamma(\alpha+2))},\nn\db
&\mfAb_0(v)\mfA_1(v)\propto
        \frac{q^{h}_1(v+\frac{i}{2}\gamma\alpha)
            q_2(v+\frac{i}{2}\gamma\alpha)}
           {[\sh(iv+u_N+\frac{\gamma}{2}\alpha)
                  \sh(iv-u_N-\frac{\gamma}{2}\alpha)]^{\frac{N}{2}}
            q^{h}_2(v+\frac{i}{2}\gamma(\alpha+2))}.\nn
\end{align}
For the largest eigenvalue,
we find  numerically the following analyticity properties:
$q_1(v+\frac{i}{2}\gamma(\alpha+1))$, 
$q_2(v+\frac{i}{2}\gamma(\alpha+2)$ 
($q_1^{h}(v+\frac{i}{2}\alpha\gamma)$, 
$q_2^{h}(v+\frac{i}{2}\gamma(\alpha-2))$)
are analytic and non-zero in the upper (lower) half plane\footnote{
In fact we also observe that $q_1^{h}(v+\frac{i}{2}\alpha\gamma)=
q_1(-v+\frac{i}{2}\gamma(\alpha+1))$ and 
$q_2^{h}(v+\frac{i}{2}\gamma(\alpha-2))=q_2(-v+\frac{i}{2}\gamma(\alpha+2))$
are valid in the largest eigenvalue case and some special 
sub-leading ones.
}, and $\lambda_2(v+\frac{i}{2}\gamma)+\lambda_3(v+\frac{i}{2}\gamma)+
\lambda_4(v+\frac{i}{2}\gamma)$ is analytic and non-zero
in the physical strip $-\frac{\gamma}{2}<\Im v<\frac{\gamma}{2}$.

Because of these analyticity properties, we can take the logarithmic
derivative and perform the Fourier transform on both sides
of~\eqref{fr}.
Explicitly we find
\begin{align}
\widehat{\mfa}_0&=
   \begin{cases}
      -\widehat{q}_1^{h}e^{-k\gamma\alpha}+
       \widehat{q}_2^{h}e^{-k\gamma(\alpha-2)}+
       E(\frac{\gamma}{2}\alpha-u_N)
       &\text{for $k>0$} \\
       -\widehat{q}_2e^{k\gamma(\alpha+2)}
       -E(-\frac{\gamma}{2}(\alpha+2)+u_N)
        &\text{for $k<0$} \\
      -\widehat{q}_1^{h}
      +\widehat{q}_2^{h}
      -\widehat{q}_2
       &\text{for $k=0$}
   \end{cases},                         \nn \db
\widehat{\mfab}_0&=
   \begin{cases}
      -\widehat{q}_2^{h}e^{-k\gamma(\alpha-2)}+
       E(\frac{\gamma}{2}(\alpha+2)-u_N)
       &\text{for $k>0$} \\
       -\widehat{q}_1e^{k\gamma(\alpha+1)}+
        \widehat{q}_2e^{k\gamma(\alpha+2)}-
        E(-\frac{\gamma}{2}\alpha+u_N)
        &\text{for $k<0$} \\
      -\widehat{q}_2^{h}
      -\widehat{q}_1
      +\widehat{q}_2
       &\text{for $k=0$}
   \end{cases},                         \nn \db
\widehat{\mfA}_1-\widehat{\mfa}_1&= 
    \begin{cases}
       \widehat{q}_1^{h}e^{-k\gamma\alpha}-
       E(\frac{\gamma}{2}\alpha-u_N)-
       E(\frac{\gamma}{2}(\alpha+2)-u_N)
       &\text{for $k>0$} \\
       \widehat{q}_1e^{k\gamma(\alpha+1)}+
       E(-\frac{\gamma}{2}\alpha+u_N)+
       E(-\frac{\gamma}{2}(\alpha+2)+u_N)
        &\text{for $k<0$} \\
       \widehat{q}_1^{h}
      +\widehat{q}_1
       &\text{for $k=0$}
   \end{cases},                         \nn \db
\widehat{\mfA}_0+\widehat{\mfA}_1&=     
     \begin{cases}
       \widehat{q}_2^{h}e^{-k\gamma\alpha}-
       E(\frac{\gamma}{2}\alpha+u_N)
       &\text{for $k>0$} \\
        \widehat{q}_1e^{k\gamma(\alpha+1)}-
        \widehat{q}_2e^{k\gamma(\alpha+2)}+
        E(-\frac{\gamma}{2}\alpha-u_N)
        &\text{for $k<0$} \\
       \widehat{q}_2^{h}
      +\widehat{q}_1
      -\widehat{q}_2
       &\text{for $k=0$}
   \end{cases},                         \nn \db
\widehat{\mfAb}_0+\widehat{\mfA}_1&=    
   \begin{cases}
       \widehat{q}_1^{h}e^{-k\gamma\alpha}-
        \widehat{q}_2^{h}e^{-k\gamma(\alpha+2)}-
       E(\frac{\gamma}{2}\alpha+u_N)
       &\text{for $k>0$} \\
        \widehat{q}_2e^{k\gamma\alpha}+
        E(-\frac{\gamma}{2}\alpha-u_N)
        &\text{for $k<0$} \\
       \widehat{q}_1^{h}-
       \widehat{q}_2^{h}+
       \widehat{q}_2
       &\text{for $k=0$}
   \end{cases},                
\label{fourier1}       
\end{align}
where we have adopted the notation $\hat f$ for the Fourier transform
of the logarithmic derivative of a function $f(v)$, 
\begin{equation}
\widehat{f}=\frac{1}{\pi}\int_{-\frac{\pi}{2}}^{\frac{\pi}{2}}
          \left\{\frac{\partial}{\partial v} \ln f(v)\right\}e^{2i k v}dv,
\qquad
\frac{\partial}{\partial v} \ln f(v)=\sum_{k=-\infty}^{\infty}
                       \widehat{f}e^{-2 i k v},
\label{deffourier}
\end{equation}
and $E(x):=i N e^{-2 k x}$.
Applying this procedure to the eigenvalue~\eqref{eigenlam},
we have
\begin{equation}
\widehat{\Lambda}=
   \begin{cases}
       \widehat{q}_1^{h}-
       E((\alpha+1)\gamma-u_N)-
       E(\alpha\gamma+u_N)
       &\text{for $k>0$} \\
       \widehat{q}_1e^{k\gamma(2\alpha+1)}+
       E(-(\alpha+1)\gamma+u_N)+
       E(-\alpha\gamma-u_N)
        &\text{for $k<0$} \\
      \widehat{q}_1^{h}
      +\widehat{q}_1^{h}
       &\text{for $k=0$}
   \end{cases}.
\end{equation}
{}From the last two equations in~\eqref{fourier1},
we find the Fourier modes  $\widehat{q}_1$, $\widehat{q}_2$, 
$\widehat{q}_1^{h}$ and $\widehat{q}_2^{h}$ 
can be expressed in terms of $\widehat{\mfA}_0$,
$\widehat{\mfAb}_0$ and $\widehat{\mfA}_1$.
Substituting them for the first three equations 
in~\eqref{fourier1} and integrating over $v$
after performing the inverse
Fourier transform~\eqref{deffourier},
we arrive at
\begin{align}
\ln\mfa_0(v)&=
\psi_N(v)+k_0*\ln\mfAb_0(v)+k_0*\ln\mfA_1(v)+\beta\mu, 
                            \nn \db           
\ln\mfab_0(v)&=
\psi_N(-v)+\overline{k}_0*\ln\mfA_0(v)+
              \overline{k}_0*\ln\mfA_1(v)+\beta\mu, \nn \db
\ln\mfa_1(v)&=\psi_N(v)+
\psi_N(-v)+\overline{k}_0*\ln\mfA_0(v)+
             k_0*\ln\mfAb_0(v)+
              k_1*\ln\mfA_1(v)+2\beta\mu, \nn\db
\psi_N(v)&=\frac{N}{2}
        \ln\frac{\sh(iv-u_N+\frac{\gamma}{2}\alpha)
                 \sh(iv+u_N-\frac{\gamma}{2}(\alpha+2))}
                {\sh(iv+u_N+\frac{\gamma}{2}\alpha)
                 \sh(iv-u_N-\frac{\gamma}{2}(\alpha+2))}.
\label{nlief}
\end{align}
where the integration contours for the convolutions 
with $\ln\mfA_0$, $\ln\mfAb_0$ and $\ln\mfA_1$
are taken by the straight line
with the imaginary part $-\delta$, 
$+\delta$ and $0$
($\delta$ is arbitrary but fixed
in the range $0<\delta<\gamma/2$), 
respectively. 
In the same way, we obtain
\begin{align}
&\ln\Lambda^{\rm max}(v)=\Psi_N(v)+
                 \zeta*\ln\mfA_0(v)+
               \overline{\zeta}*\ln\mfAb_0(v)+
              (\zeta+\overline{\zeta})*\ln\mfA_1(v), \nn \db
&\Psi_N(v)=\frac{N}{2}
        \ln\frac{\sh(iv+u_N+\gamma(\alpha+1))
                 \sh(iv-u_N-\gamma(\alpha+1))}
                {\sh(iv-u_N+\gamma(\alpha+1))
                 \sh(iv+u_N-\gamma(\alpha+1))}.
\label{eignf}
\end{align}
In these NLIE, we can take the Trotter limit $N\to\infty$
analytically.
\begin{equation}
\lim_{N\to\infty}\psi_N(v)=\psi(v),\qquad
\lim_{N\to\infty}\Psi_N(v)=\Psi(v).
\end{equation}
We have determined the integration constants 
$\beta \mu$, $2\beta\mu$ in~\eqref{nlielgst} 
and 0 in~\eqref{largest} by taking
the limit $|iv|\to\infty$~\footnote{
Note that we take this limit by making use of 
analytic continuation of the NLIEs (see Appendix D).
} and comparing the results with
\begin{equation}
\mfa_0\to\frac{e^{\beta\mu}}{e^{\beta \mu}+1},\quad
\mfab_0\to\frac{e^{\beta\mu}}{e^{\beta \mu}+1}, \quad
\mfa_1\to\frac{e^{2\beta\mu}}{2e^{\beta\mu}+1},
\qquad \Lambda\to (e^{\beta\mu}+1)^2.
\end{equation}
%

%
%
\rnc{\theequation}{C.\arabic{equation}}\setcounter{equation}{0}
%
\section*{Appendix C\,\, Derivation of  NLIE for excited
states}
%
%
%
Here we consider the excited states 
characterizing the correlations
which are discussed in the main text.
As mentioned in section~3, the excited 
states are described by certain 
distribution patterns of additional 
zeros and poles which enter the physical strip.
To derive the NLIEs, we must
explore these zeros and poles.
By taking the integration contours 
in a way that they circumvent the zeros
in the physical strip,
we can derive the NLIEs directly by 
utilizing the above results for the 
largest eigenvalue. 

\begin{figure}[t]
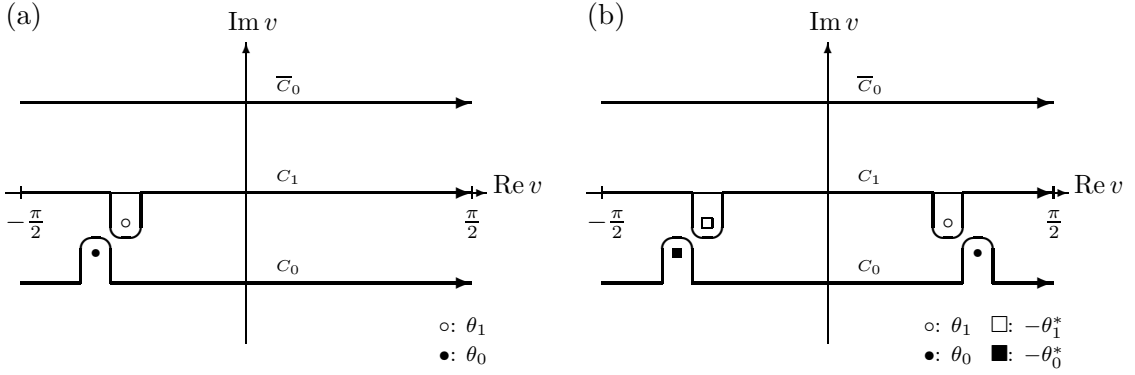

\begin{center}
\contourgr
\contourspt
\end{center}
\caption{Integration contours for (a) the one-particle Green
function
and  (b) the triplet superconducting pair correlations.
Here $\cl_0$, $\overline{\cl}_0$ and $\cl_1$ denote the
contours for $\mfA_0$, $\mfAb_0$ and $\mfA_1$, respectively.}
\label{contourgrpht}
\end{figure}

Taking the integration contours of the 
convolutions in~\eqref{nlief} as in
Fig.~\ref{contourgrpht} (and subsequent figures), we have the same 
algebraic structure of NLIEs as in~\eqref{nlief} without 
additive terms.
This is true at least in the case of the intermediate (differentiated)
version of \eqref{nlief} where $\ln\mfa$, $\ln\mfA$, $\psi_N$ are
replaced by $(\ln\mfa)'$, $(\ln\mfA)'$, $\psi'_N$. 

When integrating
the NLIEs with respect to the argument we want to make sure that the
convolutions are understood in the manner
$k\ast(\ln\mfA)'(v):=\int_{\cl}k(x)(\ln\mfA)'(v-x)dx$ with a
$v$-independent path $\cl$. This can always be achieved with $\cl$ 
starting (ending) at $-\pi/2$ ($+\pi/2$) and suitable deformations
in order to avoid the singularities. The integration with respect to
$v$ and up to an integration constant yields a set of NLIEs which has
the same algebraic structure as in the case of the largest eigenvalue
where now terms 
$k\ast\ln\mfA(v)=\int_{\cl}k(x)\ln\mfA(v-x)dx$ appear with deformed
contours. Alternatively, we can straighten the contours 
and due to Cauchy's theorem obtain additional terms from the singularities 
of the auxiliary functions (these are the ``$\sin$''-terms below).

On top of these terms we obtain a second set of additional terms.
Note that the functions $\ln\mfA_0(v)$ and
$\ln\mfA_1(v)$ are not periodic:
$\ln\mfA_0(\pi/2)-\ln\mfA_0(-\pi/2)=2\pi i$ and
$\ln\mfA_1(\pi/2)-\ln\mfA_1(-\pi/2)=2\pi i$ for
the one-particle Green functions and
$\ln\mfA_0(\pi/2)-\ln\mfA_0(-\pi/2)=4\pi i$ and
$\ln\mfA_1(\pi/2)-\ln\mfA_1(-\pi/2)=4\pi i$ for
the triplet superconducting pair correlations.
Further note that
$k\ast\ln\mfA(v)=\int_{\cl}k(x)\ln\mfA(v-x)dx =\int_{v-\cl}k(v-x)\ln\mfA(x)dx$ which in the general case (non-periodic
$\ln\mfA$) is different from $\int_{-\cl}k(v-x)\ln\mfA(x)dx$. 
For many numerical applications
we actually like to have convolutions of the last type where the auxiliary 
functions are evaluated on fixed (straight) contours. The difference 
of the two versions of convolution integrals can be worked out and
gives rise to the ``$\cos$''-terms below.
By use of
the analytic continuation of the NLIE (see Appendix~D),
one determines the integration constants in a manner similar
to the largest eigenvalue case.

We want to categorize the additional zeros and poles of the
auxiliary functions 
in the following four cases of excitations:
(1)~the one-particle Green functions
and the triplet superconducting pair correlations,
(2)~the density-density (longitudinal)
correlations,
(3)~the transversal spin-spin correlations
and (4)~the singlet superconducting 
pair correlations.
\\\\
{\bf(1) One-particle Green function and
triplet superconducting pair correlations}\\\\
%
%
First we consider the NLIE for the
one-particle Green functions 
$\langle c_{j,\sigma}^{\dagger}c_{i,\sigma}\rangle$.
In this case, the eigenvalues~\eqref{DVF}
belong to the sector $n=N-1$ and $m=N/2-1$.
{}From the numerical calculations with
finite Trotter number $N$, we observe 
additional zeros, namely a zero
 $\theta_0$ of the auxiliary function
$\mfA_0(v)$ and a  zero $\theta_1$ of
$\mfA_1(v)$ located in the lower half plane.
The zeros $\theta_0$ and $\theta_1$
are derived from 
\begin{equation}
\lambda_2\left(\theta_0+\frac{i}{2}\gamma\alpha\right)+
\lambda_3\left(\theta_0+\frac{i}{2}\gamma\alpha\right)+
\lambda_4\left(\theta_0+\frac{i}{2}\gamma\alpha\right)=0,
\quad q^h_1\left(\theta_1+\frac{i}{2}\gamma\alpha\right)=0.
\label{zgrpht}
\end{equation}
%
%
%
%
%
\begin{figure}[h]
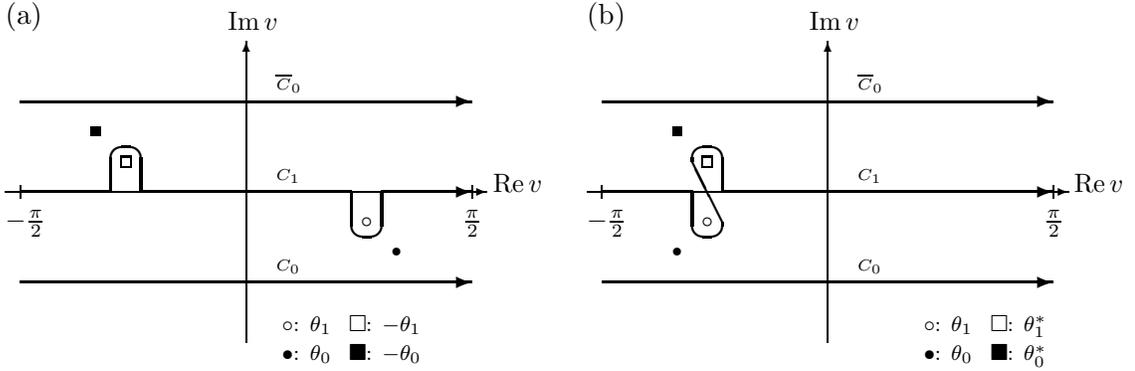

\begin{center}
\contourdd
\contoursl
\end{center}
\caption{Integration contours of the NLIE for
(a) the dominant and (b) sub-dominant contributions to
the density-density correlations.}
\label{contourdd}
\end{figure}
%
%

In the case for the triplet superconducting
pair correlations 
$\langle c_{j+1,\uparrow}^{\dagger}c_{j,\uparrow}^{\dagger}
c_{i+1,\downarrow}c_{i,\downarrow}\rangle$, the corresponding
eigenvalues~\eqref{DVF} describing these
correlations belong to the sector $n=N-2$ and $m=N/2-2$.
Numerically we observe additional parameters:
zeros $\theta_0$, $-\theta^{\ast}_0$ for the 
function $\mfA_0(v)$ and zeros
$\theta_1$, $-\theta^{\ast}_1$
function $\mfA_1(v)$
appear in the lower half plane
(the symbol $*$ denotes complex conjugation).
As for the one-particle Green function,
the zeros $\theta_0$ ($-\theta^{\ast}_0$)
and $\theta_1$ ($\theta^{\ast}_1$)
 satisfy the first
and the second equation in~\eqref{zgrpht},
respectively.
\\\\
%
{\bf(2) Density-density correlations }\\\\
%
%
Let us consider the NLIEs for the density-density
correlations $\langle n_j n_i\rangle$.
The eigenvalues~\eqref{DVF} belong
to the sector $n=N$ and $m=N/2$ which is the
same sector as for the largest eigenvalue.
These correlations are described by the distribution
patterns of a zero $\theta_0$ ($-\theta_0$) and 
a pole $\theta_1$ ($-\theta_1$) of the auxiliary function
$\mfA_0(v)$ ($\mfAb_0(v)$) appearing in the physical
strip.
In this case, these zeros  and poles are derived from 
\begin{align}
&q_2^{h}\left(\theta_0+\frac{i}{2}\gamma\alpha\right)=0,\quad
q_2\left(-\theta_0+\frac{i}{2}\gamma\alpha\right)=0,\nn \db
&q^h_1\left(\theta_1+\frac{i}{2}\gamma\alpha\right)=0,\quad
q_1\left(-\theta_1+\frac{i}{2}\gamma(\alpha+1)\right)=0.
\label{zdd}
\end{align}
In addition,
we can consider the 
sub-dominant terms described by simple 
particle-hole excitations at each Fermi 
point (see section~4).
These sub-dominant terms are determined from
zeros $\theta_0$ ($\theta^{\ast}_0$)  
for $\mfA_0(v)$ ($\mfAb_0(v)$) and
$\theta_1$, $\theta_1^{\ast}$ for 
$\mfA_1(v)$, respectively.
These zeros and poles satisfy eq.~\eqref{zdd} 
(replace  $-\theta_1$ and $-\theta_0$ by
$-\theta_1\to\theta^{\ast}_1$ and $-\theta_0\to\theta^{\ast}_0$,
respectively).
Taking the integration contours as in
Fig.~\ref{contourdd}, we arrive at NLIEs
which have the same form as in~\eqref{nlief}.
For both cases, we take the convolutions
in the ordinary way, i.e. as for the 
largest eigenvalue.\\\\
%
{\bf(3) Transversal spin-spin correlations}\\\\
%
%
%
\begin{figure}[t]
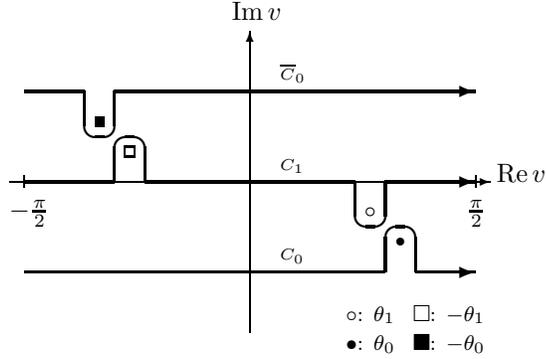

\begin{center}
\contourst
\end{center}
\caption{Integration contours of the NLIE for
the transversal spin-spin correlations.}
\label{contourst}
\end{figure}%
%
%
Let us consider the eigenvalues which
describe the transversal spin-spin correlations 
$\langle\sigma_j^{+}\sigma_i^{-}\rangle$.
The eigenvalues~\eqref{DVF} belong to the
sector $n=N$ and $m=N/2-1$.
Making use of numerical calculations,
we observe  additional zeros $\theta_0$
($-\theta_0$) and poles $\theta_1$ 
($-\theta_1$) of the auxiliary function
$\mfA_0(v)$ ($\mfAb_0(v)$) appear in  the
lower (upper) half plane.
As in case~(1), these zeros and poles get 
close to but never 
cross the real axis in the low temperature limit.
These additional zeros (poles) are derived from
\begin{align}
&\lambda_2\left(\pm\theta_0+\frac{i}{2}\gamma\alpha\right)+
\lambda_3\left(\pm\theta_0+\frac{i}{2}\gamma\alpha\right)+
\lambda_4\left(\pm\theta_0+\frac{i}{2}\gamma\alpha\right)=0,\nn\db
&q^h_1\left(\theta_1+\frac{i}{2}\gamma\alpha\right)=0,\qquad
q_1\left(-\theta_1+\frac{i}{2}\gamma(\alpha+1)\right)=0.
\label{zst}
\end{align}
We derive the NLIEs by taking the contours as
in Fig.~\ref{contourst}.
%
%
\begin{figure}[t]
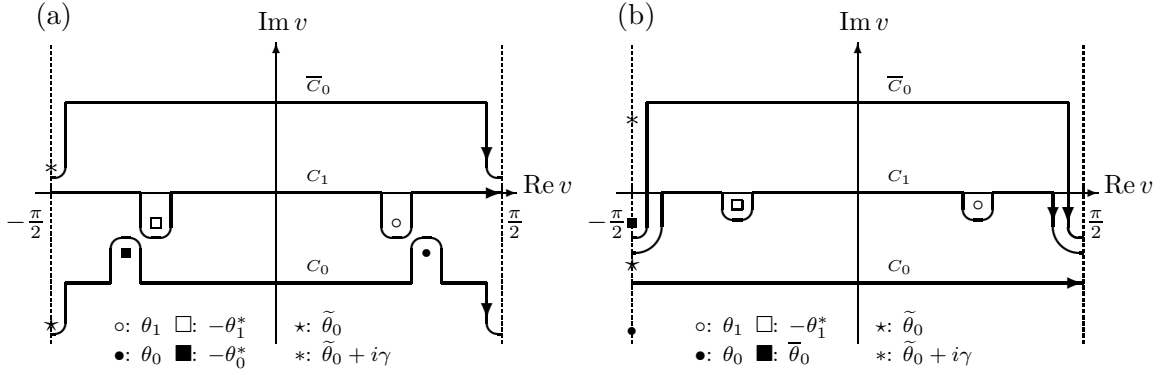

\begin{center}
\contourspsh
\contourspsl
\end{center}
\caption{Integration contours of the NLIE for
the singlet superconducting pair correlations
for $T>T_c$ (a) and for $T<T_c$ (b).}
\label{contoursps}
\end{figure}%
%
%
\\\\
%
{\bf(4) Singlet superconducting pair correlations}\\\\
%
We consider 
the singlet superconducting pair correlations
$\langle c_{j+1,\uparrow}^{\dagger}c_{j,\downarrow}^{\dagger}
c_{i+1,\uparrow}c_{i,\downarrow}\rangle$.
In this case,
the eigenvalues written as~\eqref{DVF} belong
to the sector $n=N-2$ and $m=N/2-1$.
These correlations are described by the distribution
patterns of zeros $\theta_0$, $\overline{\theta}_0$, 
$\widetilde{\theta}_0$ and poles
$\theta_1$, $-\theta^{\ast}_1$ of the auxiliary function
$\mfA_0(v)$.
Numerically we observe that the zero $\widetilde{\theta}_0$ is
located on the $\pm\pi/2$--axis ($\Re v=\pm \pi/2$) and 
moves to the real axis with decreasing temperature. 
At sufficiently high temperatures, the zeros $\theta_0$ and $\thetab_0$ 
are related by $\thetab_0=-\theta^{\ast}_0$ and 
get close to the boundary ($\pm\pi/2$--axis)
with decreasing temperature.
At a certain temperature $T_c$, the zeros $\theta_0$,
$\thetab_0$ and $\theta^{\ast}_0$ satisfy
$\theta_0=\thetab_0=\theta^{\ast}_0$.
Below $T_c$, the relation $\thetab_0=-\theta^{\ast}_0$
is no longer valid, and these zeros are located on
the $\pi/2$--axis  satisfying
the relation $\Im \thetab_0<\Im \widetilde{\theta}_0<\Im \theta_0$.
$\widetilde{\theta}_0$ and $\theta_0$ move upwards (into
$+i$ direction along the $\pi/2$--axis), 
opposite to this $\thetab_0$ moves into $-i$ direction.
In the low temperature limit, $\widetilde{\theta}_0$ comes near
to the real axis but never crosses it.
On the other hand $\theta_0$ enters  the
upper half plane and finally leaves the physical strip.
$\thetab_0$ also leaves the physical strip in the
lower half plane.
Consequently at low temperature, only $\theta_1$ and $-\theta^{\ast}_1$ 
(zeros of $\mfA_1(v)$) are circumvented by the integration contour 
of $\mfA_1(v)$
and these zeros characterize 
the massless pair-excitations (see section~4).

Due to this behavior, we must take different integration contours
for the different regimes $T>T_c$ and $T<T_c$.
These zeros are derived from
\begin{align}
&\lambda_2\left(\theta_0+\frac{i}{2}\gamma\alpha\right)+
\lambda_3\left(\theta_0+\frac{i}{2}\gamma\alpha\right)+
\lambda_4\left(\theta_0+\frac{i}{2}\gamma\alpha\right)=0,\nn \db
&\lambda_2\left(\thetab_0+\frac{i}{2}\gamma\alpha\right)+
\lambda_3\left(\thetab_0+\frac{i}{2}\gamma\alpha\right)+
\lambda_4\left(\thetab_0+\frac{i}{2}\gamma\alpha\right)=0,\nn \db
&q^{h}_2(\widetilde{\theta_0})=0, \quad
q^h_1\left(\theta_1+\frac{i}{2}\gamma\alpha\right)=0,\quad
q_1\left(-\theta_1+\frac{i}{2}\gamma(\alpha+1)\right)=0.
\label{zsps}
\end{align}
We derive the NLIEs by taking the integration contours
as in Fig.~\ref{contoursps}a  for the case $T>T_c$
and as in Fig.~\ref{contoursps}b for the case $T<T_c$ .\\\\
{\bf Additional terms in the NLIEs}\\\\
As mentioned above, we obtain the same structure of NLIEs as
in the case of the largest eigenvalue by taking the integration
contours of the convolutions in a way that they circumvent
the zeros.
%
%
To solve the NLIEs numerically, we are forced to use straight
integration contours. Due to Cauchy's theorem in this
formulation we obtain certain additional terms in the NLIEs.
In the remainder we list the additional contributions to the driving
terms in \eqref{nliegeneral} and \eqref{eigengeneral}.\\\\
%
{\bf (i) One-particle Green function}\\\\
%
%
First we consider the one-particle Green function.
The corresponding NLIEs are characterized by the zeros
$\theta_0$ and $\theta_1$.
In the low temperature limit, these parameters come near to the
real axis but never enter the upper half plane.
{}From \eqref{zgrpht}, 
we find that these parameters satisfy the subsidiary conditions
\begin{alignat}{3}
\mfA_0(\theta_0)&=\mfAb_0(\theta_0)=0,
&\qquad& \mfA_0(\theta_1)&=\infty, \nn \\
\mfA_1(\theta_1)&=0,&\qquad& \mfA_1(\theta_0)&=\infty.
\label{subgr}
\end{alignat}

Performing the Fourier transform and employing
Cauchy's theorem, we determine the NLIE~\eqref{nliegeneral} with
the following additional terms 
\begin{align}
&\overline{\varphi}^{(0)}_0(v)=
               \begin{cases}
                     \frac{\cos(v-i\gamma+i\delta)}
                     {\cos(v+i\delta)}
                 & \text{for $\Im\theta_0<-\delta$}\\[2.5mm]
                 \frac{\sin(v-\theta_0-i \gamma)}
                     {\sin(v-\theta_0)}
                & \text{for $-\delta<\Im\theta_0<0$} 
                \end{cases}, \nn\db
&\overline{\varphi}^{(1)}_0(v)=
               \begin{cases}
                 \frac{\cos(v+i\delta)}
                     {\cos(v-i\gamma+i\delta)}
                 & \text{for $\Im\theta_1<-\delta$} \\[2.5mm]
                   \frac{\sin(v-\theta_1)}
                     {\sin(v-\theta_1-i\gamma)}
                & \text{for $-\delta<\Im\theta_1<0$}   
                \end{cases}, \nn \db
&\overline{\varphi}_0(v)=\overline{\varphi}^{(0)}_0(v)
                         \overline{\varphi}^{(1)}_0(v), \quad
\varphi_0(v)=\frac{\sin(v-\theta_1+i\gamma)}{\sin(v-\theta_1)},\quad
\varphi_1(v)=\varphi_0(v)\overline{\varphi}_0(v).
\label{addgr}
\end{align}
The corresponding eigenvalues $\Lambda(v)$ are written 
as~\eqref{eigengeneral} with the term $\chi(v)$
(split into two terms $\chi(v)=\chi^{(0)}(v)\chi^{(1)}(v)$),
\begin{align}
&\chi(v)=\chi^{(0)}(v)\chi^{(1)}(v), \nn \db
&\chi^{(0)}(v)=\begin{cases}
    \frac{\cos(v+\frac{i}{2}\gamma \alpha+i\delta)}
           {\cos(v-\frac{i}{2}\gamma(\alpha + 2)+i\delta)}
      &\text{for $\Im\theta_0<-\delta$} \\[2.5mm]
  \frac{\sin(v-\theta_0+\frac{i}{2}\gamma \alpha)}
       {\sin(v-\theta_0-\frac{i}{2}\gamma(\alpha + 2))}
       &\text{for $-\delta<\Im\theta_0<0$}              
\end{cases}, \nn \db
&\chi^{(1)}(v)= \frac{\sin(v-\theta_1-\frac{i}{2}\alpha\gamma)}
     {\sin(v-\theta_1+\frac{i}{2}\gamma(\alpha+2))} \times
     \begin{cases}
       \frac{ \cos(v-\frac{i}{2}\gamma(\alpha + 2)+i\delta)}
        {\cos(v+\frac{i}{2}\gamma \alpha+i\delta)}
      &\text{for $\Im\theta_1<-\delta$} \\[2.5mm]
  \frac{\sin(v-\theta_1-\frac{i}{2}\gamma(\alpha + 2))}
      {\sin(v-\theta_1+\frac{i}{2}\alpha\gamma)}
      &\text{for  $-\delta<\Im\theta_1<0$}               
\end{cases}.
\label{addeigengr}
\end{align}
In fact the NLIE~\eqref{nliegeneral} with the additional
terms~\eqref{addgr} have two solutions. These two
solutions are related to each other by taking the
mirror image with respect to the imaginary axis.
Correspondingly, the eigenvalues $\Lambda(0)$ are
doubly degenerate in magnitude and they are complex
conjugate to each other.
The oscillatory behavior of the one-particle
Green function (referred to as $k_{\rm F}$--oscillation
at zero temperature) derives from this degeneracy.
The correlation length for the one-particle
Green function is given 
through formula~\eqref{correlation}.
\\\\
%
%
{\bf (ii) Triplet superconducting pair correlations}\\\\
%
We determine the additional terms for 
the triplet superconducting pair correlations.
As mentioned above, these correlations are characterized 
by the additional zeros $\theta_0$ and $\theta_1$.
In the low temperature limit,
these zeros come near the real axis
but never enter the upper half plane.
{}From \eqref{zgrpht}, one finds that these zeros satisfy
\begin{alignat}{3}
\mfA_0(\theta_0)&=\mfA_0(-\theta^{\ast}_0)=
\mfAb_0(\theta_0)=\mfAb_0(-\theta^{\ast}_0)=0,&\quad&
\mfA_0(\theta_1)&=\mfA_0(-\theta^{\ast}_1)=\infty, \nn \db
\mfA_1(\theta_1)&=\mfA_1(-\theta^{\ast}_1)=0,&\quad&
\mfA_1(\theta_0)&=\mfA_1(-\theta^{\ast}_0)=\infty.
\end{alignat}

Performing the Fourier transform and employing
Cauchy's theorem, we determine the NLIE~\eqref{nliegeneral} with
the following additional terms
\begin{align}
&\overline{\varphi}^{(0)}_0(v)=
               \begin{cases}
                     \frac{\cos^2(v-i\gamma+i\delta)}
                     {\cos^2(v+i\delta)}
                 & \text{for $\Im\theta_0<-\delta$} \\[2.5mm]
                   \frac{\sin(v-\theta_0-i \gamma)
                     \sin(v+\theta^{\ast}_0-i \gamma)}
                     {\sin(v-\theta_0)
                      \sin(v+\theta^{\ast}_0)}
                & \text{for $-\delta<\Im\theta_0<0$}
                 \end{cases}, \nn \db
&\overline{\varphi}^{(1)}_0(v)=
               \begin{cases}
                    \frac{\cos^2(v+i\delta)}
                     {\cos^2(v-i\gamma+i\delta)}
                 & \text{for $\Im\theta_1<-\delta$} \\[2.5mm]
                    \frac{\sin(v-\theta_1)
                          \sin(v+\theta^{\ast}_1)}
                     {\sin(v-\theta_1-i\gamma)
                      \sin(v+\theta^{\ast}_1-i\gamma)}
                & \text{for $-\delta<\Im\theta_1<0$} 
                \end{cases}, \nn \db
&\overline{\varphi}_0(v)=\overline{\varphi}^{(0)}_0(v)
                         \overline{\varphi}^{(1)}_0(v), \quad
\varphi_0(v)=\frac{\sin(v-\theta_1+i\gamma)
                   \sin(v+\theta^{\ast}_1+i\gamma)}
                  {\sin(v-\theta_1)\sin(v+\theta^{\ast}_1)}, \nn \db
&\varphi_1(v)=\varphi_0(v)\overline{\varphi}_0(v).
\label{addpt}
\end{align}
The corresponding eigenvalues $\Lambda(v)$ are written 
as~\eqref{eigengeneral} with the term $\chi(v)$,
\begin{align}
&\chi(v)=\chi^{(0)}(v)\chi^{(1)}(v), \nn \db
&\chi^{(0)}(v)=\begin{cases}
    \frac{\cos^2(v+\frac{i}{2}\gamma \alpha+i\delta)}
           {\cos^2(v-\frac{i}{2}\gamma(\alpha + 2)+i\delta)}
      &\text{for $\Im\theta_0<-\delta$} \\[2.5mm]
  \frac{\sin(v-\theta_0+\frac{i}{2}\gamma \alpha)
        \sin(v+\theta^{\ast}_0+\frac{i}{2}\gamma \alpha)}
       {\sin(v-\theta_0-\frac{i}{2}\gamma(\alpha + 2))
        \sin(v+\theta^{\ast}_0-\frac{i}{2}\gamma(\alpha + 2))}
       &\text{for $-\delta<\Im\theta_0<0$}              
\end{cases},\db
&\chi^{(1)}(v)=\frac{\sin(v-\theta_1-\frac{i}{2}\alpha\gamma)
      \sin(v+\theta^{\ast}_1-\frac{i}{2}\alpha\gamma)}
     {\sin(v-\theta_1+\frac{i}{2}\gamma(\alpha+2))
      \sin(v+\theta^{\ast}_1+\frac{i}{2}\gamma(\alpha+2))} \nn \\
   & \qquad\quad \times
     \begin{cases}
      \frac{ \cos^2(v-\frac{i}{2}\gamma(\alpha + 2)+i\delta)}
        {\cos^2(v+\frac{i}{2}\gamma \alpha+i\delta)}
      &\text{for $\Im\theta_1<-\delta$} \\[2.5mm]
  \frac{\sin(v-\theta_1-\frac{i}{2}\gamma(\alpha + 2))
        \sin(v+\theta^{\ast}_1-\frac{i}{2}\gamma(\alpha + 2))}
      {\sin(v-\theta_1+\frac{i}{2}\alpha\gamma)
       \sin(v+\theta^{\ast}_1+\frac{i}{2}\alpha\gamma)}       
      &\text{for  $-\delta<\Im\theta_1<0$}               
\end{cases}.
\label{addeigenpt}
\end{align}
\\\\
%
{\bf (iii) Density-density (longitudinal spin-spin)
correlations}\\\\
%
Next we determine the NLIE for the density-density 
correlations.
As the critical exponents of the density-density correlation 
functions are identical to those of the longitudinal
spin-spin correlations $\langle\sigma_j^{z}\sigma_i^{z}\rangle$,
the resultant NLIEs are also identical.
{}From~\eqref{zdd},
we find these zeros satisfy the subsidiary
conditions~\eqref{subdd1}.
\begin{equation}
\mfA_0(\theta_0)=\mfAb_0(-\theta_0)=0,\quad
\mfA_0(\theta_1)=\mfAb_0(-\theta_1)=\infty,\quad
\mfA_1(\pm \theta_1)=0.
\label{subdd1}
\end{equation}

Correspondingly, the additional terms in the 
NLIE~\eqref{nliegeneral} read
\begin{align}
&\overline{\varphi}^{(0)}_0(v)=
               \begin{cases}
                   \frac{\sin(v-\theta_0)\cos(v-i\gamma+i\delta)}
                    {\sin(v-\theta_0-i \gamma)\cos(v+i\delta)}
                & \text{for $\Im\theta_0<-\delta$} \\[2.5mm]
                   1
                 & \text{for $-\delta<\Im\theta_0$}
                \end{cases}, \nn \db
&\overline{\varphi}^{(1)}_0(v)= \frac{\sin(v+\theta_1-i \gamma)}
                     {\sin(v+\theta_1)} \times
               \begin{cases}
                   \frac{\cos(v+i\delta)}
                         {\cos(v-i\gamma+i\delta)}
                 & \text{for $\Im\theta_1<-\delta$} \\[2.5mm]
                    \frac{\sin(v-\theta_1)}
                     {\sin(v-\theta_1-i\gamma)}
                  & \text{for $-\delta<\Im\theta_1<0$}          
                \end{cases}. \nn \db
&\overline{\varphi}_0(v)=\overline{\varphi}^{(0)}_0(v)
                        \overline{\varphi}^{(1)}_0(v), \quad
\varphi_0(v)=\overline{\varphi}_0(-v),\quad
\varphi_1(v)=\varphi_0(v)\overline{\varphi}_0(v).
\label{adddd}
\end{align}
The additional term $\chi(v)$ in~\eqref{eigengeneral}
is determined from
\begin{align}
&\chi(v)=\chi^{(0)}(v)\chi^{(1)}(v), \nn \db
&\chi^{(0)}(v)=\begin{cases}
      \frac{\sin(v-\theta_0-\frac{i}{2}\gamma(\alpha + 2))
        \sin(v+\theta_0+\frac{i}{2}\gamma(\alpha + 2))
        \cos(v+\frac{i}{2}\gamma \alpha+i\delta)
        \cos(v-\frac{i}{2}\gamma \alpha-i\delta)}
      {\sin(v-\theta_0+\frac{i}{2}\gamma \alpha)
      \sin(v+\theta_0-\frac{i}{2}\gamma \alpha)
      \cos(v-\frac{i}{2}\gamma(\alpha + 2)+i\delta)
        \cos(v+\frac{i}{2}\gamma(\alpha + 2)-i\delta)}
      &\text{for  $\Im\theta_0<-\delta$}         \\[2.5mm]    
      1
      &\text{for $-\delta<\Im\theta_0$}
\end{cases}, \nn \db
&\chi^{(1)}(v)= \frac{\sin(v-\theta_1-\frac{i}{2}\alpha\gamma)
        \sin(v+\theta_1+\frac{i}{2}\alpha\gamma)}
      {\sin(v-\theta_1+\frac{i}{2}\gamma(\alpha + 2))
       \sin(v+\theta_1-\frac{i}{2}\gamma(\alpha + 2))} \nn \\
     &\qquad\quad\times
      \begin{cases}
      \frac{ \cos(v-\frac{i}{2}\gamma(\alpha + 2)+i\delta)
             \cos(v+\frac{i}{2}\gamma(\alpha + 2)-i\delta)}
           { \cos(v+\frac{i}{2}\gamma \alpha+i\delta)
             \cos(v-\frac{i}{2}\gamma \alpha-i\delta)}
      &\text{for $\Im\theta_1<-\delta$}       \\[2.5mm]        
  \frac{ \sin(v-\theta_1-\frac{i}{2}\gamma(\alpha + 2))
         \sin(v+\theta_1+\frac{i}{2}\gamma(\alpha + 2))}
      {\sin(v-\theta_1+\frac{i}{2}\alpha\gamma)
       \sin(v+\theta_1-\frac{i}{2}\alpha\gamma)}
      &\text{for $-\delta<\Im\theta_1<0$} 
\end{cases}. 
\label{addeignedd}
\end{align}
In the low temperature limit,
$\theta_1$ 
gets close to the real axis but does not 
cross it.
In contrast, $\theta_0$ crosses the real axis
and enters the upper half
plane.
Hence only one additional zero $\theta_1$
appears explicitly in the above NLIE,
which is characteristic of massless excitations 
(see section~4). 
The solutions to the above NLIE have the 
symmetry: $\mfab_0(v)=\mfa_0(-v)$ and
$\mfa_1(v)$ is symmetric with respect
to the imaginary axis.
The NLIEs~\eqref{adddd} have two solutions which are
connected with each other by taking the mirror image
with respect to the imaginary axis.
Hence the  corresponding eigenvalues $\Lambda(0)$
are degenerate in magnitude; they are complex conjugate
to each other.
{}Due to this degeneracy, the correlations have an
oscillation factor. 
At zero temperature,
this oscillatory behavior is the characteristic
$2k_{\rm F}$-oscillation in the density-density correlations.

The sub-dominant terms discussed above
are characterized by another distribution pattern
of zeros and poles.
Consequently, a zero $\theta_0$ ($\theta^{\ast}_0$)
and a pole $\theta_1$ ($\theta^{\ast}_1$) of the 
functions $\mfa_0(v)$ ($\mfab_0(v)$) appear in the 
physical strip. 
{}From~\eqref{zdd} (replace  $-\theta_1$ and $-\theta_0$ by
$-\theta_1\to\theta^{\ast}_1$ and $-\theta_0\to\theta^{\ast}_0$,
respectively), 
we find that these zeros and poles satisfy
the subsidiary conditions
\begin{equation}
\mfA_0(\theta_0)=\mfAb_0(\theta^{\ast}_0)=0,\quad
\mfA_0(\theta_1)=\mfAb_0(\theta^{\ast}_1)=\infty,\quad
\mfA_1(\theta_1)=\mfA_1(\theta^{\ast}_1)=0.
\label{subdd2}
\end{equation}
In the same way as before
one obtains the NLIEs characterized by the following
additional terms 
\begin{align}
&\overline{\varphi}^{(0)}_0(v)=
               \begin{cases}
                    \frac{\sin(v-\theta_0)\cos(v-i\gamma+i\delta)}
                     {\sin(v-\theta_0-i\gamma)\cos(v+i\delta)}
                & \text{for $\Im\theta_0<-\delta$} \\[2.5mm]
                   1
                 & \text{for $-\delta<\Im\theta_0$}
                \end{cases}, \nn \db
&\overline{\varphi}^{(1)}_0(v)=\frac{\sin(v-\theta^{\ast}_1-i\gamma)}
                     {\sin(v-\theta^{\ast}_1)}\times
               \begin{cases}
                     \frac{\cos(v+i\delta)}
                         {\cos(v-i\gamma+i\delta)}
                  & \text{$\Im\theta_1<-\delta$}     \\[2.5mm]
                   \frac{\sin(v-\theta_1)}
                     {\sin(v-\theta_1-i\gamma)}
                 & \text{for for $-\delta<\Im\theta_1<0$} 
                \end{cases}, \nn \db
&\overline{\varphi}_0(v)=\overline{\varphi}^{(0)}_0(v)
                         \overline{\varphi}^{(1)}_0(v), \quad
\varphi_0(v)=\overline{\varphi}^{\ast}_0(v), \quad
\varphi_1(v)=\varphi_0(v)\overline{\varphi}_0(v).
\label{addsl}
\end{align}
The term $\chi(v)$ in the eigenvalues~\eqref{eigengeneral}
is written as
\begin{align}
&\chi(v)=\chi^{(0)}(v)\chi^{(1)}(v), \nn \db
&\chi^{(0)}(v)=\begin{cases}
     \frac{\sin(v-\theta_0-\frac{i}{2}\gamma(\alpha + 2))
        \sin(v-\theta^{\ast}_0+\frac{i}{2}\gamma(\alpha + 2))
         \cos(v+\frac{i}{2}\gamma \alpha+i\delta)
        \cos(v-\frac{i}{2}\gamma \alpha-i\delta)}
       {\sin(v-\theta_0+\frac{i}{2}\gamma \alpha)
        \sin(v-\theta^{\ast}_0-\frac{i}{2}\gamma \alpha)
        \cos(v-\frac{i}{2}\gamma(\alpha + 2)+i\delta)
        \cos(v+\frac{i}{2}\gamma(\alpha + 2)-i\delta)}
      &\text{for $\Im\theta_0<-\delta$}       \\[2.5mm]        
      1
      &\text{for $-\delta<\Im\theta_0$}
\end{cases}, \nn \db
&\chi^{(1)}(v)= \frac{\sin(v-\theta_1-\frac{i}{2}\alpha\gamma)
        \sin(v-\theta^{\ast}_1+\frac{i}{2}\alpha\gamma)}
      {\sin(v-\theta_1+\frac{i}{2}\gamma(\alpha+2))
       \sin(v-\theta^{\ast}_1-\frac{i}{2}\gamma(\alpha+2))} \nn \\
   &\qquad\quad  \times
     \begin{cases}
        \frac{ \cos(v-\frac{i}{2}\gamma(\alpha + 2)+i\delta)
             \cos(v+\frac{i}{2}\gamma(\alpha + 2)-i\delta)}
           { \cos(v+\frac{i}{2}\gamma \alpha+i\delta)
             \cos(v-\frac{i}{2}\gamma \alpha-i\delta)}
     &\text{for $\Im\theta_1<-\delta$}    \\[2.5mm]
     \frac{\sin(v-\theta_1-\frac{i}{2}\gamma(\alpha + 2))
        \sin(v-\theta^{\ast}_1+\frac{i}{2}\gamma(\alpha + 2))}
      {\sin(v-\theta_1+\frac{i}{2}\alpha\gamma)
       \sin(v-\theta^{\ast}_1-\frac{i}{2}\alpha\gamma)}
      &\text{for $-\delta<\Im\theta_1<0$}
\end{cases}.
\label{addeigensl}
\end{align}
At low temperatures, the parameter $\theta_1$
(zero of the function $\mfA_1(v)$) 
moves toward the real axis but never crosses it,
whereas $\theta_0$ (zero of the function $\mfA_0(v)$) 
crosses the real axis and enters the
upper half plane.
In this limit, the NLIEs depend only 
on the two zeros (of $\mfA_1(v)$) $\theta_1$
and $\theta_1^{\ast}$, which characterize
the particle-hole excitations 
(see section~4).

Note that the solutions to the above NLIEs 
satisfy $\mfa^{\ast}_0(v)=\mfab_0(v)$ and
$\mfa_1(v)$ is symmetric with respect to
the real axis but not to the imaginary axis.
Therefore $\mfa_1(v)$ is real for $v\in{\mathbb{R}}$.
The three NLIE are also reducible as in the dominant
case.
The NLIE with~\eqref{addsl} admits two solutions.
In contrast to the above dominant case, they are
related to each other by taking the mirror image with
respect to the real axis.
The corresponding eigenvalues $\Lambda(0)$ 
are degenerate and are real.
Hence
one finds that the corresponding correlations do not
have oscillatory terms.\\\\
%
%
%
{\bf (iv) Transversal spin-spin correlations}\\\\
%
Let us determine the additional terms of the NLIEs
describing the transversal spin-spin correlations.
As mentioned above, these correlations are characterized by
the zeros and poles satisfying eq.~\eqref{zst}.
Hence we find that these zeros (poles) 
satisfy the following subsidiary conditions.
\begin{alignat}{3}
&\mfA_0(\pm\theta_0)=\mfAb_0(\pm\theta_0)=0,&\qquad&
&&\mfA_0(\theta_1)=\mfAb_0(-\theta_1)=\infty, \nn \\
&\mfA_1(\pm \theta_1)=0,&\qquad&
&&\mfA_1(\pm \theta_0)=\infty.
\label{subst}
\end{alignat}
Consequently, the additional terms are written as follows:
\begin{align}
&\overline{\varphi}^{(0)}_0(v)=
               \begin{cases}
                   \frac{\cos(v-i\gamma+i\delta)}
                        {\cos(v+i\delta)}
                 & \text{for $\Im\theta_0<-\delta$}\\[2.5mm]
                    \frac{\sin(v-\theta_0-i\gamma)}
                     {\sin(v-\theta_0)}
                & \text{for $-\delta<\Im\theta_0<0$} 
                \end{cases}, \nn \db
&\overline{\varphi}^{(1)}_0(v)= 
                     \frac{\sin(v+\theta_1-i \gamma)}
                     {\sin(v+\theta_1)}\times
               \begin{cases}
                   \frac{\cos(v+i\delta)}
                        {\cos(v-i\gamma+i\delta)}
                 & \text{for $\Im\theta_1<-\delta$}  \\[2.5mm]
                    \frac{\sin(v-\theta_1)}
                     {\sin(v-\theta_1-i\gamma)}
                  & \text{for $-\delta<\Im\theta_1<0$}         
                \end{cases}, \nn \db
&\overline{\varphi}_0(v)=\overline{\varphi}^{(0)}_0(v)
                         \overline{\varphi}^{(1)}_0(v), \quad
\varphi_0(v)=\overline{\varphi}_0(-v), \quad
\varphi_1(v)=\varphi_0(v)\overline{\varphi}_0(v).
\label{addst}
\end{align}
The term $\chi(v)$ in the eigenvalue~\eqref{eigengeneral}
is written as
\begin{align}
&\chi(v)=\chi^{(0)}(v)\chi^{(1)}(v) \nn \db
&\chi^{(0)}(v)=\begin{cases}
  \frac{\cos(v+\frac{i}{2}\gamma \alpha+i\delta)
        \cos(v-\frac{i}{2}\gamma \alpha-i\delta)}
       {\cos(v-\frac{i}{2}\gamma(\alpha + 2)+i\delta)
        \cos(v+\frac{i}{2}\gamma(\alpha + 2)-i\delta)}
      &\text{for $\Im\theta_0<-\delta$}    \\[2.5mm]
  \frac{\sin(v-\theta_0+\frac{i}{2}\gamma \alpha)
        \sin(v+\theta_0-\frac{i}{2}\gamma \alpha)}
       {\sin(v-\theta_0-\frac{i}{2}\gamma(\alpha + 2))
        \sin(v+\theta_0+\frac{i}{2}\gamma(\alpha + 2))}
      &\text{for  $-\delta<\Im\theta_0<0$}              
\end{cases}, \nn \db
&\chi^{(1)}(v)= \frac{\sin(v-\theta_1-\frac{i}{2}\alpha\gamma)
      \sin(v+\theta_1+\frac{i}{2}\alpha\gamma)}
      {\sin(v-\theta_1+\frac{i}{2}\gamma(\alpha+2))
       \sin(v+\theta_1-\frac{i}{2}\gamma(\alpha+2))}\nn \\
&\qquad\quad\times
      \begin{cases}
      \frac{ \cos(v-\frac{i}{2}\gamma(\alpha + 2)+i\delta)
             \cos(v+\frac{i}{2}\gamma(\alpha + 2)-i\delta)}
           { \cos(v+\frac{i}{2}\gamma \alpha+i\delta)
             \cos(v-\frac{i}{2}\gamma \alpha-i\delta)}
      &\text{for $\Im\theta_1<-\delta$}   \\[2.5mm]
  \frac{\sin(v-\theta_1-\frac{i}{2}\gamma(\alpha + 2))
        \sin(v+\theta_1+\frac{i}{2}\gamma(\alpha + 2))}
      {\sin(v-\theta_1+\frac{i}{2}\alpha\gamma)
       \sin(v+\theta_1-\frac{i}{2}\alpha\gamma)}
      &\text{for $-\delta<\Im\theta_1<0$}              
\end{cases}.
\label{addeigenst}
\end{align}

As in the largest eigenvalue case,
the solutions $\mfa_0(v)$ and $\mfab_0(v)$
have the symmetry $\mfab_0(v)=\mfa_0(-v)$.
Therefore one reduces the above set of 
three NLIEs to the set of only two NLIEs.
In addition,
one finds that the NLIEs~\eqref{addst} admit two
solutions which are related to each other by taking
the mirror image with respect to the
imaginary axis.
Thus the corresponding eigenvalues $\Lambda(0)$ 
are degenerate in magnitude
and are complex conjugate.
Hence, oscillatory behavior of the correlation 
functions  is observed,
referred to as $2k_{\rm F}$--oscillation for zero temperature.
\\\\
\\
%
{\bf (v) Singlet superconducting pair correlations}\\\\
%
Finally we consider the NLIE for the singlet superconducting
pair correlations characterizing the above mentioned additional
zeros (poles).
{}From~\eqref{zsps},
one sees that these zeros and poles satisfy the subsidiary conditions
\begin{align}
\mfA_0(\theta_0)&=\mfA_0(\overline{\theta}_0)=
\mfA_0(\widetilde{\theta}_0)=\mfAb_0(\theta_0)=
\mfAb_0(\overline{\theta}_0)=0,\nn \db
\mfA_0(\theta_1)&=\mfA_0(-\theta^{\ast}_1)=
\mfAb_0(\theta_1) =\mfAb_0(-\theta^{\ast}_1)=
\mfAb_0(\widetilde{\theta}_0+i\gamma)=\infty, \nn \db
\mfA_1(\theta_1)&=\mfA_1(-\theta^{\ast}_1)=0,\qquad
\mfA_1(\theta_0)=\mfA_1(\overline{\theta}_0)=\infty.
\end{align}
The additional terms in 
the NLIE~\eqref{nliegeneral} read explicitly,
\begin{align}
&\overline{\varphi}^{(0)}_0(v)=
               \begin{cases}
                   1
                & \text{for $\Im\theta_0<-\delta$} \\[2.5mm]
                    \frac{\sin(v-\theta_0-i \gamma)\cos(v+i\delta)}
                    {\sin(v-\theta_0)\cos(v-i\gamma+i\delta)}
                 & \text{for $-\delta<\Im\theta_0<0$}\\[2.5mm]
                   \frac{\cos(v+i\delta)\cos(v-i\gamma)}
                    {\cos(v-i\gamma+i\delta)\cos(v)}
                & \text{for $0<\Im\theta_0$}
                \end{cases}, \nn \db
&\overline{\varphi}^{(\overline{0})}_0(v)=
               \begin{cases}
                  1
                & \text{for $\Im\overline{\theta}_0<-\delta$} \\[2.5mm]
                    \frac{\sin(v-\thetab_0-i \gamma)\cos(v+i\delta)}
                         {\sin(v-\thetab_0)\cos(v-i\gamma+i\delta)}
                 & \text{for $-\delta<\Im\overline{\theta}_0<0$}
                \end{cases}, \nn \db
&\overline{\varphi}^{(\widetilde{0})}_0(v)=
               \begin{cases}
                     \frac{\sin(v-\widetilde{\theta}_0)}
                    {\sin(v-\widetilde{\theta}_0-i\gamma)}
                 &  \text{for $\Im\widetilde{\theta}_0<-\delta$} \\[2.5mm]
                     \frac{\cos(v+i\delta)}
                    {\cos(v-i\gamma+i\delta)}
                &\text{for $-\delta<\Im\widetilde{\theta}_0<0$}
                \end{cases}, \nn \db
&\overline{\varphi}^{(1)}_0(v)=
               \begin{cases}
                  1
                 & \text{for $\Im\theta_1<-\delta$} \\[2.5mm]
                    \frac{\sin(v-\theta_1)\sin(v+\theta^{\ast}_1)
                          \cos^2(v-i\gamma+i\delta)}
                     {\sin(v-\theta_1-i\gamma)
                      \sin(v+\theta^{\ast}_1-i\gamma)
                      \cos^2(v+i\delta)}
                  & \text{for $-\delta<\Im\theta_1<0$}           
                \end{cases}, \nn \db
&\varphi^{(0)}_0(v)=
               \begin{cases}
                   1
                & \text{for $\Im\theta_0<0$} \\[2.5mm]
                    \frac{\sin(v-\theta_0+i \gamma)
                          \cos(v)}
                    {\sin(v-\theta_0)\cos(v+i\gamma)}
                 & \text{for $0<\Im\theta_0<\delta$}\\[2.5mm]
                   \frac{\cos(v)\cos(v+i\gamma-i\delta)}
 {\cos(v+i\gamma)\cos(v-i\delta)}                    
                & \text{for $\delta<\Im\theta_0$}
                \end{cases}, \nn\db
&\varphi^{(\widetilde{0})}_0(v)=
               \begin{cases}
                   \frac {\sin(v-\widetilde{\theta}_0-i\gamma)}
                         {\sin(v-\widetilde{\theta}_0)}
                & \text{for $\Im\widetilde{\theta}_0+\gamma<\delta$} \\[2.5mm]
                    \frac{\cos(v-i\delta)}
                        {\cos(v+i\gamma-i\delta)}
                 & \text{for $\delta<\Im\widetilde{\theta}_0+\gamma$}
                \end{cases}, \nn \db
&\varphi^{(1)}_0(v)=\frac{\sin(v-\theta_1+i\gamma)
                      \sin(v+\theta^{\ast}_1+i\gamma)}
                       {\sin(v-\theta_1)\sin(v+\theta^{\ast}_1)}, \quad
\overline{\varphi}_0(v)=\overline{\varphi}^{(0)}_0(v)
                        \overline{\varphi}^{(\overline{0})}_0(v)
                        \overline{\varphi}^{(\widetilde{0})}_0(v)
                        \overline{\varphi}^{(1)}_0(v), \nn \db
&\varphi_0(v)=\varphi^{(0)}_0(v)
             \varphi^{(\widetilde{0})}_0(v)\varphi_0^{(1)}(v),
 \quad
\varphi_1(v)=\varphi_0(v)\overline{\varphi}_0(v).
\label{addphs2}
\end{align}
The additional term $\chi(v)$ in~\eqref{eigengeneral}
is determined from
\begin{align}
&\chi(v)=\chi^{(0)}(v)\chi^{(\overline{0})}(v)\chi^{(\widetilde{0})}(v)
         \chi^{(1)}(v), \nn \db
&\chi^{(0)}(v)=\begin{cases}
      1
      &\text{for  $\Im\theta_0<-\delta$}      \\[2.5mm]
  \frac{\sin(v-\theta_0+\frac{i}{2}\gamma \alpha) 
        \cos(v-\frac{i}{2}\gamma(\alpha+2)+i\delta)}
       {\sin(v-\theta_0-\frac{i}{2}\gamma(\alpha + 2))
        \cos(v+\frac{i}{2}\gamma\alpha+i\delta)}
       &\text{for $-\delta<\Im\theta_0<0$} \\[2.5mm]
       \frac{\sin(v-\theta_0-\frac{i}{2}\gamma \alpha)
              \cos(v-\frac{i}{2}\gamma(\alpha+2)+i\delta)
              \cos(v+\frac{i}{2}\gamma\alpha)
               \cos(v+\frac{i}{2}\gamma(\alpha+2))}
       {\sin(v-\theta_0+\frac{i}{2}\gamma(\alpha + 2))
         \cos(v+\frac{i}{2}\gamma\alpha+i\delta)
         \cos(v-\frac{i}{2}\gamma(\alpha+2))
         \cos(v-\frac{i}{2}\gamma\alpha)}
        &\text{for $0<\Im\theta_0<\delta$} \\[2.5mm]
        \frac{\cos(v-\frac{i}{2}\gamma \alpha-i\delta)
              \cos(v-\frac{i}{2}\gamma(\alpha+2)+i\delta)
              \cos(v+\frac{i}{2}\gamma\alpha)
            \cos(v+\frac{i}{2}\gamma(\alpha+2))}
       {\cos(v+\frac{i}{2}\gamma(\alpha + 2)-i\delta)
         \cos(v+\frac{i}{2}\gamma\alpha+i\delta)
         \cos(v-\frac{i}{2}\gamma(\alpha+2))
          \cos(v-\frac{i}{2}\gamma\alpha)}
        &\text{for $\delta<\Im\theta_0$}
\end{cases}, \nn\db
&\chi^{(\overline{0})}(v)=\begin{cases}
   1
      &\text{for  $\Im\thetab_0<-\delta$} \\[2.5mm]
   \frac{\sin(v-\thetab_0+\frac{i}{2}\gamma \alpha)
         \cos(v-\frac{i}{2}\gamma(\alpha+2)+i\delta)}
       {\sin(v-\thetab_0-\frac{i}{2}\gamma(\alpha + 2))
        \cos(v+\frac{i}{2}\gamma\alpha+i\delta)}
      &\text{for $-\delta<\Im\thetab_0<0$}              
\end{cases}, \nn \db
&\chi^{(\widetilde{0})}(v)=\begin{cases}
      1
      &\text{for  $0<\Im\widetilde{\theta}_0+\gamma<\delta$}      \\[2.5mm]
       \frac{\sin(v-\widetilde{\theta}_0-\frac{i}{2}\gamma(\alpha + 2))
            \cos(v+\frac{i}{2}\gamma(\alpha+2)-i\delta)}
         {\sin(v-\widetilde{\theta}_0+\frac{i}{2}\gamma \alpha)
             \cos(v-\frac{i}{2}\gamma\alpha-i\delta)}
      &\text{for $\Im\widetilde{\theta}_0<-\delta$,
                $\delta<\Im\widetilde{\theta}_0+\gamma$} \\[2.5mm]
       \frac{ \cos(v-\frac{i}{2}\gamma(\alpha+2)+i\delta)
              \cos(v+\frac{i}{2}\gamma(\alpha+2)-i\delta)}
            {\cos(v+\frac{i}{2}\gamma\alpha+i\delta)
             \cos(v-\frac{i}{2}\gamma\alpha-i\delta)}
      &\text{for $-\delta<\Im\widetilde{\theta}_0<0$} \\[2.5mm]
\end{cases}, \nn \db
&\chi^{(1)}(v)= \frac{\sin(v-\theta_1-\frac{i}{2}\alpha\gamma)
        \sin(v+\theta^{\ast}_1-\frac{i}{2}\alpha\gamma)}
      {\sin(v-\theta_1+\frac{i}{2}\gamma(\alpha + 2))
       \sin(v+\theta^{\ast}_1+\frac{i}{2}\gamma(\alpha + 2))} \nn \\
 &\qquad\quad \times
      \begin{cases}
  1
      &\text{for $\Im\theta_1<-\delta$}      \\[2.5mm]
  \frac{\sin(v-\theta_1-\frac{i}{2}\gamma(\alpha + 2))
         \sin(v+\theta^{\ast}_1-\frac{i}{2}\gamma(\alpha + 2))
        \cos^2(v+\frac{i}{2}\gamma\alpha+i\delta)}
      {\sin(v-\theta_1+\frac{i}{2}\alpha\gamma)
       \sin(v+\theta^{\ast}_1+\frac{i}{2}\alpha\gamma)
       \cos^2(v-\frac{i}{2}\gamma(\alpha+2)+i\delta)}
      &\text{for $-\delta<\Im\theta_1<0$}        
\end{cases}.
\label{addeignephs}
\end{align}
%
%
%
%
%
\rnc{\theequation}{D.\arabic{equation}}\setcounter{equation}{0}
%
\section*{Appendix D\,\, Analytic continuation for NLIE}
%
The above derived NLIEs are well-defined only
in the region $-\gamma+\delta<\Im v<0$ for $\ln\mfa_0(v)$,
$0<\Im v<\gamma-\delta$ for $\ln\mfab_0(v)$ and
$-\delta<\Im v<\delta$ for $\ln\mfa_1(v)$.
To determine the integration constants (which should be
determined from the limit $i|v|\to\infty$)
and the additional zeros for the excited states,
we must extend the region by using the analytic 
continuation.
We can achieve this by applying Cauchy's theorem to
the convolutions.
As a consequence, the additional terms appear due to
the residues of the kernels.
The resultant equations can be written explicitly as follows.
\begin{align}
\ln\mfa_0(v)&=
\psi(v)+k_0\stackrel{\overline{C}_0}{*}
           \ln\mfAb_0(v)+k_0\stackrel{C_1}{*}
                           \ln\mfA_1(v)+\beta\mu-
                           \varphi^{\rm c}_0(v), 
                            \nn \db           
\ln\mfab_0(v)&=
\psi(-v)+\overline{k}_0\stackrel{C_0}{*}\ln\mfA_0(v)+
              \overline{k}_0\stackrel{C_1}{*}
                          \ln\mfA_1(v)+\beta\mu-
                           \overline{\varphi}^{\rm c}_0, \nn \db
\ln\mfa_1(v)&=\psi(v)+\psi(-v)+
            \overline{k}_0\stackrel{C_0}{*}\ln\mfA_0(v)+
             k_0\stackrel{\overline{C}_0}{*}
              \ln\mfAb_0(v)+
              k_1\stackrel{C_1}{*}
               \ln\mfA_1(v)+2\beta\mu-\varphi^{\rm c}_1(v), 
\end{align}
where
\begin{align}
&\varphi^{\rm c}_0(v)=
\begin{cases}
\ln\mfAb_0(v+i\gamma)+\ln\mfA_1(v+i\gamma)
&\text{for $\Im v<-\gamma$} \\
\ln\mfAb_0(v+i\gamma)
&\text{for $-\gamma<\Im v<-\gamma+\delta$} \\
0
&\text{for $-\gamma+\delta<\Im v<0$} \\
\ln\mfA_1(v)
&\text{for $0<\Im v<\delta$} \\
\ln\mfAb_0(v)+\ln\mfA_1(v)
&\text{for $\delta<\Im v$} 
\end{cases}, \nn \db
&\overline{\varphi}^{\rm c}_0(v)=
\begin{cases}
\ln\mfA_0(v)+\ln\mfA_1(v)
&\text{for $\Im v<-\delta$} \\
\ln\mfA1(v)
&\text{for $-\delta<\Im v<0$} \\
0
&\text{for $0<\Im v<\gamma-\delta$} \\
\ln\mfA_0(v-i\gamma)
&\text{for $\gamma-\delta<\Im v<\gamma$} \\
\ln\mfA_0(v-i\gamma)+\ln\mfA_1(v-i\gamma)
&\text{for $\gamma<\Im v$} 
\end{cases}, \nn \db
&\varphi^{\rm c}_1(v)=
\begin{cases}
\ln\mfA_0(v)+\ln\mfAb_0(v+i\gamma)+\ln\mfA_1(v+i\gamma)
&\text{for $\Im v<-\gamma$} \\
\ln\mfA_0(v)+\ln\mfAb_0(v+i\gamma)
&\text{for $-\gamma<\Im v<-\gamma+\delta$} \\
\ln\mfA_0(v)
&\text{for $-\gamma+\delta<\Im v<-\delta$} \\
0
&\text{for $-\delta<\Im v<\delta$} \\
\ln\mfAb_0(v)
&\text{for $\delta<\Im v<\gamma-\delta$} \\
\ln\mfAb_0(v)+\ln\mfA_0(v-i\gamma)
&\text{for $\gamma-\delta<\Im v<\gamma$} \\
\ln\mfAb_0(v)+\ln\mfA_0(v-i\gamma)+\ln\mfA_1(v-i\gamma)
&\text{for $\gamma<\Im v$} 
\end{cases}.
\end{align}
Similarly, the NLIE for the eigenvalue can be written as 
follows.
\begin{equation}
\ln\Lambda(v)=\Psi(v)+
                 \zeta*\ln\mfA_0(v)+
               \overline{\zeta}*\ln\mfAb_0(v)+
              (\zeta+\overline{\zeta})*\ln\mfA_1(v)+\chi_0^{\rm c}(v)
              +\overline{\chi}_0^{\rm c}(v)+\chi_1^{\rm c}(v),
\end{equation}
where
\begin{align}
&\chi^{\rm c}_0(v)=
\begin{cases}
\ln\mfA_0(v+\frac{i}{2}\gamma \alpha)
&\text{for $\Im v<-\frac{\gamma}{2}\alpha-\delta$} \\
0
&\text{for $-\frac{\gamma}{2}\alpha-\delta<
             \Im v<\frac{\gamma}{2}(\alpha+2)-\delta$} \\
\ln\mfA_0(v-\frac{i}{2}\gamma (\alpha+2))
&\text{for $\frac{\gamma}{2}(\alpha+2)-\delta<\Im v$},
\end{cases}, \nn \db
&\overline{\chi}^{\rm c}_0(v)=
\begin{cases}
\ln\mfAb_0(v+\frac{i}{2}\gamma(\alpha+2))
&\text{for $\Im v<-\frac{\gamma}{2}(\alpha+2)+\delta$} \\
0
&\text{for $-\frac{\gamma}{2}(\alpha+2)+\delta<
             \Im v<\frac{\gamma}{2}\alpha+\delta$} \\
\ln\mfAb_0(v-\frac{i}{2}\gamma\alpha)
&\text{for $\frac{\gamma}{2}\alpha+\delta<\Im v$},
\end{cases}, \nn \db
&\chi^{\rm c}_1(v)=
\begin{cases}
\ln\mfA_1(v+\frac{i}{2}\gamma \alpha)+
\ln\mfA_1(v+\frac{i}{2}\gamma(\alpha+2))
&\text{for $\Im v<-\frac{\gamma}{2}(\alpha+2)$} \\
\ln\mfA_1(v+\frac{i}{2}\gamma \alpha)
&\text{for $-\frac{\gamma}{2}(\alpha+2)<
             \Im v<-\frac{\gamma}{2}\alpha$} \\
0
&\text{for $-\frac{\gamma}{2}\alpha<
             \Im v<\frac{\gamma}{2}\alpha$} \\
\ln\mfA_1(v-\frac{i}{2}\gamma \alpha)
&\text{for $\frac{\gamma}{2}\alpha<
             \Im v<\frac{\gamma}{2}(\alpha+2)$} \\
\ln\mfA_1(v-\frac{i}{2}\gamma \alpha)+
\ln\mfA_1(v-\frac{i}{2}\gamma (\alpha+2))
&\text{for $\frac{\gamma}{2}(\alpha+2)<\Im v$}.
\end{cases}.
\end{align}
%
%

%
\end{document}